\documentclass[aps,prx,superscriptaddress,twocolumn, secnumarabic, amssymb,nobibnotes, floatfix, 10pt]{revtex4-2}
\pdfoutput=1

\usepackage[T1]{fontenc}
\usepackage{lmodern}
\usepackage[nolist]{acronym}
\usepackage{microtype}
\usepackage{graphicx} 
\usepackage{grffile} 
\usepackage{lipsum} 
\usepackage[english]{babel}
\usepackage{amsmath}
\usepackage{cancel}
\usepackage{bm}
\usepackage{siunitx}

\usepackage{color}
\usepackage[percent]{overpic}
\usepackage{parskip}
\DeclareSIUnit{\ang}{\mbox{\normalfont\AA}}

\graphicspath{{./figs/}}

\begin{document}

\title{Spectral signatures of excess-proton waiting and transfer-path dynamics \\ in aqueous hydrochloric acid solutions}

\author{Florian N. Br\"unig}
\affiliation{Freie Universit\"at Berlin, Department of Physics, 14195 Berlin, Germany}

\author{Manuel Rammler}
\affiliation{Freie Universit\"at Berlin, Department of Physics, 14195 Berlin, Germany}

\author{Ellen M. Adams}
\affiliation{Ruhr Universit\"at Bochum, Department of Physical Chemistry II, 44780 Bochum, Germany}

\author{Martina Havenith}
\affiliation{Ruhr Universit\"at Bochum, Department of Physical Chemistry II, 44780 Bochum, Germany}

\author{Roland R. Netz}
\email[]{rnetz@physik.fu-berlin.de}
\affiliation{Freie Universit\"at Berlin, Department of Physics, 14195 Berlin, Germany}

\date{\today}
\begin{abstract}
Signatures of solvated excess protons in infrared difference absorption spectra, 
such as the continuum band between the water bend and stretch bands, 
have been experimentally known for a long time, 
but the theoretical  basis for linking spectral signatures with the microscopic proton-transfer mechanism  
so far 
relied on normal-mode analysis. 
We analyze the excess-proton dynamics in  ab initio molecular-dynamics simulations  of aqueous hydrochloric acid solutions
by trajectory-decomposition techniques.
The continuum band 
in the  \SIrange{2000}{3000}{cm^{-1 }} range 
is shown to be due to normal-mode oscillations of  temporary H$_3$O$^+$  complexes.
An additional  prominent peak at \SI{400}{cm^{-1}} 
 reports on the  coupling of excess-proton motion
 to the relative vibrations of the  two flanking  water molecules.
The actual  proton transfer between two water molecules, which for large water separations involves
 crossing  of a barrier and thus is not a normal mode, is
characterized by  two characteristic  time scales: Firstly, the  waiting time for transfer to occur 
 in the range of \SIrange{200}{300}{fs},
 which 
 leads to a broad weak shoulder around ~\SI{100}{cm^{-1}},
 consistent with  our experimental THz spectra. 
 Secondly, the  mean duration of a transfer event of about \SI{14}{fs},  which produces a rather well-defined spectral contribution
  around \SI{1200}{cm^{-1}} and agrees in location and width with previous  experimental mid-infrared spectra. 
       
\end{abstract}


\maketitle

\section{Introduction}

The motion of excess protons in aqueous solution is fundamental for many  biological and chemical processes.
The  excess-proton diffusivity  is significantly higher compared to other monovalent cations in water \cite{Marx2006, Agmon2016}, 
since  the excess proton exchanges its identity with water hydrogens during  the diffusion process
\cite{Tuckerman1995, Berkelbach2009}.
Grotthus hypothesized a similar  process over two centuries ago \cite{Grotthus1806, Agmon1995}, 
but a detailed understanding of the proton-transfer dynamics  between  water or other molecules 
 remains difficult to date due to the multitude of time scales involved  
and the only indirect experimental evidence.

\Ac{IR} spectroscopy in the THz and mid-\ac{IR} regimes is a powerful tool to explore the ultrafast dynamics of 
water and aqueous ion solutions. 
For example, the prominent absorption 
peak around \SI{200}{cm^{-1}} of bulk water is dominated by first-solvation-shell dynamics, 
whereas   motion involving the second solvation shell contributes most significantly below \SI{80}{cm^{-1}} (2.4 THz) \cite{Heyden2010,Heyden2012}. 
Furthermore,  so-called `rattling' modes for strongly hydrated ions lead to characteristic absorption features,
while  for weakly hydrated ions  vibrationally induced charge fluctuations are dominant \cite{Schwaab2019,Balos2020}, as suggested  by
 dissecting  simulation spectra  into contributions from different solvation shells \cite{Schienbein2017,Carlson2020}.


\ac{IR} spectroscopy has  
proven particularly useful for the study of   the ultrafast dynamics of 
the  excess proton in aqueous solution \cite{Decka2015, Thaemer2015}.
Due to their low pH value, aqueous \ac{HCl}  solutions are perfect model systems
 to study solvated excess-proton dynamics  in water.
The characteristic continuum band in the \ac{IR} absorption spectrum, located
 between the water-bending mode around  \SI{1650}{cm^{-1}} and the water-stretching mode around \SI{3300}{cm^{-1}},
  has long been known and led to the hypothesis of the Zundel state, i.e. two water molecules symmetrically 
  sharing the excess proton \cite{Zundel1968}.
This model has been  challenged by a contrasting picture, the Eigen state, which is a hydronium ion caged symmetrically
 by  three water molecules \cite{Wicke1954}. 
  Ever since these idealized structures have been proposed, 
  their relative stability has been controversially debated \cite{Komatsuzaki1994, Esser2018, Biswas2017, Dahms2017, Kulig2014, Kulig2013, Decka2015, Fournier2018, Kundu2019, Yu2019a,Carpenter2020, Calio2021}.
{
 It is now known that neither the idealized Zundel nor the Eigen states are realistic structural representations
and that the excess proton mostly resides slightly asymmetrically shared between two water molecules, 
in the `special pair' state, which geometrically can be interpreted as 
a `distorted Zundel' state  or a `distorted Eigen' state 
\cite{Dahms2017,Fournier2018,Calio2020,Kundu2019,Carpenter2020, Calio2021}.}

\begin{figure*}
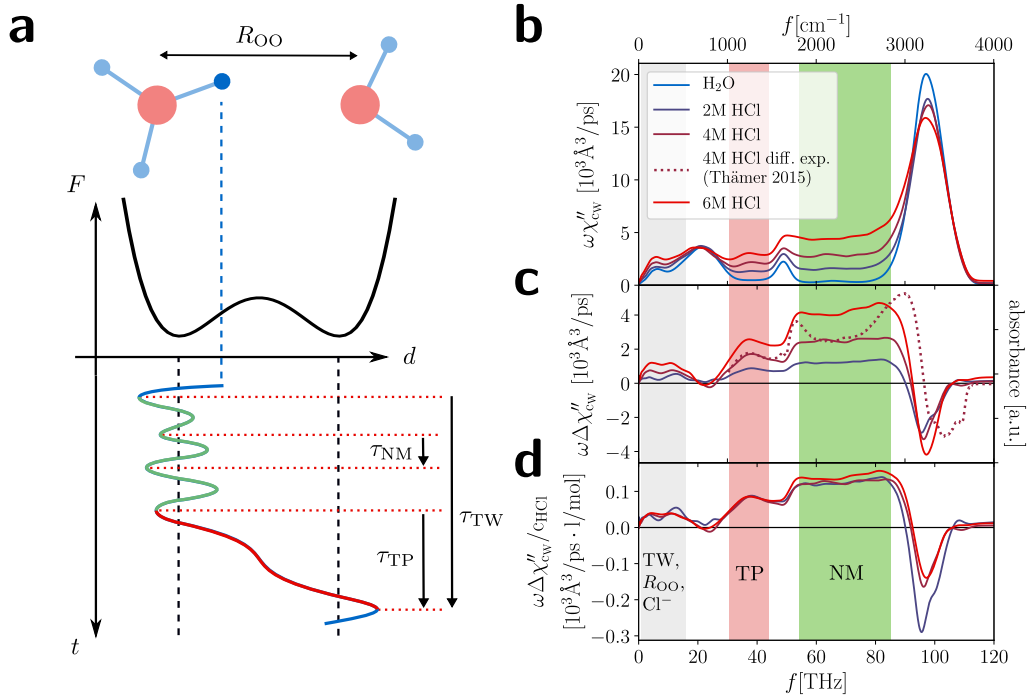

\centering
\begin{overpic}[width=0.75\textwidth]{{/../figs/fig_1_scheme_color_swap_rev}.pdf}
\put(-2,65){\huge \bfseries \sffamily a}
\put(48,65){\huge \bfseries \sffamily b}
\put(48,40){\huge \bfseries \sffamily c}
\put(48,22){\huge \bfseries \sffamily d}
\end{overpic}
\caption{{\bfseries \sffamily Time scales of excess-proton dynamics and simulated absorption spectra.}
{\bfseries \sffamily a:} Schematic trajectory of an excess proton that transfers 
between two water molecules, together with a schematic free energy profile  $F(d)$ that exhibits a barrier and is representative 
of a relatively large oxygen-oxygen separation $R_{\mathrm{OO}}$.  Three time scales characterize the proton trajectory, 
 the normal-mode vibrational period of the solvated transient  H$_3$O$^+$,  $\tau_{\mathrm{NM}}$, 
 the transfer-path time, $\tau_{\mathrm{TP}}$ and the 
 transfer-waiting time, $\tau_{\mathrm{TW}}$, where $\tau_{\rm{TW}} > \tau_{\rm{TP}} > \tau_{\rm{NM}}$. 
  {\bfseries \sffamily b:} Infrared (\acs{IR}) absorption spectra obtained from ab initio \acf{MD} simulations of pure water (blue solid line) and hydrochloric acid (\acs{HCl}) solutions at various concentrations (dark purple: \SI{2}{M}, purple: \SI{4}{M} and red: \SI{6}{M}). The spectra are divided by the water molecular number concentration $c_{\rm W}$.
{\bfseries \sffamily c:} Difference spectra between the three \ac{HCl} spectra and the water spectrum, 
obtained from the spectra in b. The purple dotted line shows an experimental difference spectrum of \ac{HCl} at \SI{4}{M}  \cite{Thaemer2015}, rescaled in height to match the simulation results.
{\bfseries \sffamily d:} The simulated difference spectra (as shown in c) divided by the \ac{HCl}  concentrations $c_{\rm HCl}$. Three distinct spectral regions are shaded in different colors, that are identified with different excess-proton dynamic processes:  transfer waiting  (TW, gray), 
transfer paths (TP, red) and normal modes (NM, green). 
The transfer-waiting time is close to the chloride-ion (Cl$^-$) rattling time and the oxygen vibrational  time in H$_5$O$_2^+$ complexes that is described by the $R_{\rm OO}$ coordinate. 
}
\label{hcl_intro}
\end{figure*}

{
While simulations can reproduce most experimental spectroscopic signatures, 
the understanding of the proton transfer mechanism requires model building  based on and
guided by simulation. 
It is generally accepted  that proton transfer  involves  consecutive transitions between  states that can 
be viewed as more Eigen-like and more Zundel-like
and  have fast interconversion times \cite{Woutersen2006, Kundu2019, Carpenter2020}. 
That the excess proton diffusion involves the crossing of free-energetic  barriers follows from the experimentally
known Arrhenius behavior of the excess proton conductivity\cite{Loewenstein1962, Luz1964, Carpenter2019, Calio2020}.}
Along the lines of the above-mentioned debate on the relative stability of Eigen and Zundel states, 
it remains discussed  whether the Zundel state is the transition state between two Eigen states or the opposite is the case, 
i.e. whether the Eigen  state is the transition state between two Zundel states \cite{Fournier2018, Kundu2019, Calio2021}. 
Theoretical models  for the spectroscopic signatures of the hydrated proton motion so far
relied on normal-mode calculations and have explained many aspects of  experimental 
 linear absorption  \cite{Xu2010,Biswas2016,Napoli2018}  as well as 
 2D \ac{IR} spectra \cite{Yu2019a, Carpenter2020}.
However, normal modes by construction cannot deal with 
the thermally activated transfer of an excess proton over a  free-energy barrier, 
since this corresponds to an unstable mode with a negative free-energy curvature along the  transfer reaction coordinate \cite{Williams1972}.
{ It is clear that such proton-barrier-transfer events  will make a sizable spectroscopic contribution,
since they involve  fast motion of a highly charged object over relatively large distances.
From this follows  that an excess-proton  transition state, which corresponds to a free energy maximum and 
thus occurs with a small probability, nevertheless can make a dominant contribution to the spectrum,
which would lead to characteristic differences between experimental spectra  and  
normal mode theory predictions.
Indeed, it has been noted that the normal-mode spectra computed from instantaneous
configurations do not  explain all experimental  spectral signatures associated with the excess proton in water
\cite{Biswas2017, Yu2019a,Carpenter2020, Calio2021}, in particular of the proton-transfer dynamics \cite{Wang2017}.
In essence, it is not clear with current theoretical methodology what the spectroscopic signature of an unstable mode is and
whether the continuum band stems just  from vibrations in metastable states or whether  transfer reactions
over barriers are involved. 
Thus, a  theoretical approach  that is complementary to normal modes  and can handle  proton-transfer
events that involve free-energy barriers is needed.}

In this study we investigate the excess-proton dynamics in aqueous \ac{HCl} solutions at ambient conditions using 
 ab initio \ac{MD} simulations at the Born-Oppenheimer level and experimental THz/\ac{FTIR} measurements.
Our simulated  \ac{IR} difference absorption spectra  compare well to our experimental data in the THz regime 
as well as to literature data in the mid-\ac{IR} regime.
By projecting the excess-proton dynamics onto the  two-dimensional coordinate system
spanned by  the proton position along the axis connecting the two closest water oxygens $d$ and the oxygen distance $R_{\rm OO}$ \cite{Huggins1971, Marx1999}, the excess-proton  trajectories and their spectral signatures are subdivided into three contributions with distinct time scales, as illustrated in fig.~\ref{hcl_intro}a.
The fastest time scale, $\tau_{\rm NM}$, reflects vibrations when the excess proton transiently forms a solvated H$_3$O$^+$ molecule which  is asymmetrically solvated in a special pair.
It is well captured by a normal-mode description and has been amply discussed in  literature \cite{Wolke2016, Wang2017, Biswas2017, Yu2019a, Carpenter2020}.
The other two  spectral signatures, stemming from proton transfer  events, are the focus of this study.

\begin{figure*}
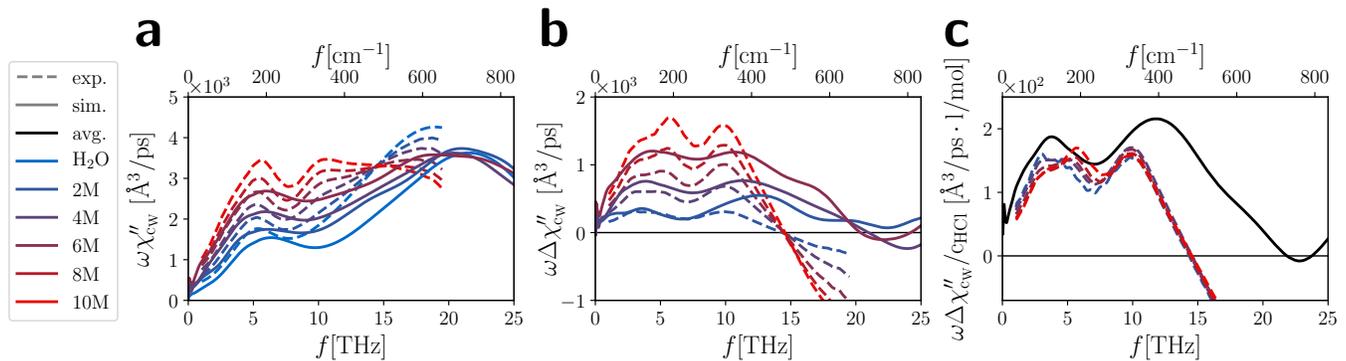

\centering
\begin{overpic}[width=\textwidth]{{/../figs/fig_2_3cols}.pdf}
\put(10,24){\huge \bfseries \sffamily a}
\put(40,24){\huge \bfseries \sffamily b}
\put(70,24){\huge \bfseries \sffamily c}
\end{overpic}
\caption{{\bfseries \sffamily Experimental absorption spectra.}
{\bfseries \sffamily a:} Experimental THz/\acl{FTIR} (THz/\acs{FTIR}) absorption spectra of \ac{HCl} solutions at various concentrations (colored broken lines for \SIrange{2}{10}{M}), compared to literature data of pure water \cite{Bertie1996} (blue broken line).
The experimentally measured extinction spectra have been  converted into energy absorption spectra using the Kramers-Kronig
relation, no amplitude adjustment is used in the comparison with the ab initio spectra (solid lines).
{\bfseries \sffamily b:} Experimental and ab initio \acl{MD} (\ac{MD}) difference spectra derived
from the results  given in a and plotted in the respective colors and line styles.
{\bfseries \sffamily c:} The experimental difference spectra (shown in b) are 
divided by the \ac{HCl} concentration $c_{{\rm HCl}}$ (colored broken lines) and 
 compared  to the  average of the simulated difference spectra after dividing by $c_{{\rm HCl}}$ (black solid line).}
\label{hcl_exp}
\end{figure*}

Let us briefly describe how  proton transfer  gives rise to two distinct
spectroscopically relevant  time scales that cannot be captured by normal-mode analysis.
In general, the transfer of a particle with mass $m$ over
an energy barrier with negative curvature $k<0$  corresponds to an unstable mode. 
The dynamics of such a barrier-crossing  is not characterized  by a vibrational time scale,
 which according to a harmonic oscillator model could erroneously  be written as
 $\tau \simeq \sqrt{|k|/m} \simeq \sqrt{U_0/(L^2 m)}$, 
 where $U_0$ is the barrier height and $L$ is the barrier width,
but rather by two other time scales, namely  the mean transfer-waiting
 time $\tau_{\rm TW}$ and the mean transfer-path time $\tau_{\rm TP}$. 
 $\tau_{\rm TW}$ is the average waiting time before a transfer
 event occurs  and $\tau_{\rm TP}$ is the average time of the actual transfer path over the energy barrier. 
 The former one scales exponentially with the barrier height $U_0$, $\tau_{\rm TW}\sim \exp{(U_0/k_BT)}$ \cite{Kramers1940, Williams1972}, 
 whereas the latter one scales logarithmically with the barrier height $U_0$,
 $\tau_{\rm TP}\sim \log{(U_0/k_BT)}$ \cite{Hummer2004,Chung2009,Kim2015,Cossio2018}.
The time scales $\tau_{\rm TW}$ and $\tau_{\rm TP}$ are  directly obtained from our 
 simulated excess proton  trajectories using a multidimensional  path analysis. 
While the transfer-waiting
 time $\tau_{\rm TW}$  of  aqueous proton-transfer events has been studied  recently \cite{Roy2020}, 
the  identification of both $\tau_{\rm TW}$ and $\tau_{\rm TP}$ in simulated and experimental spectra is a main result of this work. 
We find a mean transfer-waiting  time of 
$\tau_{\rm TW}=$ \SIrange{200}{300}{fs} depending on HCl
concentration, which in our experimental THz spectra shows up as  a broad weak shoulder around ~\SI{100}{cm^{-1}}, 
that is partially overlaid by the absorption due to   rattling chloride anions at about 
 \SI{150}{cm^{-1}} \cite{Schwaab2019, Schienbein2017}. 
 The mean transfer-path time, from simulations obtained as 
 $\tau_{\rm TP}=$ \SI{14}{fs},  produces  a spectroscopic signature around \SI{1200}{cm^{-1}}, which is well captured in experimental mid-\ac{IR} spectra \cite{Thaemer2015, Dahms2017, Fournier2018, Kundu2019, Carpenter2020}.
 {
 Note that in our simulations  the proton transfer becomes barrier-less for small water separation and thus 
 includes the highly anharmonic normal-mode vibration of  Zundel-like configurations, which have also been analyzed theoretically  \cite{Wolke2016, Wang2017, Biswas2017, Yu2019a} .}
  In the THz regime our experimental difference spectra show an additional  prominent peak at \SI{400}{cm^{-1}},
   in good agreement with our simulation data, which is demonstrated  to be caused by   the coupling of the excess proton motion to the relative oscillations of the two flanking   water molecules in transient H$_5$O$_2{}^+$ complexes.
 
  {
Proton transfer events between 
 water molecules are frequently followed by a few immediate back-and-forth transfer events, which is a consequence of non-Markovian effects \cite{Kappler2018} 
 that have to do with the slowly changing solvation structure around the excess proton.
 Although these transfer events are therefore not always productive in the sense that they lead
 to large-scale diffusion of the excess proton, they nevertheless give rise to pronounced 
 experimental spectroscopic signatures and therefore cannot be excluded from the analysis. }
 
   {
While the good agreement between our simulated and experimental   spectra supports  our 
 chosen simulation methodology, it is clear that our classical treatment of  nuclei motion 
is a drastic approximation  and therefore some of the agreement might be due to fortuitous cancellation 
of errors. Interestingly, previous studies found no significant differences between \ac{IR} 
spectra computed from simulations with and without NQEs  below \SI{3000}{cm^{-1}} \cite{Biswas2016, Napoli2018},
which might suggest that  quantum-mechanical zero-point motion influences the excess-proton dynamics
less than  the instantaneous excess-proton distribution.
We discuss quantum nuclear effects and basis set issues in a separate section before the Conclusions and 
in the Methods section. 
There we also compare our simulation
results for proton diffusion coefficients and radial distribution functions with previous reports
and discuss other observables
that have been used in literature to characterize excess-proton transfer dynamics, such as identity correlation functions \cite{Hassanali2013, Arntsen2021, Calio2021}, the hydrogen-bond asymmetry around hydronium ions \cite{Napoli2018}
and the number of hydrogen bonds hydronium ions participate in \cite{Tse2015, Biswas2016, Fischer2019}. 
}

\section{Results and discussion}

\subsection*{Infrared and THz spectra of HCl solutions}

Within linear spectroscopy, the    energy  absorption rate of  incident light  with frequency $f = \omega/(2 \pi)$ 
 is proportional to the imaginary part of the dielectric susceptibility and given by  $\omega \chi''(\omega)$.
\ac{IR} power spectra are obtained from ab initio \ac{MD} simulations of water (blue solid line) and \ac{HCl} 
solutions at three concentrations between \SIrange{2}{6}{M} (purple to red solid lines) and are shown in fig.~\ref{hcl_intro}b.
The spectra are divided by the water molecular number concentration $\omega \chi''_{c_{\rm W}} = \omega \chi'' /c_{\rm W}$.
Simulation details are provided in the Methods section.
All \ac{IR} spectra show the characteristic features of pure water spectra, which are the prominent OH-stretching peak around \SI{3300}{cm^{-1}}, the HOH-bending mode around \SI{1650}{cm^{-1}} and librational modes in the far \ac{IR} regimes between \SI{200}{cm^{-1}} and \SI{800}{cm^{-1}}.
The \ac{IR} spectra of HCl solutions  additionally show a broad continuum between the bending and the stretching peaks, from \SIrange{2000}{3000}{cm^{-1}}, and a broad peak at around \SI{1200}{cm^{-1}}, both of which are commonly interpreted as to reflect the 
excess-proton dynamics \cite{Thaemer2015,Fournier2018,Daldrop2018a}.
Furthermore, additional features are observed below \SI{800}{cm^{-1}}, 
that are shown in fig.~\ref{hcl_exp} in comparison to our experimental THz spectra
and will be discussed further below.

The simulated difference spectra  in fig.~\ref{hcl_intro}c (solid lines) 
clearly demonstrate three distinct regions (color shaded), that relate to distinct  time scales of 
the excess-proton dynamics and will in this work be identified as transfer-waiting (TW, gray), transfer-path (TP, red) 
and normal-mode contributions (NM, green).
We obtain rather  good agreement with the  experimental difference spectrum for  \SI{4}{M} HCl \cite{Thaemer2015}, 
which was scaled to match the height of  the simulated IR \SI{1200}{cm^{-1}} peak, 
see SI section \ref{spectraSection} for a comparison of different 
experimental data.
Our simulated \SI{4}{M} difference spectrum in fig.~\ref{hcl_intro}c does not reproduce  the local maximum of the experimental difference spectra around \SI{1750}{cm^{-1}}, which is interpreted as acid-bend band, i.e. a blue shift of the bending mode in H$_3$O$^+$  compared to water,
and also not the shape of the experimental acid-stretch signature around \SI{3000}{cm^{-1}} \cite{Thaemer2015,Biswas2017,Dahms2017,Fournier2018,Carpenter2018}.
 The reason for this disagreement is unclear, 
 we note that the normalization of spectra when calculating difference spectra is a subtle issue, 
 see  SI section \ref{normalizationSection} for a discussion.
 
Fig.~\ref{hcl_intro}d shows that the three simulated HCl difference spectra divided by the HCl  concentrations $c_{{\rm HCl}}$ are nearly indistinguishable.    {
This clearly indicates that the spectroscopic features are due to  single-proton dynamics and that 
collective proton  effects as well as proton-chloride coupling effects, which would scale non-linearly in the HCl concentration, 
are minor. 
This is an important finding and  justifies our theoretical analysis of single excess-proton motion in this work.
}

\begin{figure*}
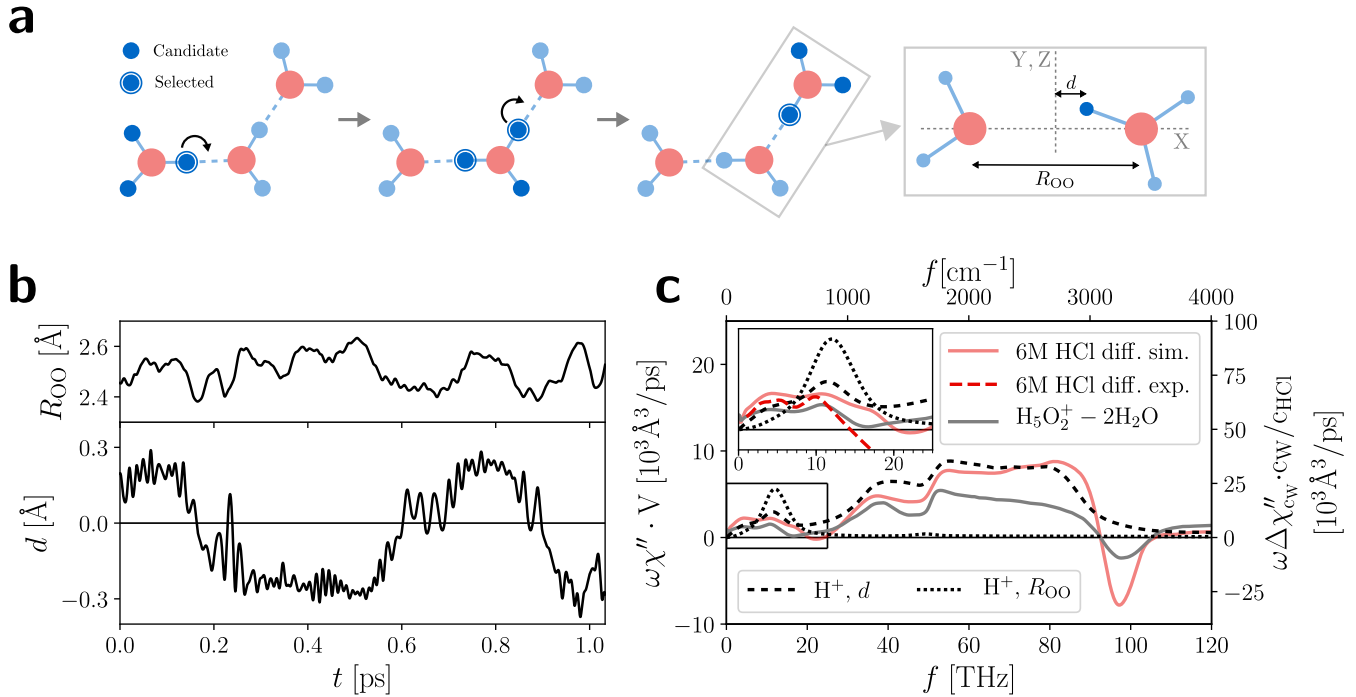

\begin{overpic}[width=\textwidth]{{/../figs/fig_3}.pdf}
\put(0,50){\huge \bfseries \sffamily a}
\put(0,30){\huge \bfseries \sffamily b}
\put(48,30){\huge \bfseries \sffamily c}
\end{overpic}
\caption{{\bfseries \sffamily Excess-proton trajectories and spectra.}
{\bfseries \sffamily a:} Illustration of the method used to extract continuous  excess-proton trajectories from ab initio \acl{MD} (\ac{MD}) simulations. 
From the three protons in a hydronium ion, the one that transfers to a neighboring water is identified as excess proton. 
 To the right  the coordinates $d$ and $R_{\rm OO}$ are defined.
{\bfseries \sffamily b:} Example trajectories of the $d$ and $R_{\rm OO}$ coordinates.
{\bfseries \sffamily c:} Power spectra of the $d$ and $R_{\rm OO}$  coordinates are shown as  broken and   dotted  black lines,
respectively (vertical axis on the left). Also shown are the
simulated difference spectrum of the \SI{6}{M} \ac{HCl} solution, divided by the \ac{HCl} concentration $c_{{\rm HCl}}$ and multiplied by the water concentration $c_{\rm W}$ (red solid line),
 and the simulated difference spectrum between a transient H$_5$O$_2{}^+$ complex in HCl solution and two hydrogen-bonded  water molecules in pure water (gray solid line),
  using the vertical axis on the right.  
  The inset shows a zoom into the THz regime, which additionally
   shows our  \SI{6}{M} experimental THz/\ac{FTIR} difference spectrum (red broken line).}
\label{hcl_intro2}
\end{figure*}

In order to investigate the intermolecular vibrational dynamics  of  water, solvated protons and chloride ions,  
we experimentally measure  THz absorption spectra
 for HCl concentrations of \SI{2}{M}, \SI{4}{M}, \SI{6}{M}, \SI{8}{M} and \SI{10}{M}.
For comparison to simulation data the experimentally measured extinction spectra 
 are converted into energy absorption spectra using the Kramers-Kronig relation,
details are described in the Methods section and in SI section \ref{kkSection}.
The experimental THz/\ac{FTIR} spectra  are shown in fig.~\ref{hcl_exp}a in the range \SIrange{0}{650}{cm^{-1}} (colored broken lines) together with  a literature spectrum of pure water (blue broken line) and are compared to the available simulated spectra (solid lines).
Again, all experimental and simulated spectra are divided by the respective water concentration $c_{\rm w}$.
One notes the  good  agreement between the experimental and simulation spectra below \SI{400}{cm^{-1}}, 
which is noteworthy since the spectral amplitudes are not  rescaled or adjusted. 
However, the reason of  the  disagreement for larger wave numbers  is not clear.
All spectra show a prominent peak at \SI{200}{cm^{-1}}.
Difference spectra of the experimental data with respect to the pure water spectrum  are shown in fig.~\ref{hcl_exp}b (broken lines)
and again compared to the available simulated difference spectra (solid lines). 
Two peaks dominate the difference spectra, one around  \SIrange{150}{200}{cm^{-1}} and one at \SI{400}{cm^{-1}}. 
The experimental difference spectra scale linearly with HCl concentration,
which is demonstrated  in fig.~\ref{hcl_exp}c, where the difference spectra are divided by the HCl
 concentrations $c_{{\rm HCl}}$. For comparison, the simulated difference spectra divided by $c_{\rm HCl}$, 
already  presented in fig.~\ref{hcl_intro}d, are averaged over the three HCl concentrations  and shown as a black solid line.  
The linear scaling of the experimental
spectra with HCl  concentration reconfirms  that  the difference spectra are related to single-ion behavior
and that collective ion effects  are negligible, in agreement with previous observations \cite{Decka2015, Thaemer2015}. 
In essence,  two different processes at \SIrange{150}{200}{cm^{-1}} and \SI{400}{cm^{-1}} are clearly indicated by our experimental
 and simulated spectra and will be interpreted by our spectral  trajectory-decomposition techniques. 
In the remainder we analyze exclusively  the \SI{6}{M} solution, which provides the best proton statistics.

\subsection*{Excess-proton trajectories and spectra}

   {
Excess protons constantly change  their identity as they move through the HCl solution.
Each identity change introduces a spurious discontinuity in the excess-proton trajectory, which 
does not actually correspond to charge transport and therefore  is spectroscopically irrelevant.
In order to extract continuous excess-proton trajectories from our simulations,
we use a  dynamic criterion as illustrated in fig.~\ref{hcl_intro2}a. 
That our extracted excess-proton trajectories are spectroscopically meaningful we will a posteriori
demonstrate by comparison of spectra calculated from excess proton trajectories  with spectra
calculated from the complete simulation system.}
Each proton is assigned to its closest oxygen atom at each time step. 
Whenever three protons are assigned to the same oxygen, thereby forming a hydronium ion, all of them are registered as excess-proton candidates. 
   {
That means, for the generation of continuous excess-proton trajectories, we do not select the hydrogen with the largest separation from the oxygen, which would lead to fast switching of the excess proton identity, the so-called `special pair dance' of hydronium  with its surrounding water molecules \cite{Markovitch2008, Calio2021}.}
Rather, if during the simulation an excess-proton candidate  becomes assigned to a different oxygen
and thus transfers to a neighboring water, 
it is selected as an excess proton for the entire time during which it was part of any hydronium ion \cite{Calio2021}. 
   {
Note that the spectral effects of the rattling of the excess-proton candidates within one hydronium ion, i.e. the `special pair dance', are in some of our calculations below included by taking into account the flanking water molecules in the calculation of spectra, but do not show significant spectral signatures.}
Excess protons that are coordinated with a chloride anion as the second nearest neighbor are neglected from our analysis. 
   {
This does not influence our excess-proton spectra, 
 since even for the highest acid concentration of \SI{6}{M},
  only 5\% of all configurations are of this type, 
  as demonstrated in SI section \ref{radialDistSection}.
Note, however,  that the fraction of protons coordinated with chloride ions  increases significantly at higher concentrations \cite{Baer2014}.}
Our  procedure for calculating continuous excess-proton trajectories  is discussed in further detail in SI section \ref{selectionSection}.

The excess-proton trajectories  are described by the two-dimensional  coordinate system defined within 
local transient H$_5$O$_2{}^+$ complexes  consisting of the excess proton and its 
two nearest water molecules, as illustrated in the right part of fig.~\ref{hcl_intro2}a.
The coordinates are the instantaneous distance between the two oxygen atoms $R_{\rm OO}$ and the excess proton's distance 
 from the midplane  $d$  \cite{Marx1999}.
 The state for $d=0$, where the excess proton is  in the middle between the oxygens, will be later used to  define
  the transition state of the proton transfer between the two flanking water molecules.
Fig.~\ref{hcl_intro2}b shows an example excess-proton trajectory  from our ab intio \ac{MD} simulations in terms
of  the $R_{\rm OO}$ and $d$ coordinates. 
While the motion along the two coordinates is strongly correlated, as we will show later, 
the   $d$ trajectory shows fast oscillatory components that are much weaker  in the   $R_{\rm OO}$ trajectory.

The power spectra of the excess-proton trajectories, averaged over all excess protons in the solution,
 are shown in fig.~\ref{hcl_intro2}c for the 
  $d$ coordinate  as a black broken line and for the $R_{\rm OO}$ coordinate 
 as a black dotted line, in the calculations we 
 assume a  bare charge of $\SI{1}{e}$  for the excess proton (left axis). 
We compare with  the simulated difference spectrum of the \SI{6}{M} \ac{HCl} solution (red solid line),
which is  multiplied by the water concentration $c_{\rm W}$ and divided by HCl  concentration $c_{{\rm HCl}}$
and thus is normalized per excess proton (right axis). 
The qualitative agreement between the two spectra (black broken and red solid lines) is very good up to an overall scaling  factor of roughly four,
 which reflects polarization enhancement due to  neighboring water molecules.  
 The good agreement indicates that the difference spectrum of an  HCl solution
  is proportional to the  spectrum of the highly \ac{IR}-active excess-proton  in terms of its coordinate $d$  \cite{Roy2020}.
In other words, the HCl-solution difference spectrum reports on the excess-proton motion relative to the two flanking water oxygens
and can therefore be used to investigate proton-transfer dynamics. 
In contrast, the dynamics of $R_{\rm OO}$, i.e. the vibrations of  the  water molecules  in the H$_5$O$_2{}^+$ complex,
black dotted line in fig.~\ref{hcl_intro2}c,
gives rise to a single spectral feature around  \SI{400}{cm^{-1}} which is present in all other calculated spectra and 
in particular also in our experimental spectra, as discussed below.

To check for the effect of the two water molecules that flank the excess proton  on the difference spectrum,
we also calculate  the \ac{IR} spectrum of transient H$_5$O$_2{}^+$ complexes, 
as also done  by \cite{Kulig2013, Daly2017} and discussed in detail in SI section \ref{crossCorrSection}. 
To construct a difference spectrum,
we subtract from the H$_5$O$_2{}^+$ spectrum the spectrum of  hydrogen-bonded
water-molecule pairs obtained from the pure-water  ab initio \ac{MD} simulation.
The resulting  difference spectrum in fig.~\ref{hcl_intro2}c   (gray solid line, right scale)
is reduced by a factor of roughly two compared to
 the difference spectrum of the entire HCl solution (red solid line)
but otherwise agrees  in shape rather nicely. 
Compared to the spectrum of the isolated excess proton (broken line, left scale) we observe 
an amplification by a factor of roughly two, but no essential spectral shape change. 
   {
We conclude that the flanking water molecules and in particular the `special pair dance' with further solvating water molecules does not modify the spectrum of the excess proton in an essential way. }
The amplification of the complete HCl-solution difference spectrum  compared to  the H$_5$O$_2{}^+$  difference spectrum
(red  and gray solid lines, respectively) we rationalize by polarization enhancement effects of water molecules
that solvate the H$_5$O$_2{}^+$ complex.

A few  spectral contributions that are not included in the excess proton power spectrum 
(black broken line in fig.~\ref{hcl_intro2}c) deserve  mentioning:
i) Dynamics orthogonal to the connecting axis of the oxygens are shown to be small in SI section \ref{crossCorrSection}.
ii) The chloride motion is shown below to contribute only slightly and at low frequencies to the spectrum. 
iii) The translation and rotation  of the internal H$_5$O$_2{}^+$ coordinate system relative to the lab frame  is  in 
SI section \ref{rotationSection} shown to only give a small spectral contribution.
By comparison, we thus conclude that the  \ac{IR} difference spectrum
between HCl solutions and pure water
reports very faithfully on  the excess-proton dynamics,
apart from an overall amplification factor.
Turning this around, a more in-depth analysis of the excess-proton spectrum will allow us to decipher the signatures of 
the HCl-solution difference spectrum.

\begin{figure*}
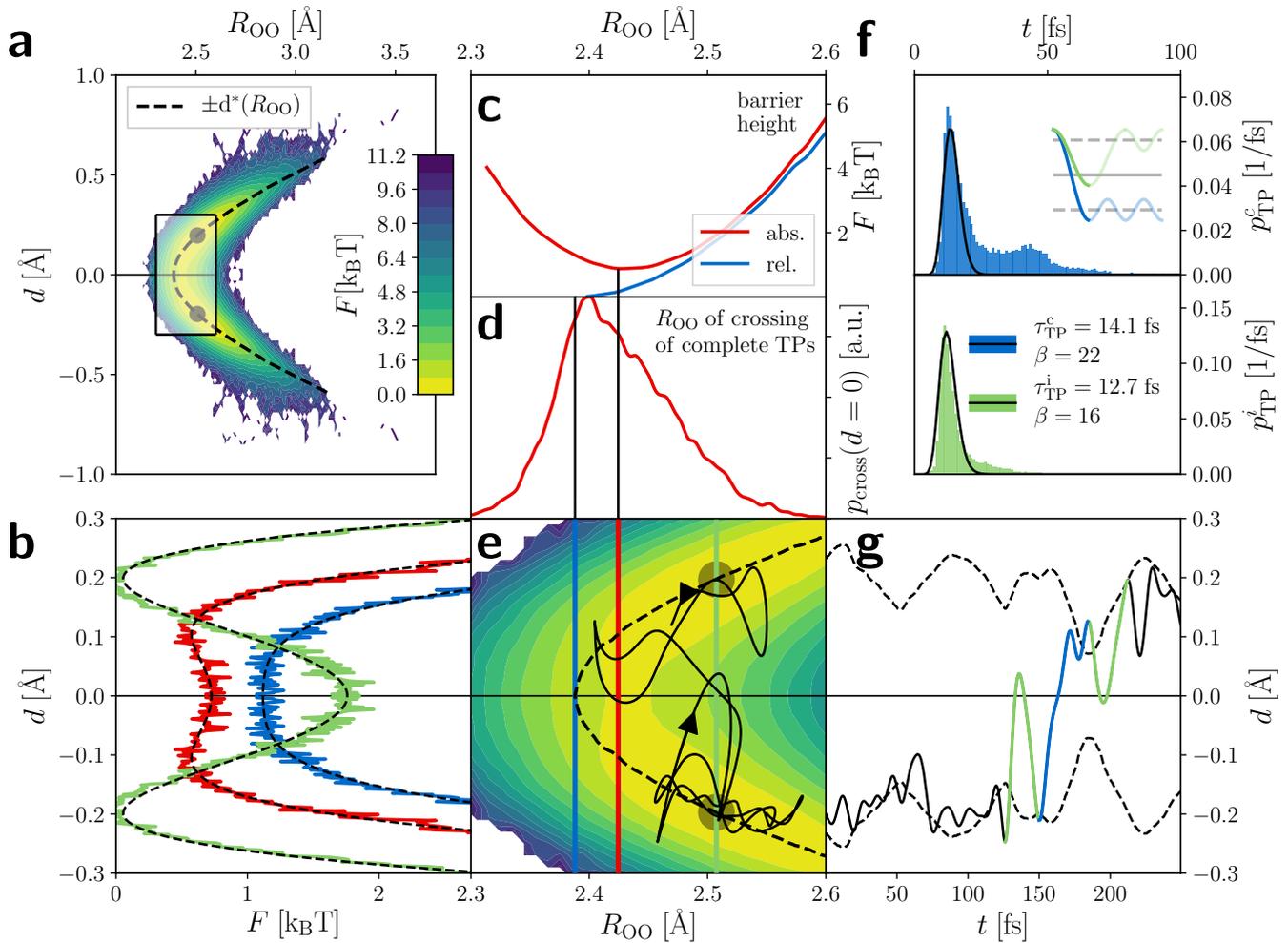

\centering
\begin{overpic}[width=\textwidth]{{/../figs/fig_4_rev}.pdf}
\put(0,70){\huge \bfseries \sffamily a}
\put(36.5,65){\huge \bfseries \sffamily c}
\put(36.5,48){\huge \bfseries \sffamily d}
\put(0,31){\huge \bfseries \sffamily b}
\put(36.5,31){\huge \bfseries \sffamily e}
\put(66,70){\huge \bfseries \sffamily f}
\put(66,31){\huge \bfseries \sffamily g}
\end{overpic}
\caption{
{\bfseries \sffamily 2D excess-proton trajectory analysis.}
{\bfseries \sffamily a:} The two-dimensional free energy of the excess protons in \SI{6}{M} \ac{HCl} solution for the 
$(d,R_{\rm OO})$ coordinates defined in the inset of fig.~\ref{hcl_intro2}a. The shaded area  is enlarged and shown in e. The gray dots denote the positions of the global minima of the 2D free energy. 
 The  minima for fixed  $R_{\rm OO}$, i.e. the  most likely proton locations, $d^*(R_{\rm OO})$  are indicated by  a black broken line.
{\bfseries \sffamily b:} Cuts through the free energy in a for $R_{\rm{OO}}=\SI{2.39}{\ang}$, where the barrier just vanishes (blue solid line), $R_{\rm{OO}}=\SI{2.42}{\ang}$, where the absolute barrier height is minimal (red solid line) and $R_{\rm{OO}}=\SI{2.51}{\ang}$,
for which the global minima of the 2D free energy are obtained (green solid line).
{\bfseries \sffamily c:} The absolute free-energetic barrier height at $d=0$   (red solid line) and the barrier height relative to 
the minima at fixed $R_{\rm OO}$ located at  $d^*(R_{\rm{OO}})$ (blue solid line).
{\bfseries \sffamily d:} Distribution of $R_{\rm{OO}}$ positions at which complete transfer paths cross the $d=0$ midplane.
{\bfseries \sffamily e:} Zoom into the free energy shown in a. The vertical colored lines indicate the cuts through the free energy shown in b. 
An example trajectory from the ab initio \acl{MD} (\ac{MD}) simulation is shown as a black solid line.
{\bfseries \sffamily f:} Time-length distributions of complete (blue, times defined from $d^*$ to $-d^*$) 
and incomplete transfer  paths (green, times defined from $d^*$ to $d=0$ ). 
Fits according to eq.~\eqref{eq:pTP} are shown as black solid lines.
{\bfseries \sffamily g:} Time course of the example trajectory along $d$, same as shown in e, 
with complete and incomplete transfer paths indicated as blue and green lines, respectively. 
The  two branches of the $R_{\rm OO}$-dependent 
minimal energy position  $d^*[R_{\rm{OO}}(t)]$ are shown as black broken lines.
}
\label{hcl_f}
\end{figure*}

\subsection*{2D excess-proton trajectory analysis}

Figure~\ref{hcl_f}a shows the two-dimensional (2D) free energy  for \SI{6}{M} HCl 
obtained from the negative logarithm of the  distribution function of the continuous excess-proton trajectories
  as a function of the coordinates $d$ and $R_{\rm OO}$,
a blow up of the shaded area is given in fig.~\ref{hcl_f}e. 
By definition, the free energy is symmetric with respect to the midplane at $d=0$, which separates two global 
 minima at $R_{\rm{OO}}=\SI{2.51}{\ang}$ and $d=\pm\SI{0.2}{\ang}$. These minima, highlighted as gray dots in fig.~\ref{hcl_f}a, 
 correspond to states where the  excess proton is asymmetrically shared between the two flanking
 water molecules. The transition between these minima, i.e. the proton transfer, 
 is therefore a barrier-crossing process in the two-dimensional plane spanned by $R_{\rm OO}$ and $d$.

Figure~\ref{hcl_f}b shows cuts through the free energy along $d$,
each fitted to a quartic expression  
$F(d)=F_{d=0}(1+\beta d^2+\gamma d^4)$ shown as black broken lines. 
Details are reported in SI section \ref{feFitSection}.
For negative $\beta$,   two minima at $d^* = \pm \sqrt{-\beta/(2\gamma)}$ are separated by a barrier at $d=0$,
which determines the  optimal proton asymmetry  for a given value of  $R_{\rm{OO}}$ and defines the 
parabolic function $d^*(R_{\rm{OO}})$, 
which is plotted as a black broken line in figs.~\ref{hcl_f}a and E.
The cuts  in fig.~\ref{hcl_f}b are shown for $R_{\rm{OO}}=\SI{2.39}{\ang}$, 
where the barrier just vanishes (blue solid line), $R_{\rm{OO}}=\SI{2.42}{\ang}$, where the absolute barrier height is minimal, 
which contains the  transition state  in the $d$-$R_{\rm OO}$ plane  at $d=0$ (red solid line), and $R_{\rm{OO}}=\SI{2.51}{\ang}$, 
 which contains the global minima of the 2D free energy  (green solid line).

The absolute  free energy of the barrier at $d=0$ is plotted in fig.~\ref{hcl_f}c as a red solid line 
and compared to the barrier height relative to the $R_{\rm OO}$-dependent minima at 
$d^*(R_{\rm{OO}})$  (blue solid line).
The minimal absolute barrier free energy of $0.9\,k_BT$ (red line), located at $R_{\rm{OO}}=\SI{2.42}{\ang}$, 
defines the transition state;
for $R_{\rm{OO}}=\SI{2.51}{\ang}$, for which the most probable excess-proton state is obtained,
the barrier has a moderate height of  $1.8\,k_BT$, suggesting that proton transfer is not excluded  for this value
of  $R_{\rm{OO}}$. 
Note that for  $R_{\rm{OO}}<\SI{2.39}{\ang}$ the relative barrier height vanishes and 
thus a symmetrically shared excess proton is most likely.

\begin{figure*}
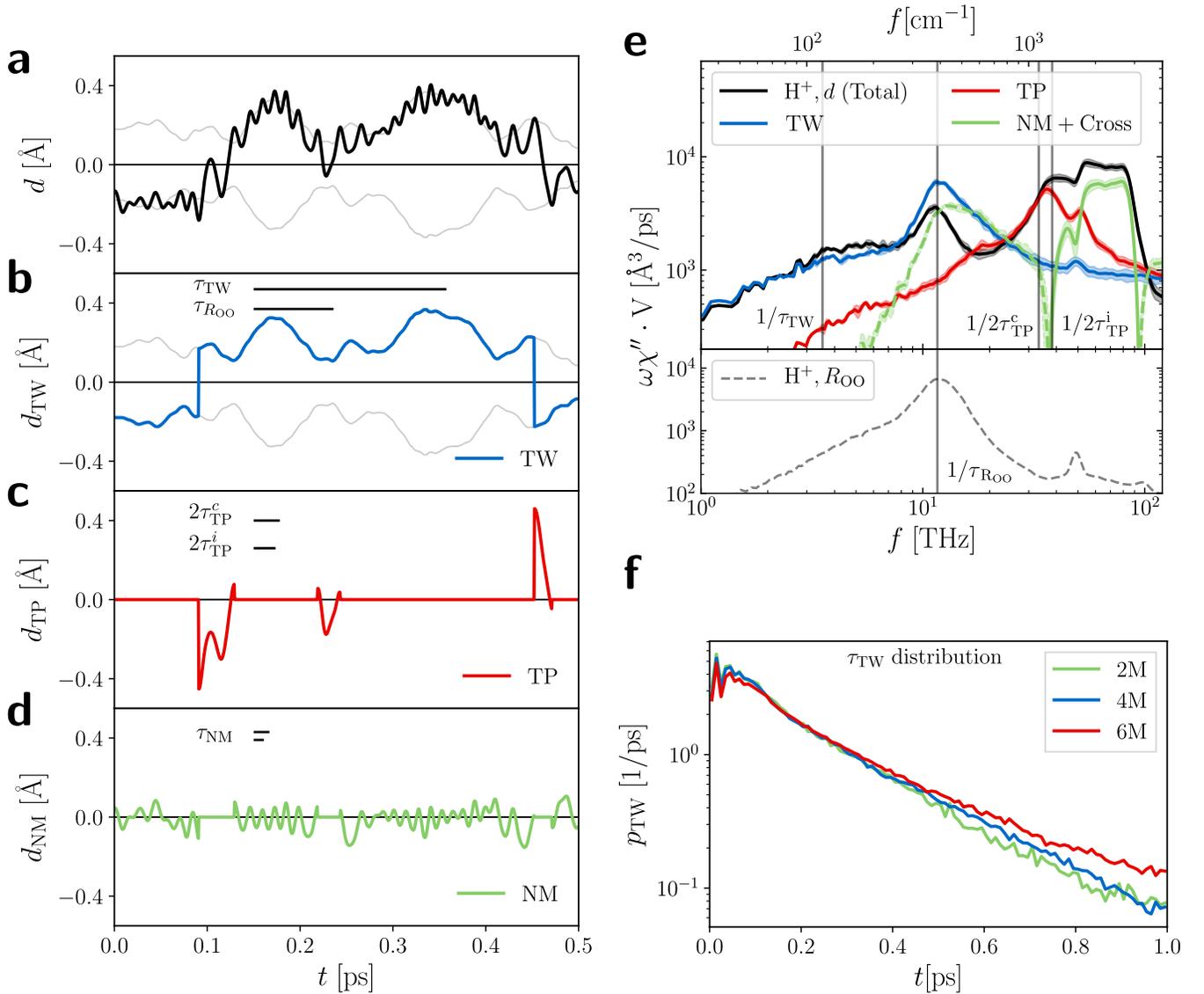

\centering
\begin{overpic}[width=\textwidth]{{/../figs/fig_5_rev}.pdf}
\put(0,78.5){\huge \bfseries \sffamily a}
\put(0,60){\huge \bfseries \sffamily b}
\put(0,42){\huge \bfseries \sffamily c}
\put(0,23.5){\huge \bfseries \sffamily d}
\put(52,80){\huge \bfseries \sffamily e}
\put(52,35){\huge \bfseries \sffamily f}
\end{overpic}
\caption{{\bfseries \sffamily Spectral signatures of proton transfer.}
{\bfseries \sffamily a--d:} Decomposition of an excess-proton trajectory  $d(t)$ (black solid line in a) into transfer-waiting  $d_{\rm TW}$ (blue solid line in b), transfer-path $d_{\rm TP}$ (red solid line in c) and normal-mode contributions $d_{\rm NM}$ (green solid line in d). 
The time course of the $R_{\rm OO}$-dependent most likely proton  positions at $d^*[R_{\rm{OO}}(t)]$ are shown as thin gray solid lines in a and b. 
{\bfseries \sffamily e:} \ac{IR} spectrum of the excess-proton motion projected onto $d$, shown as a black solid line, and \ac{IR} spectra of the contributions shown in a--c in the respective color (the green broken line denotes negative values). The power spectrum of the $R_{\rm OO}$ coordinate is shown as a gray broken line in the lower panel. The inverse of the characteristic time scales $\tau_{\rm TW}$, $\tau_{R_{\rm{OO}}}$,
$2 \tau^c_{\rm TP}$ and  $2 \tau^i_{\rm TP}$  are shown as thin vertical solid gray lines. {\bfseries \sffamily f:} 
Distributions of the transfer-waiting first-passage times
of excess protons, see main text for details.
}
\label{hcl_decomp}
\end{figure*}

Next, to decompose the excess proton trajectories into segments
where the excess proton  moves around  the local free energy minima
and where a transfer across the midplane happens,  transfer path start and end points need to be defined. 
 For this we use the  most likely proton location $d^*(R_{\rm OO})$ (black broken lines in figs.~\ref{hcl_f}a, e and g). 
The start of a transfer path 
is thus defined as the last crossing of $d^*(R_{\rm{OO}})$ on one side of the 
midplane at $d=0$ and the end of a transfer path as 
the first crossing of $d^*(R_{\rm{OO}})$ on the other side of the 
midplane at $d=0$.
The transfer paths are  slightly extended forward and backward in time to the points where the velocity along $d$ vanishes, 
the so-called turning points, in order to be consistent  with the analytical theory presented in \cite{Bruenig2021}. 
An example trajectory    is shown  in fig.~\ref{hcl_f}e (thin black line) in the 
$(d,R_{\rm OO})$ plane,
  the corresponding time-dependent position $d$ is given in fig.~\ref{hcl_f}g, 
where  the transfer path is highlighted in blue. 
Many attempts to transfer are unsuccessful and lead to incomplete transfer paths, 
where the excess proton crosses the mid-plane $d=0$ but does not reach to the minimal free energy state
$ d^*(R_{\rm{OO}})$ on the other side. Two incomplete transfer paths are shown in green in fig.~\ref{hcl_f}g 
and for consistency  are also extended  to their  turning points.
Transfer-path-time distributions are given 
 in fig.~\ref{hcl_f}f,
 with our definitions used, 
incomplete transfer paths turn out to be  slightly faster.
  In total, there are about as many incomplete ($n=14877$) as complete transfer  paths  ($n=14357$), 
  meaning that about half of all excess protons reaching the midplane $d=0$ actually  transfer from one water molecule to 
  the other.
The main peaks in the distributions are  fitted by the Erlang distribution \cite{Cox1977}
 \begin{equation}
 p_{\mathrm{TP}}(t)=\frac{t^{\beta-1}}{(\beta-1)!} \left(\frac{\beta}{\tau_{\rm TP}}\right)^{\beta} e^{-\beta t/\tau_{\rm TP}},
 \label{eq:pTP}
 \end{equation}
with the mean transfer-path time defined by $\tau_{\rm TP}$, shown as black solid lines in fig.~\ref{hcl_f}f, the fit parameters are given in the legend.

   {
The distribution of transition states in fig.~\ref{hcl_f}d, i.e. the $R_{\rm{OO}}$ position at which complete transfer paths  cross the midplane at $d=0$, 
is rather broad and peaks slightly below $R_{\rm{OO}}=\SI{2.42}{\ang}$, the most probable excess-proton position at $d=0$. 
Most paths, in fact  77\%, cross for  $R_{\rm{OO}}>\SI{2.39}{\ang}$, i.e., for values of $R_{\rm OO}$   where a  barrier along 
the  $d$ coordinate is present.
This means that the dominant mechanism for  proton transfer is not one where 
the proton waits until the oxygen-oxygen separation $R_{\rm OO}$ reaches   small values so that the remaining barrier along 
$d$ is small or absent. Rather, protons cross the $d=0$ midplane for a broad distribution of oxygen-oxygen separations $R_{\rm OO}$
and by doing so overcome substantial free-energy barriers. This reverberates that a normal-mode analysis
cannot account for all aspects of  proton transfer  in HCl solutions.}

\subsection*{Spectral signatures of proton transfer}

In order  dissect the  excess-proton spectrum in fig.~\ref{hcl_intro2}c (black broken line)
into contributions that have to do with  proton-transfer events and those that do not,
the excess-proton trajectories $d(t)$ are decomposed into three parts according to 
\begin{equation}
d(t) = d_{\rm{TW}}(t) + d_{\rm{TP}}(t) + d_{\rm{NM}}(t).
\end{equation}
To illustrate this decomposition, fig.~\ref{hcl_decomp}a shows part of an example excess-proton trajectory, $d(t)$  (black line), together with the
most likely excess-proton positions $d^*[R_{\rm{OO}}(t)]$  (thin gray lines);
the deviations between the black and gray  lines visualize excess-proton motion relative to the oxygen it is bound to.
We define the transfer-waiting  contribution
as  $d_{\rm{TW}}(t)\equiv d^*[R_{\rm{OO}}(t)]$  projected onto 
  the closer branch of $d^*[R_{\rm{OO}}(t)]$,
shown as a blue solid line in fig.~\ref{hcl_decomp}b.
Thereby, $d_{\rm{TW}}(t)$  reflects the proton transfer jumps and also contains the water motion.
 The transfer-path contribution $d_{\rm{TP}}(t)$ in fig.~\ref{hcl_decomp}c (red solid line)
 is defined as $d_{\rm{TP}}(t) = d(t)-d_{\rm{TW}}(t)$ during complete and incomplete transfer  paths 
 (as defined  in fig.~\ref{hcl_f}e and g) and is zero elsewhere,
 it describes the excess-proton motion during transfer processes.
Finally, by subtracting $d_{\rm{TW}}(t)$ and  $d_{\rm{TP}}(t)$ from  $d(t)$,
we are left with the oscillations around $d^*[R_{\rm{OO}}(t)]$ when the excess proton is not
undergoing a transfer, 
which constitute the normal-mode contribution $d_{\rm{NM}}(t)$ in fig.~\ref{hcl_decomp}d (green solid line).
Different or more detailed  excess-proton trajectory  decompositions are certainly conceivable,
the usefulness of the present scheme follows from its spectral decomposition properties.

In fig.~\ref{hcl_decomp}e  the excess-proton spectrum,
(black solid line)  is decomposed as
 \begin{equation}
 \omega \chi'' = \omega \chi''_{\rm{TW}} +  \omega \chi''_{\rm{TP}} +  \omega \chi''_{\rm{NM}}.
 \end{equation} 
 The power spectra of the transfer-waiting, $ \chi''_{\rm{TW}}$ (blue line),
 and transfer-path contributions, $ \chi''_{\rm{TP}}$ (red line), 
 are computed from the $d_{\rm{TW}}(t)$ and $d_{\rm{TP}}(t)$ trajectories
 using the Wiener-Kintchine theorem  (see Methods section for details).
All cross-correlation contributions   are included in the normal-mode contribution, 
$ \chi''_{\rm{NM}}$ (green line).

The   normal-mode spectrum $\omega \chi''_{\rm{NM}}$ in fig.~\ref{hcl_decomp}e
accounts  for the continuum band located  between \SIrange{2000}{3000}{cm^{-1}},
it is in fact amenable to normal-mode analysis \cite{Biswas2017, Yu2019a, Calio2021} but  by construction does not include the 
proton-transfer dynamics. 
The range of the dominant normal-mode time scales included in $\omega \chi''_{\rm{NM}}$,
$\tau_{\rm NM}=\,$\SIrange{11}{17}{fs}, follows from the spectral width of the continuum band, taken to be 
$f=\,$\SIrange{2000}{3000}{cm^{-1}} in fig.~\ref{hcl_decomp}e, via $\tau_{\rm NM}=1/f$.

The transfer-path spectrum $\omega \chi''_{\rm{TP}}$ in fig.~\ref{hcl_decomp}e shows a pronounced peaked around \SI{1200}{cm^{-1}}.
An analytical model calculation shows that the peak in the transfer-path spectrum
is related to the mean transfer-path time as $f_{\rm TP}= 1/(2\tau_{\rm TP})$ \cite{Bruenig2021}.
Taking the results  from  the fits  in fig.~\ref{hcl_f}f,
yielding $\tau^c_{\rm TP}=\SI{14.1}{fs}$ for complete  and $\tau^i_{\rm TP}=\SI{12.7}{fs}$ for incomplete transfer paths, 
we predict  $f_{\rm TP}^c=\SI{1170}{cm^{-1}}$  and $f_{\rm TP}^i=\SI{1300}{cm^{-1}}$, 
indicated in fig.~\ref{hcl_decomp}e as vertical lines
and which bracket the transfer-path peak very nicely.

\begin{table}
	\centering
	\begin{tabular}{l|r|r|r}
		conc. & \SI{2.0}{M} & \SI{4.0}{M}  & \SI{6.0}{M} \\ \hline
		$\tau_{\rm{TW}}$  \ \ [fs]\ & $208 \pm 6$ & $229 \pm 4$ & $283 \pm 4$ \\
		$\tau_{\rm{Cl}^-}$ \ [fs]\ & $241 \pm 3$ & $249 \pm 3$ & $ 231 \pm 3$ \\
		$\tau_{R_{\rm {OO}}}$ [fs]\ & $81 \pm 3$ & $82 \pm 2$ & $86 \pm 2$ \\		
		$\tau^c_{\rm{TP}}$  \ \ [fs]\ & $14.08 \pm 0.08$ & $14.34 \pm 0.05$ & $14.10 \pm 0.04$ \\
		$\tau^i_{\rm{TP}}$  \ \ [fs]\ & $12.65 \pm 0.09$ & $12.80 \pm 0.05$ & $12.74 \pm 0.05$ \\
		$\tau_{\rm NM}$ \,\ [fs] & \multicolumn{3}{c}{$(11\pm1)-(17\pm1)$}
	\end{tabular}
	\caption{
{\bfseries \sffamily Characteristic time scales of excess-proton transfer dynamics.} 
The standard errors of the transfer-waiting time $\tau_{\rm{TW}}$, transfer-path times $\tau^i_{\rm{TP}}$ and $\tau^c_{\rm{TP}}$ are estimated from the variances of the fitted distributions.
The errors of the normal-mode times $\tau_{\rm NM}$, oscillation time of the two flanking water molecules $\tau_{R_{\rm {OO}}}$ 
and rattling time of chloride ions $\tau_{\rm{Cl}^-}$ are estimated from the resolution of the underlying spectra.
}
	\label{tab:mfpts}
\end{table}

\begin{figure}
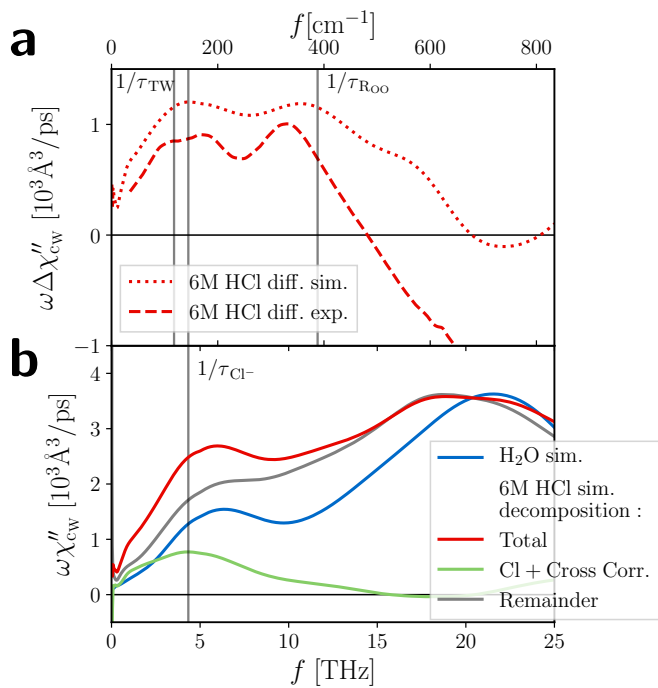

\centering
\begin{overpic}[width=0.48\textwidth]{{/../figs/fig_6_rev}.pdf}
\put(-2,92){\huge \bfseries \sffamily a}
\put(-2,46){\huge \bfseries \sffamily b}
\end{overpic}
\caption{
{\bfseries \sffamily Absorption spectra in the THz regime.}
{\bfseries \sffamily a:} 
Experimental THz/\acl{FTIR} (THz/\acs{FTIR}) difference spectra of \SI{6}{M} \ac{HCl} solution (red broken line) compared to the difference spectrum from ab initio \acl{MD} (\ac{MD}) simulations (red dotted line). {\bfseries \sffamily b:} 
The simulated \SI{6}{M} \ac{HCl} spectrum (red solid line) is decomposed into a chloride-ion contribution (green solid line, including cross correlations) and a remainder (gray solid line). For comparison the simulated pure-water spectrum (blue solid line) is also shown. 
}
\label{hcl_thz}
\end{figure}

The  transfer-waiting spectrum $\omega \chi''_{\rm{TW}}$ in fig.~\ref{hcl_decomp}e  exhibits
 a peak around \SI{400}{cm^{-1}}, a shoulder  around \SI{100}{cm^{-1}}
 and a slow decay for lower frequencies.
The peak  around \SI{400}{cm^{-1}} (\SI{12}{THz}) is caused by oscillations of the oxygen-oxygen separation, 
$R_{\rm OO}$, which couple to the proton position $d$ via the most likely
proton position  $d^*[R_{\rm{OO}}(t)]$; 
in simple terms, the proton vibrates with the water molecule it is bound to. 
The oxygen vibrational time scale $\tau_{R_{\rm OO}} = \SI{86}{fs}$,
 indicated  in fig.~\ref{hcl_decomp}b, follows 
from the peak of the power spectrum of $R_{\rm{OO}}$ around \SI{400}{cm^{-1}}, 
which is  plotted in the lower panel of fig.~\ref{hcl_decomp}e
and agrees perfectly with the peak in $\omega \chi''_{\rm{TW}}$. 
This peak is in fact also well visible in our experimental THz/\ac{FTIR} difference spectra, 
shown again in fig.~\ref{hcl_thz}a as a broken red line for a  \SI{6}{M} HCl solution,
the dotted red line shows the corresponding simulated  difference spectrum. 
Note that this  translational vibration of two water oxygens 
in the transient H$_5$O$_2^+$ complex is about 
twice as fast as the translational vibration  of two hydrogen-bonded water molecules in pure water, 
which gives rise to the well-known \ac{IR} signature around \SI{200}{cm^{-1}}, 
shown as a blue solid line in fig.~\ref{hcl_thz}b obtained from pure-water simulations  \cite{Carlson2020}.
This frequency shift is the reason why the water-vibration peak appears prominently in the difference spectra in fig.~\ref{hcl_thz}a.

The shoulder  in  $\omega \chi''_{\rm{TW}}$ around \SI{100}{cm^{-1}} 
is related to the transfer waiting time 
 $\tau_{\rm TW}$, which is the average time 
 between two consecutive  complete  proton-transfer events,
 as predicted from an analytically solvable barrier-crossing model \cite{Bruenig2021}.
  In fig.~\ref{hcl_decomp}f we show  distributions of the transfer-waiting first-passage times, i.e. distributions  of the time difference  between 
crossing the most likely proton position   $d^*[R_{\rm{OO}}(t)]$ on one side of the midplane $d=0$
and crossing $d^*[R_{\rm{OO}}(t)]$ on the other side of the midplane $d=0$ for the first time, 
 for the three HCl concentrations.
 { The distributions are essentially exponential in nature, which  means that  transfer events occur at a roughly constant rate and reflects the stochastic nature of the process.}
The mean of these first-passage  distributions defines  the transfer-waiting time $\tau_{\rm TW}$, 
which is given  in tab.~\ref{tab:mfpts} and increases with rising HCl  concentration.
This indicates that hydronium ions have a slightly longer life time at higher HCl concentrations. 
In contrast,  both complete and incomplete  transfer-path times  interestingly show no dependence on the HCl concentration.
The  inverse of the transfer-waiting time $\tau_{\rm TW}=\SI{280}{fs}$ for \SI{6}{M}, 
which is shown in fig.~\ref{hcl_decomp}e as a vertical line, 
is located at \SI{120}{cm^{-1}} (\SI{3.5}{THz}) and 
corresponds well to the  position of the shoulder
 in $\omega \chi''_{\rm{TW}}$, which confirms the connection between the transfer-waiting time and the spectroscopic signature
 around \SI{100}{cm^{-1}} that is predicted by analytical theory \cite{Bruenig2021}.
 We note that the total length of the continuous proton-transfer trajectories are roughly twice as long as the 
 mean transfer-waiting times, meaning that typically a few back-and-forth proton-transfer events occur in each trajectory
 (see SI section \ref{fptSection} for more details). 

The characteristic time scales of each contribution, 
i.e. the transfer-waiting time $\tau_{\rm TW}=\SI{280}{fs}$, 
the water-oxygen vibrational time $\tau_{R_{\rm OO}} = \SI{86}{fs}$, 
the transfer-path times $\tau^c_{\rm TP}=\SI{14.1}{fs}$ and $\tau^i_{\rm TP}=\SI{12.7}{fs}$ 
and the normal-mode times $\tau_{\rm NM}=\,$\SIrange{11}{17}{fs}
 are unambiguously extracted from the simulations and characterize
 both the trajectory contributions in the time domain in   fig.~\ref{hcl_decomp}a--d,
where they are included  as horizontal black bars,
and also the  different spectral contributions
in  fig.~\ref{hcl_decomp}e.

{
We comment on the subtle spectral features in the range \SIrange{1400}{1800}{cm^{-1}} in fig.~\ref{hcl_decomp}e, where small but distinct peaks are revealed in the different spectral contributions. 
The transfer-waiting contribution (blue line) peaks at about \SI{1650}{cm^{-1}}, hinting to a weak coupling to an unperturbed water bending mode of the flanking water molecules. 
The transfer-path contribution (red line) peaks at \SI{1750}{cm^{-1}}, the location of the experimental acid bend signature, which suggests that the acid bend couples particularly to the transfer path motion of the excess proton.
Note, that even though the acid bend is primarily produced by the excess-proton motion orthogonal to the $d$ coordinate, this contribution to the isotropic spectrum is largely compensated by motion the flanking water molecules, as shown in SI section \ref{crossCorrSection}.
The normal-mode contribution (green line) peaks around \SI{1500}{cm^{-1}}, consistent with previously calculated normal-mode spectra of Eigen-like solvated proton structures \cite{Yu2019a}.}

So far we have concentrated on the excess-proton spectral contribution and not discussed the chloride contribution.
The  decomposition of the total simulated 6M HCl spectrum in fig.~\ref{hcl_thz}b  (red solid line) 
into the chloride contribution (green solid line, including all cross correlations) and the  remainder (gray solid line)
demonstrates a prominent chloride peak around \SI{150}{cm^{-1}}, which translates to a corresponding time scale of 
$\tau_{\rm Cl^-}=\SI{185}{ps}$ and is due to the rattling of a chloride in its hydration cage \cite{Schwaab2019,Balos2020}.
This peak is also seen in the simulated difference spectrum in fig.~\ref{hcl_thz}a  (red dotted line) 
and is slightly shifted to larger frequencies in the experimental difference spectrum (red broken line). 
 At \SIrange{100}{200}{cm^{-1}} the remainder contribution  in fig.~\ref{hcl_thz}b  (gray solid line)  
 is  significantly stronger than the pure-water spectrum (blue line), 
 indicating that a process related to excess-proton motion significantly contributes in this wavenumber range. 
 We suggest  that this  process is  the excess-proton transfer-waiting contribution, which   
 in fig.~\ref{hcl_decomp}e is shown to produce a broad shoulder around \SI{100}{cm^{-1}}.

  {
 \subsection*{Alternative methods for simulation and characterization of excess-proton dynamics}}

{
It is known that nuclear quantum effects (NQEs), specifically zero-point effects, 
significantly influence  distributions of excess protons in water \cite{Marx1999, Napoli2018, Calio2021}.
While the techniques for simulating NQEs have significantly advanced in recent years, 
the accurate calculation of dynamical properties, which is the focus of this study,
 remains an active field of research \cite{Markland2018}. 
There are some subtle questions how the  time-dependent polarization  correlation functions of decomposed excess-proton trajectories,
which form the basis of our spectroscopic  analysis, 
would be extracted from  simulation data encompassing NQEs,
in particular, it is unclear whether the common approaches taken by centroid or ring-polymer \ac{MD} are directly applicable.
Most of all,
it is noteworthy that previous studies found no significant differences 
between \ac{IR} spectra computed from simulations with and without NQEs
below \SI{3000}{cm^{-1}} 
 \cite{Biswas2016, Napoli2018}, 
 which could mean that the neglect of NQEs might have less severe consequences for time-dependent correlation 
 functions than it has for spatial distributions of excess protons.
We therefore interpret the good 
agreement between simulated and experimental spectra in the THz regime in fig.~\ref{hcl_exp}c and 
in the mid-\ac{IR} regime in fig.~\ref{hcl_intro}c as a validation of  our chosen simulation techniques.
Besides, our neglect of NQEs allows us to generate long trajectories that improve statistics and therefore the quality of our spectra, particularly down in the THz regime.}

 {
In SI section \ref{radialDistSection} we extract radial distribution functions involving excess protons and chloride ions from our simulation trajectories  and obtain  good agreement with previous simulations and experimental data \cite{Baer2014, Fulton2010, Xu2010, Calio2020, Fischer2019}.

 In SI section \ref{fptSection} we show that our
transfer-waiting time distributions in fig.~\ref{hcl_decomp}f  correspond quite closely 
to hydronium-oxygen continuous-identity auto-correlation functions \cite{Arntsen2021},
 which were previously  introduced  to characterize  proton-transfer  time scales. 
The same correlation functions, but with the fast  back-and-forth excess-proton transfer events between the same two water molecules of 
the transient H$_5$O$_2{}^+$ complexes removed, have  been used to interpret the long time scales of uni-directional 
proton transfer, which reflects signatures  observed in 2D \ac{IR} experiments at time scales of 
\SIrange{1}{2}{ps} \cite{Kundu2019,Carpenter2020, Arntsen2021}.
We note that  back-and-forth proton-transfer events that occur within transient H$_5$O$_2{}^+$ complexes 
are  spectroscopically relevant  and therefore must be included in the prediction of spectra.
The longer  time scale of uni-directional proton transfer at  \SIrange{1}{2}{ps} 
contributes  in linear absorption spectra    at   frequencies below \SI{1}{THz}.

In SI section \ref{diffusionSection} we determine the diffusivities of excess protons and water molecules 
and obtain similar results as previous simulations \cite{Xu2010,Calio2020,Biswas2016,Tse2015,Arntsen2021,Fischer2019}. 
Compared to the experimental data, the absolute diffusivities  of excess protons $D_{\rm H^+}$ 
and water oxygens $D_{\rm O}$ are smaller by a factor of about three, 
but their ratios, given by 
$D_{\rm H^+}/D_{\rm O} = (3.0 \pm 0.8), (4.3 \pm 1.3), (3.8 \pm 0.8)$ for \SI{6}{M}, \SI{4}{M}, \SI{2}{M} HCl, 
respectively,  are albeit large errors in satisfactory  agreement with the experimental ratios of about 
$D_{\rm H^+}/D_{\rm O} = 1.5, 2.3, 3.0$ for similar concentrations \cite{Dippel1991}.}

{
It has been shown that  proton-transfer events
are caused by subtle structural changes in the excess-proton solvation 
 environment, 
 for example the hydrogen-bond structure in the second solvation shell \cite{Tse2015, Biswas2016, Napoli2018, Fischer2019}. 
Based on our simulations, we  confirm that the suggested hydrogen-bond asymmetry coordinate 
can indeed be used to predict the excess proton that is most likely to transfer to a neighboring
 water among the three transfer candidates within a hydronium ion \cite{Napoli2018}.  
 Along these lines, the presence of a fourth water molecule that forms a hydrogen-bond to a  hydronium ion 
 weakly correlates with back-and-forth transfer  behavior \cite{Tse2015, Biswas2016, Fischer2019}. 
These findings are presented in detail in SI section \ref{hbSection}.
}

\section{Conclusions}
We show that the spectroscopic signature of  proton-transfer dynamics between two water molecules in \acl{HCl} (\ac{HCl}) solutions
can be investigated by trajectory decomposition  into transfer-waiting (characterized by the time scale 
$\tau_{\rm{TW}}$), transfer-path
(characterized by $\tau_{\rm{TP}}$)  and normal-mode contributions (characterized by $\tau_{\rm{NM}}$).
The decomposition is performed in the  two-dimensional coordinate system that is spanned by the 
excess-proton position and the  oxygen-oxygen distance of the two flanking water molecules
and operates both in the time domain as well as in the frequency domain. 
The  coupling of the  excess-proton motion to  the relative oscillations of the  two flanking water molecules
produces a fourth spectral proton-dynamics contribution (characterized by $\tau_{ R_{\rm OO}}$).
The dynamics of each of the four contributions are described by distinct time scales with  the ordering
 $\tau_{\rm{TW}} > \tau_{R_{\rm OO}} > \tau_{\rm{TP}} > \tau_{\rm{NM}}$ and therefore contribute with distinct peaks to the excess-proton \ac{IR} spectrum. 
 Our experimental THz/\ac{FTIR} difference spectra resolve the slowest time scales, 
 $\tau_{\rm{TW}}$ and $\tau_{ R_{\rm OO}}$, of which the former one is overlaid by an additional spectral contribution due to
 rattling of  the chloride ions, characterized by yet another time scale $\tau_{\rm Cl^-}$, which is close to $\tau_{\rm{TW}}$.
 Mid-\ac{IR} experimental difference spectra from  literature on the other hand are compatible with
  our predicted  spectra  associated with the $\tau_{\rm{TP}}$ and $\tau_{\rm NM}$ time scales.

In contrast to the transfer-path time $\tau_{\rm{TP}}$, the transfer-waiting  time $\tau_{\rm{TW}}$ 
shows a weak dependence on the HCl concentration. 
{
Possible reasons include ionic screening and repulsion effects between neighboring excess-protons, but also entropic effects due to the reduced number of accepting water molecules have been discussed \cite{Yuan2019, Calio2020, Carpenter2019}. 
This is consistent with experimental results showing a decrease of the excess-proton diffusivity with increasing HCl concentration \cite{Dippel1991}, which is reproduced in our simulations  (see SI section \ref{diffusionSection}).
One should note that the transfer-waiting times in local transient H$_5$O$_2{}^+$  complexes include back-and-forth proton-transfer events, 
which do not contribute to the long-time excess-proton diffusion \cite{Arntsen2021, Calio2020, Tse2015, Fischer2018, Xu2010}
but nevertheless have a pronounced  spectroscopic signature \cite{Bruenig2021}.}

{
Our results nicely complement   recent normal-mode calculations. 
We find that  the continuum band stems from normal-mode vibrations of less symmetric, i.e. more Eigen-like, 
configurations of the excess proton, which are strongly influenced   by the oxygen-oxygen separation $R_{\rm OO}$ \cite{Thaemer2015,Biswas2017,Yu2019a,Carpenter2020}.
On the other hand, our transfer-path signature, which is dominated by a broad absorption around \SI{1200}{cm^{-1}}, 
shows striking similarity with normal-mode spectra computed for more symmetric, i.e. more Zundel-like, configurations of the excess proton \cite{Thaemer2015,Biswas2017,Yu2019a,Carpenter2020}.
It is not implausible that normal modes for small separations of the two flanking water molecules 
show similar spectroscopic signatures as the  transfer paths we extract from our simulation trajectories. 
Yet, in a normal-mode picture the interconversion between the metastable Zundel-like and Eigen-like excess-proton states 
cannot be explained consistently, 
even though the importance of this process for a complete description of the mid-\ac{IR} signatures was acknowledged several times \cite{Thaemer2015,Napoli2018,Kundu2019,Yu2019a,Carpenter2020}.
In fact, the broad distribution of this interconversion time scale is demonstrated  by the transfer-waiting time distribution 
and also gives rise to a distinct spectral signature, that we identify in the THz regime.}

{
A recent study employing similar simulation techniques has decomposed the proton power spectra with respect to the proton asymmetry coordinate and thereby reached similar conclusions to ours: Eigen-like configurations give rise to the continuum band while Zundel-like configurations 
 dominantly contribute around \SI{1200}{cm^{-1}} \cite{Roy2020}. 
That study also  determined the  proton-transfer time scale using two-dimensional transition state theory and Marcus theory of ion pairing
and finds this time scale to be concentration dependent, in agreement with our and previous observations.}

In summary,
in many  theoretical treatments, 
only normal modes of meta-stable or stable states are  assumed to produce spectral contributions. 
Any spectral mode is therefore interpreted as being due to a meta-stable state, consequently,  
 broad spectral modes are often  interpreted as reflecting a wide collection of normal modes with slightly different frequencies.
In this paper we show that  transfer and  barrier-crossing  events of charged particles  as well as  transfer paths 
 create spectral features by a mechanism that is very different from a normal-mode picture
 and that these spectral features  are broadened by the stochastic nature of the transfer  dynamics \cite{Bruenig2021}.
 In fact,  the strength or frequency of a spectral feature does not allow to 
 tell whether it is caused by normal-mode oscillations in a stable or meta stable state
 or whether it is caused  by transfer or barrier-crossing dynamics.

\section*{Methods}
\subsection*{Computational methods: ab initio MD simulations}

The Born-Oppenheimer ab initio \ac{MD} simulations of pure water and  \ac{HCl} solutions 
at three different concentrations were performed with the CP2K 7.1 software package  using a polarizable double-zeta basis set for the valence electrons, optimized for small molecules and short ranges (DZVP-MOLOPT-SR-GTH,
with the exception of the chloride anions, that were modeled including diffuse functions in the aug-DZVP-GTH basis set), dual-space pseudopotentials, the BLYP exchange-correlation functional, and D3 dispersion corrections \cite{Kendall1992,VandeVondele2005,VandeVondele2007,Grimme2010,Kuhne2020}. The cutoff for the plane-wave representation was \SI{400}{Ry}. The system parameters are summarized in tab.~\ref{tab:sim_params}.

Before production,  each system was equilibrated in classical \ac{MD} simulations for \SI{200}{ps} under NPT conditions at atmospheric pressure and \SI{800}{ps} under NVT conditions at \SI{300}{K}, using the GROMACS 2020.5 software \cite{Abraham2015} with the SPC/E water model \cite{Berendsen1987}. The force fields for $\rm{Cl^-}$ and $\rm{H_3O^+}$ were taken from \cite{Bonthuis2016}.
The ab initio \ac{MD} simulations were subsequently performed using a time step of \SI{0.5}{fs} under NVT conditions at \SI{300}{K} by coupling the system to a CSVR thermostat with a time constant of \SI{100}{fs} \cite{Bussi2007}.

Dipole moments were  obtained after Wannier-center localization of the electron density at a time resolution of \SI{4}{fs}. At each time step the Wannier centers were assigned to the molecule of their closest oxygen or chloride ion. 
Water molecules were assembled by assigning each proton to the closest oxygen nucleus, thereby forming either water or hydronium ions. For the hydronium ions, all protons were treated as excess-proton candidates and further processed based on a dynamical criterion as discussed in the main text and SI section \ref{selectionSection}. The dipole moments $\mathbf{p}$ follow as a sum over the respective position vectors $\mathbf{r}_i$ and charges $q_i$ ($q=\SI{2}{e}$ for Wannier centers, and reduced core charges for nuclei), $\mathbf{p} = \sum_i q_i \mathbf{r_i} $ for the whole or desired sub systems.

Linear response theory relates the dielectric susceptibility $\chi(t)$ to the equilibrium autocorrelation of the dipole moment $C(t)=\sum_D \langle \mathbf{p} (t)\mathbf{p}(0)\rangle$, reading in Fourier space
\begin{align}
\label{linearResponseFT}
\chi(\omega) = \frac{1}{V  k_BT \epsilon_0 D}\left( C(0) - i \frac{\omega}{2} \widetilde C^+(\omega) \right),
\end{align}
with system volume $V$, thermal energy $k_BT$, vacuum permittivity $\epsilon_0$ and $D$ being the number of Cartesian dimensions of the polarization vector $\mathbf{p}$. \ac{IR} spectra can therefore be calculated straight-forwardly from sufficiently sampled trajectories of the ab initio \ac{MD} simulation data using eq.~\eqref{linearResponseFT} and the Wiener-Kintchine relation, derived in SI section \ref{WienerKintchineSection} as
\begin{align}
\label{eqM:WienerKintchine}
C(t) = \frac{1}{2 \pi (L_t-t)} \int_{-\infty}^{\infty} d\omega\ e^{-i\omega t} \mathbf{\tilde p} (\omega) \mathbf{\tilde p}^*(\omega),
\end{align}
where $\tilde p (\omega)$ is the Fourier-transformed dipole-moment trajectory with length $L_t$ and the asterisk denotes the complex conjugate.
Alternatively, for charged subsystems, as in case of the chloride ions, the computation using the time derivative of the polarization, i.e. the current $\mathbf{j}=\frac{d}{dt} \mathbf{p}(t)$ is preferable
\begin{align}
C(t) = \frac{1}{2 \pi (L_t-t)} \int_{-\infty}^{\infty} d\omega\ \frac{e^{-i\omega t}}{\omega^2} \mathbf{\tilde j}(\omega) \mathbf{\tilde  j} ^*(\omega).
\end{align}

Quantum corrections have previously been addressed  \cite{Ramirez2004},
but were not applied here.

\begin{table}
	\centering
	\begin{tabular}{l|l|l|l|l}
		conc. & \SI{0.0}{M} & \SI{2.0}{M} & \SI{4.0}{M}  & \SI{6.0}{M} \\ \hline
		$V^{\frac{1}{3}}$ & \SI{19.73}{\ang} & \SI{20.25}{\ang} &  \SI{20.25}{\ang} &  \SI{20.23}{\ang}\\
		$N_{\mathrm{H_2O}}$ & $256$ & $258$ & $244$  & $224$ \\
		$N_{\mathrm{H^+}}, N_{\mathrm{Cl^-}}$ & $0$ & $10$ & $20$  & $30$ \\
		$\tau$ & \SI{201}{ps}  & \SI{52}{ps} & \SI{84}{ps}  & \SI{84}{ps} \\
		$\Delta t_{\rm WC}$ & \SI{2}{fs} & \multicolumn{3}{ c }{\SI{4}{fs}}
	\end{tabular}
	\caption{{\bfseries \sffamily Parameters of the ab initio \acl{MD} simulations.}}\label{tab:sim_params}
\end{table}

Since the Wannier-center localization time step $\Delta t_{\rm WC} = \SI{4}{fs}$ is larger
 than the original simulation time step $\Delta t = \SI{0.5}{fs}$, the analysis is performed on two types of trajectories stemming from the same simulations: one set of trajectories containing the electronic degrees of freedom and another set of trajectories of higher time resolution but only containing nuclei positions. 
 This higher resolution data is used for calculation of excess proton spectra and kinetics.

All spectra were smoothed by a convolution with a Gaussian kernel of varying width, depending on their respective resolution. We used a standard deviation of $\SI{55}{cm^{-1}}$, $\SI{20}{cm^{-1}}$, and $\SI{50}{cm^{-1}}$ for bulk spectra, the H$_5$O$_2{}^+$ complex difference spectrum in fig.~\ref{hcl_intro2}c and excess-proton spectra, respectively. Experimental data was smoothed using a standard deviation of $\SI{3}{cm^{-1}}$.

To address the quality of the chosen basis set, shorter simulations at \SI{6}{M} were performed using the non-short range basis set (DZVP-MOLOPT-GTH) as well as a triple-zeta doubly polarizable (TZV2P-GTH) basis set.
Spatial correlations in the data are compared in SI section \ref{basisSetSection}.
While the coordination of excess protons with chloride ions slightly increases when the more elaborate basis sets are used, no significant differences in correlations between excess protons and oxygen nuclei are found, which are the focus of this study. 

\subsection*{Experimental methods: THz absorption measurements}
THz spectroscopic measurements in the \SIrange{30}{650}{cm^{-1}}  frequency range were done with a commercial Fourier Transform spectrometer (Bruker Vertex 80v, Germany) equipped with a mercury light source and a liquid helium cooled bolometer detector (Infrared Laboratories, Germany). Spectra are an average of 128 scans with a resolution of \SI{2}{cm^{-1}}. The liquid sample cell is composed of diamond windows (Diamond Materials GmbH, Germany) in which a Kapton spacer of approximately \SI{13}{\micro\metre} was placed between the windows to fix the sample thickness. The exact thickness of the sample cell was determined from the etaloning pattern of the empty sample cell. Temperature of the sample was held constant at $20.0\pm\SI{0.2}{^{\circ}C}$ by an external chiller. 
The measured frequency dependent extinction coefficient, $\alpha_\text{solution}(\omega)$, is 
determined using the Beer-Lamber law
\begin{equation}
	\alpha_\text{solution}(\omega)=\frac{1}{d}\ln\left(\frac{I_{water}(\omega)}{I_{solution}(\omega)}\right)+\alpha_\text{water}(\omega),
	\label{eq:AlphaSolution}
\end{equation}
where $d$ is the sample thickness, $I_{water}(\omega)$ and $I_{solution}(\omega)$ are the experimental transmitted intensities of the water reference and the sample. $\alpha_\text{water}(\omega)$  is the extinction coefficient of bulk water and is taken from literature \cite{Bertie1996}. 
The extinction coefficient $\alpha(\omega)$ is converted to the absorption spectrum, proportional to the imaginary part of the dielectric susceptibility $\chi''(\omega)$, by fitting the spectra and performing a Kramers-Kronig transform as presented in SI section \ref{kkSection}.  

The difference absorption spectra of  HCl solutions with respect to pure water
and normalized with respect to the water concentration are given by

\begin{equation}
\omega\Delta\chi''_{c_\text{w}}(\omega)= \frac{1}{c_\text{w}} \omega \chi''_\text{solution}(\omega)-\frac{1}{c_\text{w}^\text{0}}\omega \chi''_\text{water}(\omega),
	\label{eq:DeltaAlpha}
\end{equation}   

where $c_\text{w}$  and  $c_\text{w}^\text{0}$ are the concentration of water in the aqueous HCl solutions  and bulk water, respectively,  determined from the solution density at room temperature.

\subsection*{Data availability}
The datasets generated and analyzed during the current study are available from the corresponding authors on reasonable request.

\begin{acknowledgments}
We gratefully acknowledge  computing time on the HPC clusters at the physics department and ZEDAT, FU Berlin, as well as the computational resources provided by the North-German Supercomputing Alliance (HLRN) under project bep00068. 
This work was funded by the Deutsche Forschungsgemeinschaft (DFG) under Germany's Excellence Strategy – EXC 2033 – 390677874 – RESOLV
and via grants SFB 1078, project C1 and SFB 1349, project C4.
MH acknowledges funding from the ERC Advanced Grant 695437 THz Calorimetry,
RRN acknowledges funding from the ERC Advanced Grant 835117 NoMaMemo.

\end{acknowledgments}

\subsection*{Author contributions}
F.N.B. and R.R.N. conceived the theory and designed the simulations. F.N.B. performed the ab intio \ac{MD} simulations. F.N.B. and M.R. analyzed the data and designed the figures.
E.A. and M.H. carried out the THz measurements and analyzed the data.
All authors discussed the results, analyses and interpretations. F.N.B., M.H.  and R.R.N. wrote the paper with input from all authors.

\subsection*{Competing interests}
The authors declare no competing interests.

\begin{acronym}[Bash]
\acro{FTIR}{Fourier-transform infrared}
\acro{HCl}{hydrochloric acid}
\acro{IR}{infrared}
\acro{MD}{molecular dynamics}
\end{acronym}
\bibliography{bibliography.bib}

\end{document}


\title{Supplementary material: \\ Spectral signatures of excess-proton waiting and transfer-path dynamics in aqueous hydrochloric acid solutions}

\author{Florian N. Br\"unig}
\affiliation{Freie Universit\"at Berlin, Department of Physics, 14195 Berlin, Germany}

\author{Manuel Rammler}
\affiliation{Freie Universit\"at Berlin, Department of Physics, 14195 Berlin, Germany}

\author{Ellen M. Adams}
\affiliation{Ruhr Universit\"at Bochum, Department of Physical Chemistry II, 44780 Bochum, Germany}

\author{Martina Havenith}
\affiliation{Ruhr Universit\"at Bochum, Department of Physical Chemistry II, 44780 Bochum, Germany}

\author{Roland R. Netz}
\email[]{rnetz@physik.fu-berlin.de}
\affiliation{Freie Universit\"at Berlin, Department of Physics, 14195 Berlin, Germany}

\date{\today}

\pacs{}

\maketitle

\newpage

\section{Experimental infrared spectra of aqueous hydrochloric acid solutions}
\label{spectraSection}

\begin{figure}[!hb]
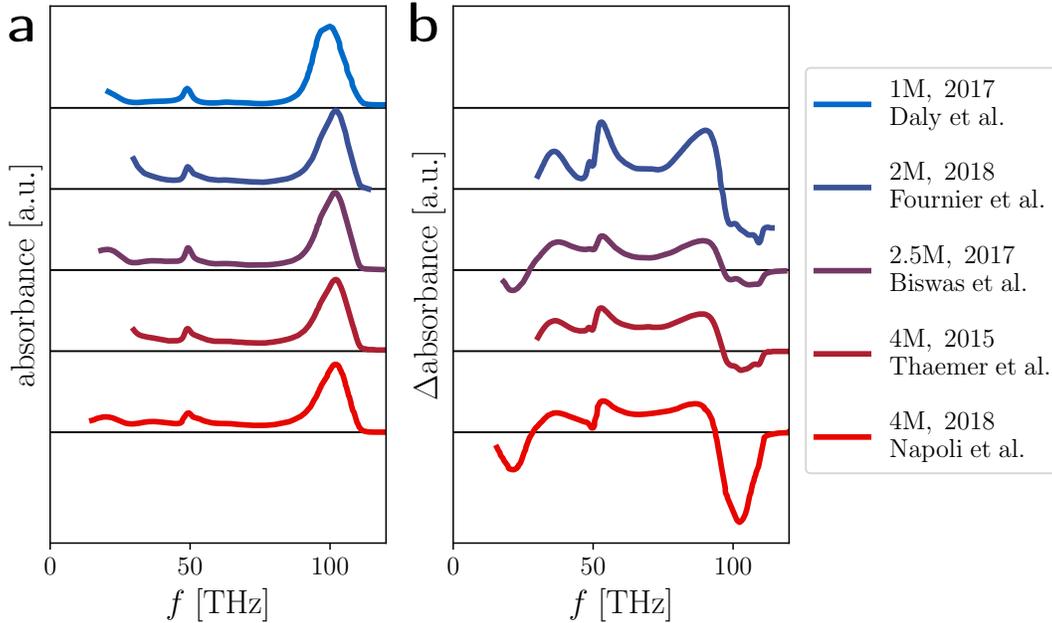

	\centering
	\begin{overpic}[width=.8\linewidth]{{/../figs/exp_specs}.pdf}
	\put(1,55){\huge \bfseries \sffamily  a}
	\put(38,55){\huge \bfseries \sffamily  b}	
	\end{overpic}
	\caption{Experimental \ac{IR} absorption spectra (\textbf{a}) and corresponding difference spectra (\textbf{b}) of aqueous \ac{HCl} at various concentrations, taken from \cite{Daly2017,Fournier2018,Biswas2017,Thaemer2015,Napoli2018}. The data was not published with an absolute scale and is thus scaled arbitrarily. 
No difference spectrum was published for the spectrum at \SI{1}{M} \cite{Daly2017}.
}\label{fig:exp_specs}
\end{figure}

Fig.~\ref{fig:exp_specs}a shows experimental \ac{IR} absorption spectra and fig.~\ref{fig:exp_specs}b shows difference spectra of aqueous \ac{HCl} solutions at various concentrations.
All data is taken directly as published in the literature. 
The proton continuum band  is clearly  visible in all difference spectra, albeit with slight differences in shape. In comparison to our calculated difference spectra in fig.~\ref{hcl_intro}b in the main text, the features at $\SI{40}{THz}$, $\SI{50}{THz}$ and $\SI{90}{THz}$ are more pronounced.

Indeed difference spectra depend on the normalization applied to the underlying \ac{HCl} solution and pure water spectra which is further elucidated in section \ref{normalizationSection}. 
Normalizations commonly used for spectra calculated from simulations employ sample volume or concentration of water molecules. 
In experimental setups, however, these parameters are often not reported. 
Of the four publications we took difference spectra from, only Napoli et al. explained the applied normalization scheme as normalization ``to a unit area over the range shown''\cite{Napoli2018}. 
In light of these subtleties, comparisons between difference spectra from different sources should be considered with care.

\newpage

\section{Normalization of infrared difference spectra of aqueous hydrochloric acid solutions}
\label{normalizationSection}

\begin{figure}[hb]
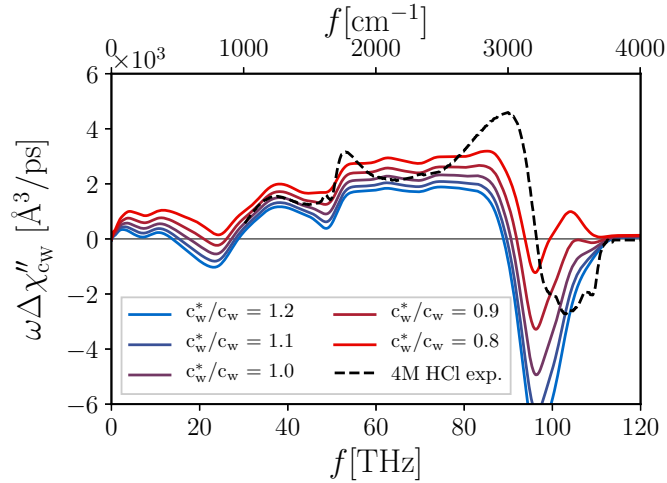

	\centering
	\begin{overpic}[width=.5\linewidth]{{/../figs/total_spectra_4M_diff_varScale-eps-converted-to}.pdf}
	\end{overpic}
\caption{Illustration of the effect of variations of the normalization protocol used for difference spectra of the \SI{4}{M} \ac{HCl} solution, obtained from ab initio \ac{MD} simulations. The difference spectra are defined according to eq.~\eqref{eq:DeltaAlphaSI} and the water concentrations for the acid solution $c^*_{\rm w}$ are varied (colored solid lines) with ratios reported in the legend. For comparison experimental data for a \SI{4}{M} \ac{HCl} solution is shown as a black broken line \cite{Thaemer2015}. }
\label{fig:normalization}
\end{figure}

The normalization of the difference spectra is subtle and in many studies not well documented as discussed in the previous section \ref{spectraSection}. 
In fig.~\ref{fig:normalization}, we illustrate how small deviations in the normalization constants can lead to drastic variations among the difference spectra.
Throughout this work difference spectra $\omega\Delta\chi''_{c_\text{w}}(\omega)$ are defined as normalized by the water concentration $c_{\rm w}$ of the aqueous \ac{HCl} solutions and  by the concentration $c_\text{w}^\text{0}$ of the reference water spectrum
\begin{equation}
\omega\Delta\chi''_{c_\text{w}}(\omega)= \frac{1}{c_\text{w}} \omega \chi''_\text{solution}(\omega)-\frac{1}{c_\text{w}^\text{0}}\omega \chi''_\text{water}(\omega),
	\label{eq:DeltaAlphaSI}
\end{equation}   
according to eq.~\eqref{eq:DeltaAlpha} from the methods section. 
In fig.~\ref{fig:normalization}, the difference spectra of the \SI{4}{M} \ac{HCl} solution, obtained from ab initio \ac{MD} simulations in this study, are shown as a blue solid line and compared to difference spectra, that are obtained if instead slightly varied water concentrations for the acid solution $c^*_{\rm w}$ are used in the normalization.
The data illustrates, that even in the moderate range of $c^*_{\rm w}/c_{\rm w}=\SIrange{0.8}{1.2}{}$, some features of the difference spectra, like the dip at the position of the water bend around \SI{1650}{cm^{-1}} (\SI{50}{THz}), or the features in the OH stretch above \SI{3000}{cm^{-1}} (\SI{90}{THz}), are significantly changed. 
These features therefore have to be considered with great care.
However, the dominant spectroscopic features, like the continuum band between \SIrange{2000}{3000}{cm^{-1}} and the peak around \SI{1200}{cm^{-1}} (\SI{35}{THz}), that are associated with the excess proton dynamics and are at the focus of this study, are modified in height but not in shape and seem therefore very robust with respect to the normalization protocol.

\clearpage

\section{Kramers-Kronig estimate of complex index of refraction}
\label{kkSection}

Our THz spectroscopic experiments measure the frequency dependent extinction coefficient, $\alpha_\text{solution}(\omega)$, defined by the Beer-Lambert law
\begin{equation}
	\alpha_\text{solution}(\omega)=\frac{1}{d}\ln \left(\frac{I_{water}(\omega)}{I_{solution}(\omega)}\right)+\alpha_\text{water}(\omega),
	\label{eq:AlphaSolution}
\end{equation}       
where $d$ is the sample thickness, $I_{water}(\omega)$ and $I_{solution}(\omega)$ are the experimental transmitted intensities of the water reference and the sample. $\alpha_\text{water}(\omega)$  is the extinction coefficient of bulk water and is taken from literature \cite{Bertie1996}. For comparison to the simulated data the power loss, i.e. the absorption spectrum,
proportional to  the imaginary part of the dielectric susceptibility $\chi''(\omega)$, has to be computed from the experimental extinction coefficient.

The extinction coefficient $\alpha(\omega)$ is related to $n''(\omega)$, the imaginary part of the index of refraction, via
\begin{align}
\label{eq:alpha}
\alpha(\omega) = \frac{2 \omega n''(\omega)}{c},
\end{align}
where $c$ is the speed of light in vacuum.

The complex dielectric susceptibility $\chi(\omega)$ is related to the complex index of refraction $n(\omega)$ as \cite{Jackson1999}
\begin{align}
n^2(\omega)&=[n'(\omega)+i\,n''(\omega)]^2 \\
&= 1+\chi(\omega)= 1+\chi'(\omega)+i \chi''(\omega),
\end{align}

from which follows
\begin{align}
\chi'(\omega)=n'^2(\omega)-n''^2(\omega) -1\\
\label{eq:chiFromN}
\chi''(\omega)=2n'(\omega)n''(\omega).
\end{align}
The missing real part of the index of refraction $n'(\omega)$ is related to the imaginary part $n''(\omega)$ by the Kramers-Kronig relation \cite{Lucarini2005}
\begin{align}
\label{eq:KK}
n'(\Omega)  = n'(\infty) + \frac{2}{\pi} \text{p.v.} \int_0^{\infty} \frac{\omega n''(\omega)}{\omega^2-\Omega^2} d\omega,
\end{align}
where p.v. denotes the Cauchy principal value and $n'(\infty)$ is the real index of refraction at the upper boundary of the integral.
In practice, since experimental data is only available in a limited frequency range, extrapolation beyond the data range may improve the Kramers-Kronig inversion and a so-called `anchor point', obtained from an independent measurement, can be used to replace $n'(\infty)$ \cite{Goplen1980}. For measurements in the \ac{IR} regime, a good estimate for $n'(\infty)$ is obtained by using the real index of refraction measured in the visible \cite{Mayerhofer2019a}.

\begin{figure}[b]
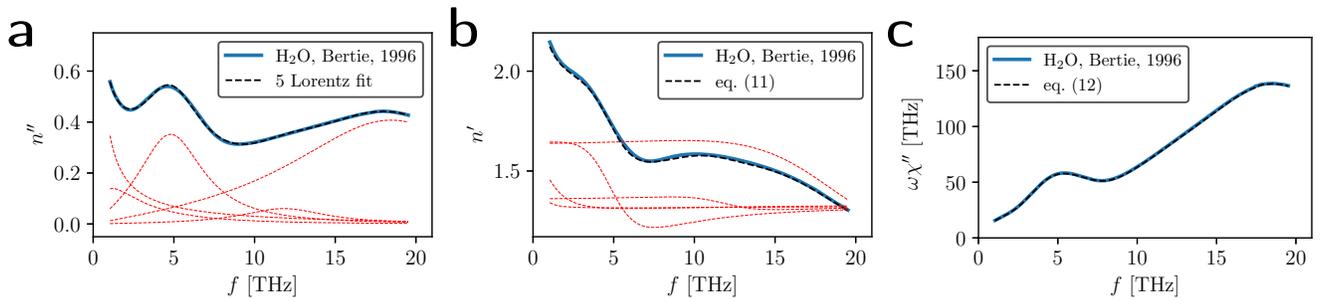

	\centering
	\begin{overpic}[width=.32\linewidth]{{/../figs/kImag_fit-eps-converted-to}.pdf}
	\put(-2,61){\huge \bfseries \sffamily  a}
	\end{overpic}
	\begin{overpic}[width=.32\linewidth]{{/../figs/nReal_fit-eps-converted-to}.pdf}
	\put(-2,61){\huge \bfseries \sffamily  b}
	\end{overpic}
	\begin{overpic}[width=.32\linewidth]{{/../figs/chiImag_fit-eps-converted-to}.pdf}
	\put(-2,61){\huge \bfseries \sffamily  c}
	\end{overpic}		
	\caption{Example of the calculation of the absorption spectrum $\omega \chi''$ from refractive index data. The blue solid lines show literature data of water \cite{Bertie1996}. 
{\bfseries \sffamily  a:} Imaginary part of the index of refraction $n''$ and a fit (black broken line) as a sum of five Lorentz-type fit functions eq.~\eqref{eq:LorentzAnsatz1} (red broken lines). 
{\bfseries \sffamily  b:} Real part of the index of refraction $n'$ and the estimate (black broken line) obtained from the Kramers-Kronig transform eq.~\eqref{eq:kkFitnReal} of the fit in a. The real parts of the individual fit functions according to eq.~\eqref{eq:LorentzAnsatz2} are shown as red broken lines, shifted by the offset $n'_{\rm lit.}=1.33$. {\bfseries \sffamily  c:} Absorption spectrum $\omega \chi''$ and the estimate (black broken line) that is computed from the original data for $n''$ in a and the estimate for $n'$ in b.}
\label{fig:kkBenchmark}
\end{figure}

\begin{figure}[t]
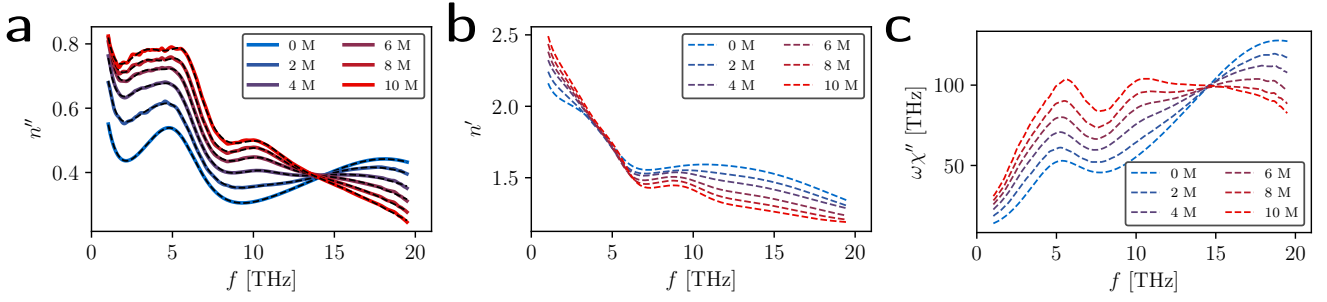

	\centering
	\begin{overpic}[width=.32\linewidth]{{/../figs/kImag_Exp_fit}.pdf}
	\put(-2,61){\huge \bfseries \sffamily  a}
	\end{overpic}
	\begin{overpic}[width=.32\linewidth]{{/../figs/nReal_Exp_fit}.pdf}
	\put(-2,61){\huge \bfseries \sffamily  b}
	\end{overpic}
	\begin{overpic}[width=.32\linewidth]{{/../figs/chiImag_Exp_fit}.pdf}
	\put(-2,61){\huge \bfseries \sffamily  c}
	\end{overpic}		
	\caption{Calculation of the absorption spectra $\omega \chi''$ from experimental THz extinction coefficient spectra of aqueous \ac{HCl} solutions at various concentrations.
{\bfseries \sffamily  a:} Imaginary part of the index of refraction $n''$ computed directly from the experimental absorption spectra (colored solid lines) and fits (black broken lines). The pure water data (blue solid line, \SI{0}{M}) is taken from literature \cite{Bertie1996}. Each fit is a sum of five Lorentz-type fit functions eq.~\eqref{eq:LorentzAnsatz1}. 
{\bfseries \sffamily  b:} Estimates of the real part of the index of refraction $n'$ (colored broken lines) obtained from the fits in a according to eq.~\eqref{eq:kkFitnReal}. {\bfseries \sffamily  c:} Absorption spectra $\omega \chi''$ (colored broken lines) that are computed from the original data for $n''$ in a and the estimates for $n'$ in b.}
	\label{fig:kkExpData}
\end{figure}

We perform the Kramers-Kronig transform using a set of Ansatz functions for the index of refraction, each consisting of a real part $l'(\omega)$ and an imaginary part $l''(\omega)$, which follow approximately from a Lorentz oscillator model for the dielectric susceptibility \cite{Mayerhofer2019a}
\begin{align}
\label{eq:LorentzAnsatz1}
l''(\omega) = \frac{\beta \omega}{(\omega_0^2-\omega^2)^2 + \beta^2 \omega^2} \\
\label{eq:LorentzAnsatz2}
l'(\omega) = \frac{\omega_0^2-\omega^2}{(\omega_0^2-\omega^2)^2 + \beta^2 \omega^2},
\end{align}
and extrapolate beyond the range of the available experimental data. We fit the imaginary index of refraction obtained from the experimental extinction coefficient $\alpha(\omega)$ via eq.~\eqref{eq:alpha} as
\begin{align}
\label{eq:kkFitkImag}
n''(\omega)  = \sum_i A_i l''_i(\omega),
\end{align}
where the $l''_i(\omega)$ contain further fit parameters $\omega_{0,i}$ and $\beta_i$.
The real index of refraction is then given by
\begin{align}
\label{eq:kkFitnReal}
n'(\omega)  = n'_{\rm lit.} + \sum_i A_i l'_i(\omega) ,
\end{align}
where the constant real index of refraction in the visible $n'_{\rm lit.}=1.33$ is taken from the literature, which has been shown to be in general a good approximation for biological matter \cite{Mayerhofer2019a}. The absorption spectrum, $\omega \chi''(\omega)$, which is eventually used to compare to the simulation data, is computed using eqs.~\eqref{eq:alpha} and \eqref{eq:chiFromN} \cite{Schienbein2017} as
\begin{align}
\omega \chi''(\omega) = c\, \alpha(\omega) n'(\omega),
\end{align}
where $\alpha(\omega)$ is the original data and $n'(\omega)$ is parametrized according to eq.~\eqref{eq:kkFitnReal}. As a benchmark example, fig.~\ref{fig:kkBenchmark} illustrates the result of this procedure applied to literature data of water \cite{Bertie1996} in the regime \SIrange{10}{600}{cm^{-1}}, for which the experimental THz spectra are obtained. 
In fig.~\ref{fig:kkExpData}a the fits to the experimental THz data of aqueous \ac{HCl} solutions in the range \SIrange{2}{10}{M} (colored solid lines) are shown as black broken lines, that are used to estimate the real part of the refractive index, shown in fig.~\ref{fig:kkExpData}b as broken colored lines. Eventually the resulting absorption spectra, shown in fig.~\ref{fig:kkExpData}c, are used to compare to simulation data in the main text.

\clearpage

\section{Spatial correlations}
\label{radialDistSection}

Spatial correlations in liquids can be quantified with \acp{RDF}, $g_{ab}(r)$, which denote the average density of particles of type $a$ at a relative distance $r$ around particles of type $b$. The local average density of $a$ relative to $b$ at distance $r$ is given as $\rho_a g_{ab}(r)$ where $\rho_a = N_a/V$ is the global density of $a$. The \ac{RDF} is computed from simulation data of two particle species with  numbers $N_a$ and $N_b$ that are located at positions $\vec{r}_i$ and $\vec{r}_j$ \cite{Hansen2013}
\begin{align}
\label{eq:rdf}
g_{ab}(r) = \frac{V} {4 \pi r^2 N_{a} N_{b} } \sum_{i=1}^{N_a} \sum_{j=1}^{N_b}
               \langle \delta(|\vec{r}_i - \vec{r}_j| - r) \rangle.
\end{align}
Average coordination numbers of species $a$ within distance $r$ of particles of species $b$ are obtained by integration over $g_{ab}(r)$
\begin{align}
\label{eq:irdf}
N_{ab}(r) = \rho_a\, \int_0^r \!\!dr' 4\pi r'^2 g_{ab}(r').
\end{align}

In this section, the spatial correlations of the excess protons in the \ac{HCl} solutions are analyzed using \acp{RDF}, specifically also the correlations with the chloride ions.
Excess protons are identified in the simulation data as the remaining protons after the water molecules are assembled for each oxygen atom with the closest two hydrogen atoms at each time step of the simulation.
This procedure is consistent with the one of section \ref{diffusionSection} and not excluding protons close to the chloride ions as done for the analysis of the proton-transfer statistics in section \ref{selectionSection} and in the main text. 
Hydronium ions are defined for each excess proton together with the water molecule of the closest oxygen atom.

\begin{figure}[hb]
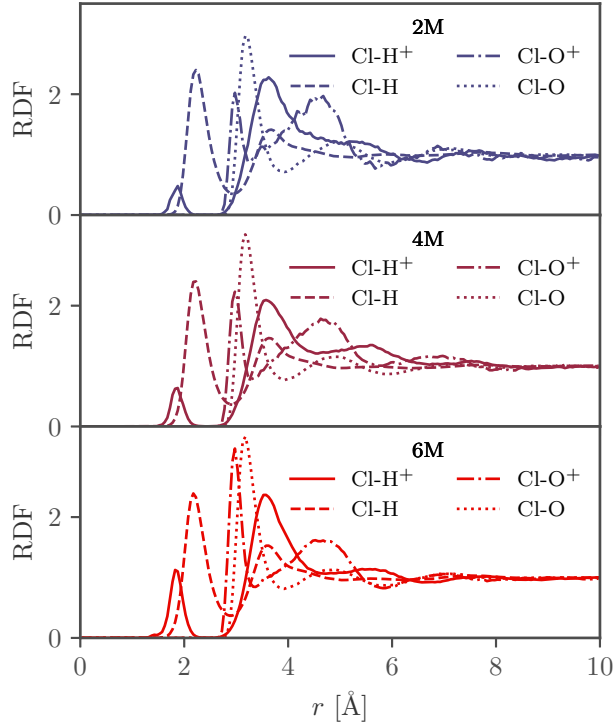

	\centering
	\begin{overpic}[width=.49\linewidth]{{/../figs/protons_rdfs_cl_all-eps-converted-to}.pdf}
	\end{overpic}	
	\caption{Spatial correlations between the chloride ions and other nuclei in terms of \aclp{RDF} (\acsp{RDF}) as obtained from ab initio \ac{MD} simulations of \ac{HCl} solutions at various concentrations. \acp{RDF} are shown for the excess protons (H$^+$) as solid lines, all hydrogen nuclei (H), including excess protons, as broken lines, the oxygen nuclei of the water molecules (O) as dotted lines and the oxygen nuclei of the hydronium ions (O$^+$) as dash-dotted lines.}
\label{fig:rdfs_cl}
\end{figure}

\subsection{Correlations with the chloride ions}
\label{clRdfSection}

In fig.~\ref{fig:rdfs_cl} the \acp{RDF} of the chloride ions are shown for the three \ac{HCl} concentrations with respect to the excess protons as solid lines, all hydrogen nuclei, including excess protons, as broken lines, the oxygen nuclei of the water molecules as dotted lines and the oxygen nuclei of the hydronium ions as dash-dotted lines.
For all \ac{HCl} concentrations, the general features of the \acp{RDF} are comparable. But, there are a few concentration-dependent trends.
The \acp{RDF} of the excess protons around the chloride ions (solid lines) show two dominant peaks, a small one that is located at around $d_{\rm Cl-H^+}=$ \SI{1.8}{\ang} and thereby left of the major peak of the hydrogen nuclei \acp{RDF} (broken lines) at around $d_{\rm Cl-H}=$ \SI{2.2}{\ang}. 
This peak is important as it corresponds to excess protons coordinated directly with the chloride ion, i.e. the excess proton sits in between the chloride ion and the oxygen atom of the respective hydronium ion. 
This contact ion pair is considered an important intermediate in the solvation of the \ac{HCl} \cite{Baer2014}.
Yet, the excess protons are still part of the hydronium ion, since an even smaller distance of around \SI{1.4}{\ang} would be expected for the covalent bond to the chloride atom, which appears at much higher concentrations than \SI{6}{M} \cite{Baer2014}.
The peak height at $d_{\rm Cl-H^+}=$ \SI{1.8}{\ang} (solid lines) clearly increases with \ac{HCl} concentration in fig.~\ref{fig:rdfs_cl}, but even for the largest concentration of \SI{6}{M} it remains smaller than the second and largest peak in the \acp{RDF} at about \SI{3.5}{\ang}.
This peak is located to the right of the first and dominant peak in the \acp{RDF} of the oxygen nuclei of the hydronium ions (dash-dotted lines) at around $d_{\rm Cl-O^+}=$ \SI{3.0}{\ang}, which occurs at a slightly smaller distance than the respective peak of all oxygen nuclei at around $d_{\rm Cl-O}=$ \SI{3.1}{\ang}. 

All of the presented data is in good agreement with results by \citet{Baer2014} from comparable ab initio simulations, who additionally show data for much higher concentrations up to \SI{16}{m} ($\SI{16}{mol/kg}\simeq \SI{11.7}{M}$) and successfully reproduced extended X-ray absorption fine structure (EXAFS) experimental measurements \cite{Fulton2010}.
Specifically, they report $d_{\rm Cl-O^+}=$ \SI{2.96}{\ang} and $d_{\rm Cl-O}=$ \SI{3.14}{\ang} in excellent agreement with experimental EXAFS data \cite{Fulton2010}. 
Data comparing various exchange-correlation functionals used in the ab initio simulation also consistently reproduced the results presented in fig.~\ref{fig:rdfs_cl} with no significant dependence on the type of functional (except for the PBE functional) and albeit large statistical uncertainties in some data \cite{Fischer2019}.

When interpreting the \acp{RDF}, the actual densities of the species need to be put into perspective.
The global densities of chloride ions, excess protons and hydronium ions are equal by definition.
However, for the \SI{2}{M} \ac{HCl} solution data set, the ratio of water molecules to chloride ions or respectively excess protons is about 24.8 and drops to 11.2 for the \SI{4}{M} and 6.5 for the \SI{6}{M} data sets. These numbers would have to be divided by two, assuming that water molecules are equally and exclusively solvating all chloride ions and excess protons.
It is therefore evident, that for the \SI{6}{M} solution, hydronium ions are necessarily residing already in the first hydration shell of the chloride ions.
More precisely, the average coordination numbers around the chloride ions are obtained by integration over the first peak of each \ac{RDF} in fig.~\ref{fig:rdfs_cl} and use of eq.~\eqref{eq:irdf}.
The obtained coordination numbers are reported in tab.~\ref{tab:cl_coord} and are generally in good agreement with previous simulation and experimental results \cite{Baer2014,Fulton2010}.
The average number of hydronium ions in the first hydration shell of any chloride ion increases from 0.13 for \SI{2}{M} to 0.52 for \SI{6}{M}, see third column in tab.~\ref{tab:cl_coord}. 
Previously, \citet{Baer2014} interpreted this significant increase in the coordination number with an increase of contact ion pairs, where the excess proton is shared between the hydronium oxygen and the chloride ion.
However, the coordination number of the excess protons around chloride ions does not increase as much, only from 0.007 for \SI{2}{M} to 0.054 for \SI{6}{M}, see first column in tab.~\ref{tab:cl_coord}. 
Therefore, the excess protons of the hydronium ions in the first solvation shell of the chloride ions mostly point away from chloride ions and are predominantly coordinated with other water molecules, rather than with chloride ions.
However, it is noteworthy that the chloride-excess-proton coordination is affected, when a different basis set is considered in the simulations, which is shown below in subsection \ref{basisSetSection} for simulations at \SI{6}{M}. 
In particular, an increase of the coordination of the excess protons with the chloride ions is observed when considering the MOLOPT or TZV2P basis sets. 
Therefore, while the above results have to be considered with care when comparing to experimental data and additional analysis with more elaborate basis sets may be useful, the analysis performed in this study is self-consistent for the MOLOPT-SR basis set.

\begin{table}
\caption{Average coordination numbers $N_{ab}$ around the chloride ions as obtained from the first peak of the \acp{RDF} in fig.~\ref{fig:rdfs_cl} and eq.~\eqref{eq:irdf}. 
Errors of the simulation data are estimated from the resolution of \acp{RDF}. 
}
\begin{tabular}{c|r r r r}
& \quad Cl-H$^+$ ($r<$ \SI{2.5}{\ang}) & \quad Cl-H ($r<$ \SI{3.0}{\ang})  & \quad Cl-O$^+$ ($r<$ \SI{3.5}{\ang})  & \quad Cl-O ($r<$ \SI{3.5}{\ang})  \\
\hline
\hline
\SI{2}{M} & $0.007 \pm 0.001$ & $5.64 \pm 0.07$ & $0.13 \pm 0.01$ & $4.8 \pm 0.1$ \\
\SI{4}{M} & $0.020 \pm 0.001$ & $5.42 \pm 0.07$ & $0.26 \pm 0.01$ & $4.6 \pm 0.1$ \\
\SI{6}{M} & $0.054 \pm 0.001$ & $5.12 \pm 0.07$ & $0.51 \pm 0.01$ & $4.2 \pm 0.1$ \\
\hline
\SI{6}{M}, MOLOPT  & $0.091 \pm 0.001$ & $5.0 \pm 0.1$ & $0.56 \pm 0.02$ & $4.0 \pm 0.2$ \\
\SI{6}{M}, TZV2P & $0.115 \pm 0.001$ & $5.0 \pm 0.2$ & $0.63 \pm 0.02$ & $4.0 \pm 0.2$ \\
\hline
\SI{2.5}{m} \cite{Baer2014} & & & 0.17 & 5.82  \\
\SI{6}{m} \cite{Baer2014} &  &  & 0.42 & 5.21 \\
\SI{10}{m} \cite{Baer2014} & &  & 0.71 & 4.67 \\
\SI{16}{m} \cite{Baer2014} &  &   & 1.05 & 3.99 \\
\hline 
\SI{6}{m} \cite{Baer2014, Fulton2010} &  &  & $0.7 \pm 0.2$ & $5.1 \pm 0.5$ \\
\SI{10}{m} \cite{Baer2014, Fulton2010} & &  & $1.0 \pm 0.3$ & $4.8 \pm 0.5$ \\
\SI{16}{m} \cite{Baer2014, Fulton2010} &  &  &  $1.6 \pm 0.3$ & $4.2 \pm 0.5$ \\
\end{tabular}
\label{tab:cl_coord}
\end{table}

Taken together, this data indicates that the excess protons in aqueous \ac{HCl} solutions mostly reside inside hydronium ions within our definition, and additionally stay away from the chloride ions, i.e. are located between two oxygen atoms and not between a chloride and an oxygen atom.
This holds true even for moderately high concentrations up to \SI{6}{M} considered in this study, but necessarily breaks down at very high concentrations when the water becomes saturated with \ac{HCl} \cite{Baer2014}.
The data confirms the high solubility of excess protons in water, due to energetic but also entropic effects \cite{Calio2020, Carpenter2019}.
Furthermore, it justifies the focus on proton transfer between water molecules and the excess-proton selection and identification scheme in section \ref{selectionSection}  to analyze the excess-proton spectral signatures of the \ac{HCl} solutions.
Note, that already the linear trend of the \ac{HCl} difference spectra with concentration shown in fig.~\ref{hcl_intro}d in the main text indicates that in the range up to \SI{6}{M} considered in the ab initio \ac{MD} simulations in this study, the effect of the chloride ions on the excess-proton spectra and dynamics is negligible. 
Furthermore, \citet{Napoli2018} previously showed that \ac{IR} spectra of excess protons coordinated with chloride ions show much weaker spectral signatures as compared to the average spectra of all the excess protons.

\subsection{Correlations between oxygen and hydrogen atoms}

\begin{figure}[tb]
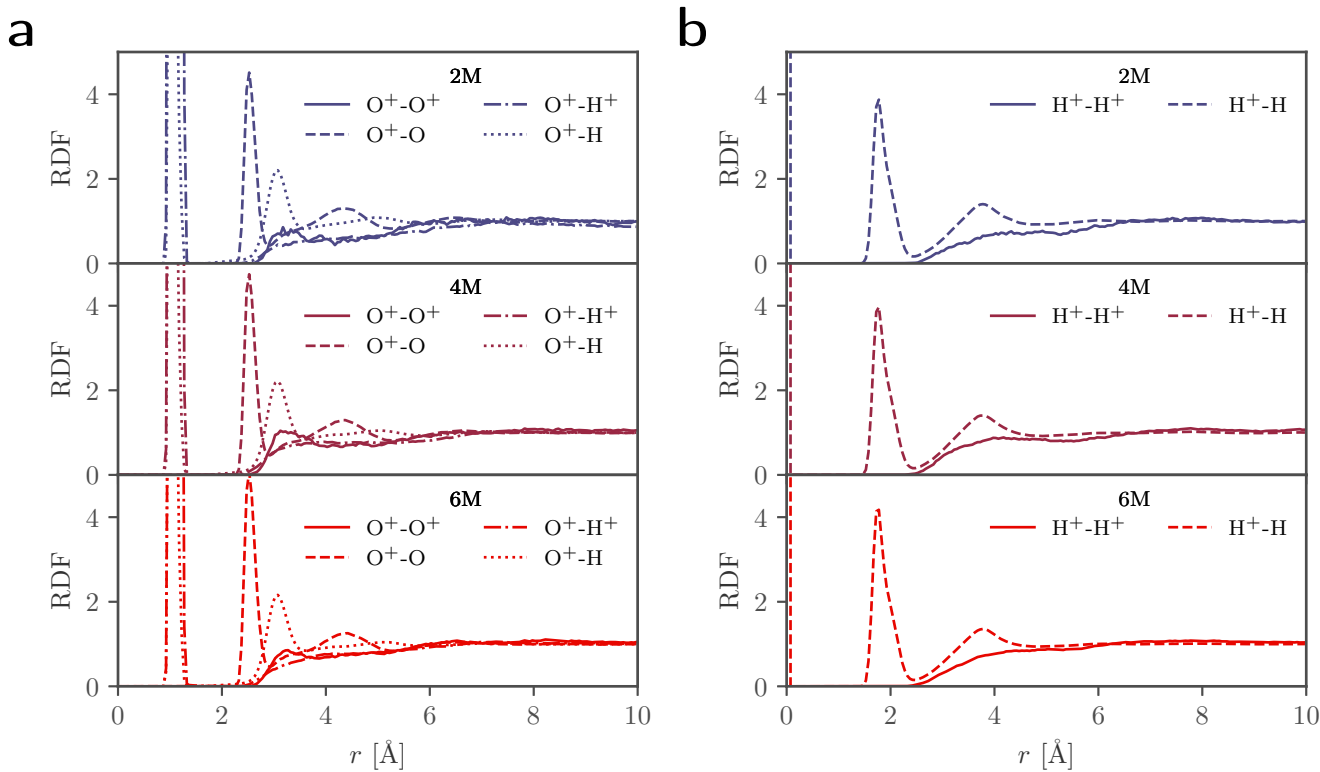

	\centering
	\begin{overpic}[width=.49\linewidth]{{/../figs/protons_rdfs_o+_all-eps-converted-to}.pdf}
	\put(-2,97){\huge \bfseries \sffamily  a}
	\end{overpic}	
	\begin{overpic}[width=.49\linewidth]{{/../figs/protons_rdfs_h+_all-eps-converted-to}.pdf}
	\put(-2,97){\huge \bfseries \sffamily  b}
	\end{overpic}	
	\caption{Spatial correlations between the oxygen atoms of hydronium ions (a) and excess protons (b) with respect to other nuclei in terms of \aclp{RDF} (\acsp{RDF}) as obtained from ab initio \ac{MD} simulations of \ac{HCl} solutions at various concentrations. 
{\bfseries \sffamily  a:} \acp{RDF} are shown for the oxygen nuclei of the hydronium ions (O$^+$) as solid lines, for the oxygen nuclei of the water molecules (O) as broken lines, for the the excess protons (H$^+$) as dash-dotted lines and all hydrogen nuclei (H), including excess protons, as dotted lines.
{\bfseries \sffamily  b:} \acp{RDF} are shown for the excess protons (H$^+$) as solid lines and all hydrogen nuclei (H), including excess protons, as broken lines.
}
\label{fig:rdfs_ho}
\end{figure}

In fig.~\ref{fig:rdfs_ho}a the spatial correlations of the hydronium ions and in fig.~\ref{fig:rdfs_ho}b of the excess protons themselves are discussed. 
Comparable results for aqueous \ac{HCl} solutions were previously shown for various exchange-correlation functionals used in ab initio simulations \cite{Fischer2019} and for different self-consistent iterative multistate empirical valence bond (SCI-MS-EVB) simulations \cite{Xu2010,Calio2020}.

In fig.~\ref{fig:rdfs_ho}a the \acp{RDF} of the oxygen nuclei of the hydronium ions are shown with respect to the oxygen nuclei of other hydronium ions (O$^+$) as solid lines, for the oxygen nuclei of the water molecules (O) as broken lines, for the the excess protons (H$^+$) as dash-dotted lines and all hydrogen nuclei (H), including excess protons, as dotted lines.
Next to the trivial peaks of the dash-dotted as well as the dotted lines at below \SI{1.0}{\ang}, belonging to the hydrogen nuclei which are part of the hydronium ions itself,
a clear peak of the broken lines at $d_{\rm O^+-O}=$ \SI{2.5}{\ang} indicates the water oxygen nuclei in the first hydration shell, each of which are candidates to form a transient H$_5$O$_2{}^+$ complex, i.e. the special pair, together with the hydronium ion.
This peak is roughly consistent with the most probable oxygen-oxygen distance of the transient H$_5$O$_2{}^+$ complex, that is seen in the free energy in fig.~\ref{hcl_f}a in the main text.
A second, much smaller, peak at about \SI{4.2}{\ang} indicates the water molecules in the second hydration shell of the hydronium ion.
Boths peaks at about $d_{\rm O^+-O}=$ \SI{2.5}{\ang} and at \SI{4.2}{\ang} are clearly shown in data for various exchange-correlation functionals \cite{Fischer2019}, as well as for SCI-MS-EVB simulations \cite{Xu2010, Calio2020}. 
Beyond that, a decomposition of the first peak into three components was used to support the picture of an asymmetric Eigen state \cite{Calio2021}.

Each peak is accompanied by peaks of the dotted lines at slightly larger distances, which indicate the hydrogen atoms belonging to the water molecules in the respective hydration shells.
Additionally, the dotted lines show a weak and broad shoulder below roughly \SI{2.5}{\ang}, the origin of which is further discussed in section \ref{hbSection}.
A slight relative maximum of the solid lines at $d_{\rm O^+-O^+}=$ \SI{3.00}{\ang} but with a magnitude below one indicates a metastable structure between two hydronium ions. 
However, note that there is no such signature in the dash-dotted lines, i.e. no apparent correlation with the excess proton of the nearest hydronium ion. 
The peak at $d_{\rm O^+-O^+}=$ \SI{3.00}{\ang} was also observed by \citet{Calio2020} and at $d_{\rm O^+-O^+}=$ \SI{3.20}{\ang} by \citet{Xu2010}.

In fig.~\ref{fig:rdfs_ho}b the \acp{RDF} of the excess protons are shown with respect to the other excess protons (H$^+$) as solid lines and all hydrogen nuclei (H), including excess protons, as broken lines.
The main peak of the broken lines at $d_{\rm H^+-H}=$ \SI{1.8}{\ang} indicates the mean distances between hydrogen atoms within the same hydronium ion. 
The peak shows a shoulder at around \SI{2.1}{\ang} possibly related to spatial correlations between the excess proton and two hydrogen atoms in the neighboring water molecule of the transient H$_5$O$_2{}^+$ complex, i.e. of the special pair. 
The second peak of the broken lines at \SI{3.8}{\ang} is related to other water molecules in the first hydration shell, that form \acp{HB} with the hydronium ion to which the excess proton is assigned.
No spatial correlations are apparent in the \acp{RDF} between excess protons, shown as solid lines.
In contrast to that, \citet{Xu2010} observed a slight peak at $d_{\rm H^+-H^+}=$ \SI{4.20}{\ang}, indicative of spatial correlations between two excess protons.

In summary, the \acp{RDF} of the hydronium ions as well as of the excess protons presented in figs.~\ref{fig:rdfs_ho}a and b indicate no spatial correlations between excess protons or hydronium ions and other excess protons.
While there appears a slight relative maximum in the \ac{RDF} between hydronium ions itself, this may simply be related to closest-packing effects in the liquid. 

\subsection{Results for different basis sets}
\label{basisSetSection}

\begin{figure}[hb]
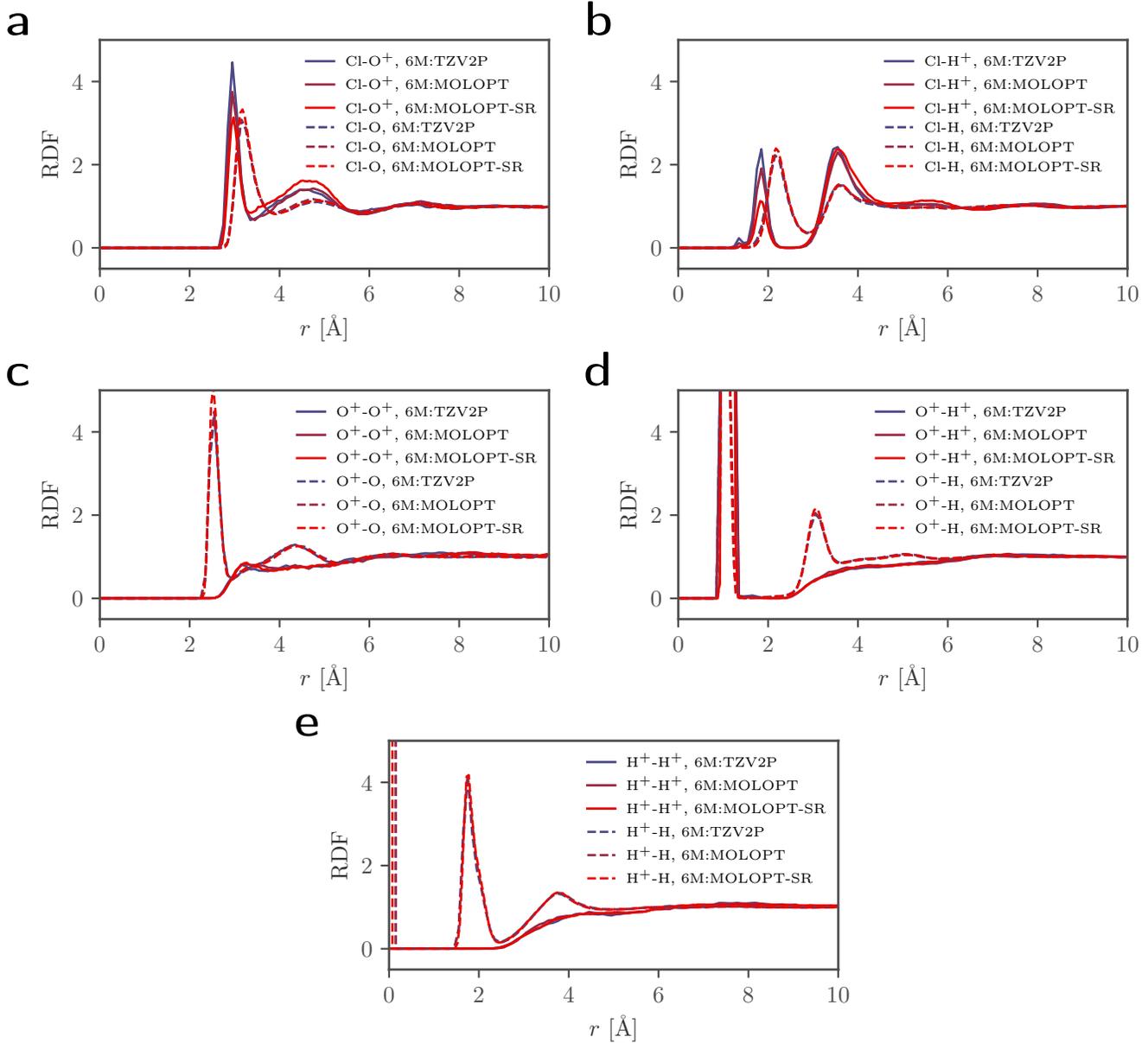

	\centering
	\begin{overpic}[width=.49\linewidth]{{/../figs/protons_rdfs_compareBasis_conc_cl-o+-eps-converted-to}.pdf}	
	\put(-2,58){\huge \bfseries \sffamily  a}
	\end{overpic}	
	\begin{overpic}[width=.49\linewidth]{{/../figs/protons_rdfs_compareBasis_conc_cl-h+-eps-converted-to}.pdf}
	\put(-2,58){\huge \bfseries \sffamily  b}
	\end{overpic}	
	\begin{overpic}[width=.49\linewidth]{{/../figs/protons_rdfs_compareBasis_conc_o+-o+-eps-converted-to}.pdf}
	\put(-2,58){\huge \bfseries \sffamily  c}
	\end{overpic}	
	\begin{overpic}[width=.49\linewidth]{{/../figs/protons_rdfs_compareBasis_conc_o+-h+-eps-converted-to}.pdf}
	\put(-2,58){\huge \bfseries \sffamily  d}
	\end{overpic}		
	\begin{overpic}[width=.49\linewidth]{{/../figs/protons_rdfs_compareBasis_conc_h+-h+-eps-converted-to}.pdf}
	\put(-2,58){\huge \bfseries \sffamily  e}
	\end{overpic}		
	\caption{Spatial correlations between the chloride ions and oxygen atoms (a), chloride ions and hydrogen atoms (b), oxygen atoms of hydronium ions and oxygen atoms (c) or hydrogen atoms (d),  and excess protons (e)  with respect to other nuclei in terms of \aclp{RDF} (\acsp{RDF}) as obtained from ab initio \ac{MD} simulations of \ac{HCl} solutions at \SI{6}{M} for various basis sets considered in the simulation. 
Data obtained for the TZV2P basis set is shown in blue, for the MOLOPT basis set in purple and for the MOLOPT-SR basis set in red.
{\bfseries \sffamily  a,c:} \acp{RDF} are shown for the oxygen nuclei of the hydronium ions (O$^+$) as solid lines, for the oxygen nuclei of the water molecules (O) as broken lines.	
{\bfseries \sffamily  b,d,e:} \acp{RDF} are shown for the excess protons (H$^+$) as solid lines and all hydrogen nuclei (H), including excess protons, as broken lines.
}
\label{fig:rdfs_bs}
\end{figure}

To address the quality of the simulations for the chosen MOLOPT-SR basis set, additional simulations of \ac{HCl} solutions at \SI{6}{M} with the DZVP-MOLOPT-GTH and TZV2P-GTH basis sets are performed \cite{VandeVondele2007, VandeVondele2005}.
For both cases the chloride ions are modeled including diffuse functions by using the aug-DZVP-GTH and aug-TZV2P-GTH basis sets, respectively. 
Otherwise the setups are equivalent to the MOLOPT-SR simulations and each simulation is run for \SI{28}{ps} under NVT condition.
The spatial correlations in each data set are then compared by computing \acp{RDF}, the results of which are presented in figs.~\ref{fig:rdfs_bs}a--e.
The correlations of the oxygen atoms and the hydrogen atoms among each other, that are most important for the solvation structure of the excess protons and therefore the focus of this study as discussed above, show no significant dependence on the choice of basis set as is seen from the near-perfect agreement of the \ac{RDF} curves throughout figs.~\ref{fig:rdfs_bs}c--e.
Regarding the chloride ions, some differences in the \acp{RDF} in figs.~\ref{fig:rdfs_bs}a and b are clearly discernible though.
Most importantly, the first major peaks in the \acp{RDF} between chloride ions and oxygen atoms of hydronium ions (solid lines in fig.~\ref{fig:rdfs_bs}a) as well as between chloride ions and excess protons (solid lines in fig.~\ref{fig:rdfs_bs}b) increase when the MOLOPT or TZV2P basis sets are used.
Additionally, a second much smaller peak is visible at a closer distance for both simulations with the MOLOPT and TZV2P basis sets (solid blue and purple lines in fig.~\ref{fig:rdfs_bs}b), indicating that contact pairs between chloride ions and excess protons appear rarely throughout the simulations. 
This means that the role of the chloride spectator, that is argued in section \ref{clRdfSection} to be of minor importance, may be underestimated in simulations using the MOLOPT-SR basis set.
In section \ref{clRdfSection} the coordination number of the excess protons around chloride ions at \SI{6}{M} was found to be 0.054 from integration over the first major peak of the \ac{RDF} up to \SI{2.5}{\ang} (solid red line in fig.~\ref{fig:rdfs_bs}b and first column in  tab.~\ref{clRdfSection}). 
When considering the MOLOPT basis set, this value increases to 0.091 and for the TZV2P basis set to 0.115, meaning that on average each excess proton is for 9\% or respectively 12\% of the time coordinated with a chloride ion as the second nearest neighbour (or rarely even as the nearest neighbor) in a configuration that is not well described by a transient H$_5$O$_2{}^+$ complex which forms the basis of the analysis performed in the main text.

\newpage

\section{Excess-proton identification}
\label{selectionSection}

The identification of excess protons from trajectory data is a non-trivial task since the notion of an excess proton itself is subtle. While for the idealized structure of a Zundel cation it is possible to identify one proton that is qualitatively different from the others, in that it is symmetrically shared between two oxygen atoms, such a discrimination is impossible for an ideal symmetric Eigen cation from a configuration at a single time. Certainly it is possible to simply determine the protons that are the least closely associated to any oxygen atom in the simulation at each time step and call them excess protons. This is a perfectly reasonable choice when examining static properties of excess protons, as for example the \acp{RDF} in section \ref{radialDistSection}. \citet{Marx1999} used a similar definition for the calculation of excess proton probability distributions in ab initio \ac{MD} simulations of protonated water.

In this work, however, we are interested in dynamical properties of excess protons. Since we aim to track and examine protons during their transfer between water molecules, including when they stay close to an oxygen atom, a selection based on a static geometric criterion would not suffice. Due to fluctuations of protons inside a hydronium molecule, also called `special pair dance' \cite{Markovitch2008, Calio2021}, a criterion based on instantaneous interatomic distances would lead to the disruption of otherwise continuous proton-transfer trajectories. Instead, similar to \citet{Calio2021}, we record the trajectories of protons that could potentially transfer from one water to a neighboring one and later discard those that did not. In fact, we are selecting \textit{protons that transfer from one water molecule to another} and call them excess protons for simplicity.

\subsection*{Selection of transfer candidates} 

Ignoring the possibility of spontaneous autoionization of water molecules, proton transfer can only occur from one water molecule to another if the former has an extra proton associated to it, i.e. if it is a hydronium molecule. 
For the identification of hydronium molecules we use a geometric criterion: Each oxygen atom gets assigned its closest two protons which together form water molecules. 
The remaining protons are then assigned to the water molecule with the closest oxygen atom and together make up a hydronium molecule.
All protons that belong to such a hydronium molecule are considered \textit{candidates} for proton transfer. 
Furthermore, we exclude those candidates whose distance to their closest chloride ion is smaller than that to their second closest oxygen atom. 
These protons reside between a chloride ion and an oxygen atom (as opposed to residing between two oxygen atoms as most hydrogen atoms do) and do not exhibit proton transfer between water molecules.

\begin{figure}[tb]
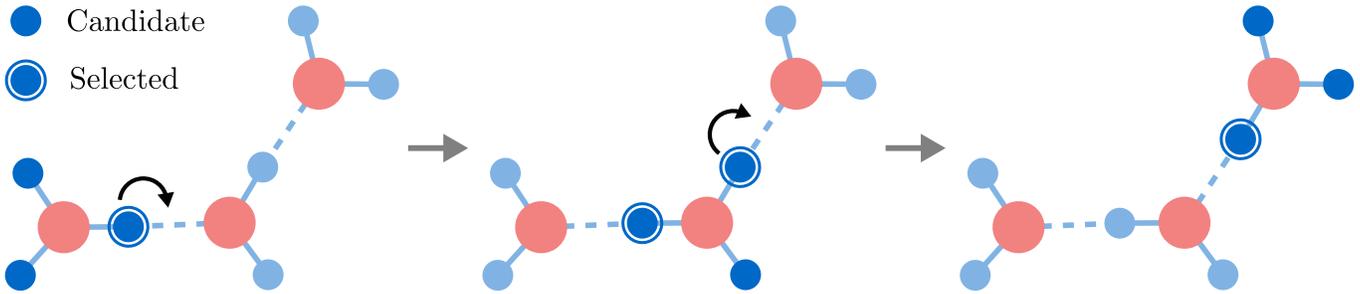

	\centering
	\begin{overpic}[width=1\linewidth]{{/../figs/candidate_illu}.pdf}
	\end{overpic}
\caption{Illustration of three consecutive snapshots during a proton transfer process. Transfer candidates and selected protons are highlighted. 
It becomes apparent why two of the protons belonging to one hydronium molecule are selected at the same time as transfer protons (see second snapshot). 
Such overlapping trajectories lead to the fact that our conditions select more protons than there are excess protons in the simulation.}\label{fig:overlap}
\end{figure}

$5.5\%$, $11.1\%$ and $16.4\%$ of all protons fulfill these conditions in the $\SI{2}{M}$, $\SI{4}{M}$ and $\SI{6}{M}$ simulation, respectively. 
This means that on average $28.9$, $56.4$ and $78.4$ protons, respectively, are being selected as transfer candidates at each time step, which is a little less than three times the number of excess protons $N_{\mathrm{H^+}} = 10, 20$ and 30 in each simulation. 
This is to be expected, since each excess proton corresponds to one hydronium molecule which in turn holds three protons that are \textit{part of $H_3O^+$}. 
The difference to $3 N_{\mathrm{H^+}}$ is due to the excluded protons that are located between an oxygen atom and a chloride ion.

\subsection*{Elimination of trajectories without transfer}

The conditions explained above define proton trajectories that potentially contain proton transfers between two water molecules. 
In order to further reduce the number of protons to those that actually contribute to proton transfer, the trajectories are filtered by another - dynamical - condition that selects only those trajectories which contain a transfer, defined by a change of the proton's closest oxygen atom (the same criterion was used by \citet{Calio2021}). 
This last condition leaves $\sim 50\%$ of the previously selected protons in each simulation that actually contribute to the analysis, which is still more than $N_{\mathrm{H^+}}$ selected protons per time step because some trajectories overlap. 
This is the case since during the lifetime of a hydronium molecule, two of its protons might be tracked at the same time. 
One proton which transfers to initially create the respective hydronium molecule and one that transfers away from it later on. 
Figure~\ref{fig:overlap} illustrates such a situation.

\clearpage

\section{Decomposition of infrared spectra of transient H$_5$O$_2{}^+$ complexes}
\label{crossCorrSection}

\begin{figure}[!ht]
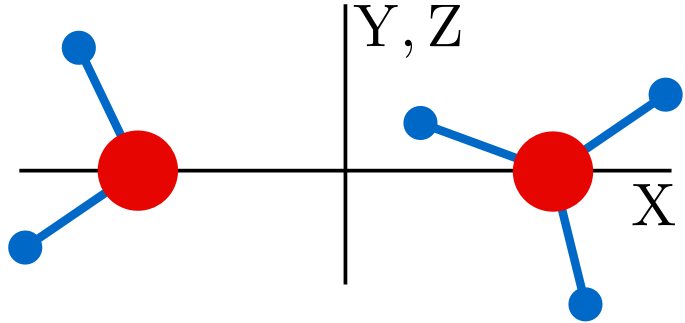

	\centering
	\begin{overpic}[width=.5\textwidth]{{/../figs/coords}.pdf}
	\end{overpic}
\caption{Internal coordinate system of the transient H$_5$O$_2{}^+$ complexes, allowing for a distinction of movements perpendicular and parallel to the connecting axis between the water oxygens (referred to as O-O axis).}\label{fig:coordinates}
\end{figure}

To allow for a decomposition of \ac{IR} spectra of transient H$_5$O$_2{}^+$ complexes into contributions perpendicular and parallel to the connecting axis of the water oxygens, the dynamics of the transient H$_5$O$_2{}^+$ complexes are described in a co-moving internal coordinate system as introduced in the main text and shown in fig.~\ref{fig:coordinates}. The x-axis is defined as the axis connecting both oxygen atoms (in the following referred to as O-O axis), the coordinate origin being located in the middle between the oxygens. The yz-plane lies perpendicular to that axis. A transformation to this coordinate system thus involves a time-dependent translation and rotation of the trajectory $\mathbf r_i(t)$ in the laboratory frame of each atom inside a H$_5$O$_2{}^+$ complex and can be described by
\begin{equation}\label{eq:coordinates}
	\mathbf{r'}_i(t) = \mathbf{M}_{\phi(t),\theta(t)}\,[\mathbf{r}_i(t) - \mathbf{r}_0(t)],
\end{equation}
where $\mathbf{r}_0(t)$ is the trajectory of the oxygen-oxygen midpoint and $\mathbf{M}_{\phi(t),\theta(t)}$ is a rotation matrix setting both oxygen atoms on the x-axis. These operations do of course have an influence on the resulting spectra which is discussed in section \ref{rotationSection}.

The internal coordinate system further defines $d$, the projection of the excess proton location onto the O-O axis, and $R_{\mathrm{OO}}$, the distance between the two oxygen atoms according to
\begin{align*}
R_{\mathrm{OO}}(t) &= |x'_{\mathrm{O}_1}(t)-x'_{\mathrm{O}_2}(t)|\\
d(t) &= x'_{\mathrm{H^+}}(t).
\end{align*}

Using such an internal coordinate system, we can decompose the spectra of protons and water molecules inside the transient H$_5$O$_2{}^+$ complexes as well as their cross-correlation spectra into their contribution along and perpendicular to the O-O axis, respectively.

For that purpose, we extract the trajectories of the closest two water molecules to each previously extracted excess proton trajectory (see section \ref{selectionSection}), resulting in trajectories of transient H$_5$O$_2{}^+$ complexes. Since the water molecules involve valence electrons, we use the localized-Wannier-center trajectory data for this, resulting in trajectories with a slightly lower time resolution of $\SI{4}{fs}$ than for the nuclei-only data. In order to be able to decompose the trajectories into parts perpendicular and parallel to the O-O axis, we describe the multidimensional H$_5$O$_2{}^+$ trajectories using the internal coordinate system introduced above and calculate spectra accordingly.

\begin{figure}[!ht]
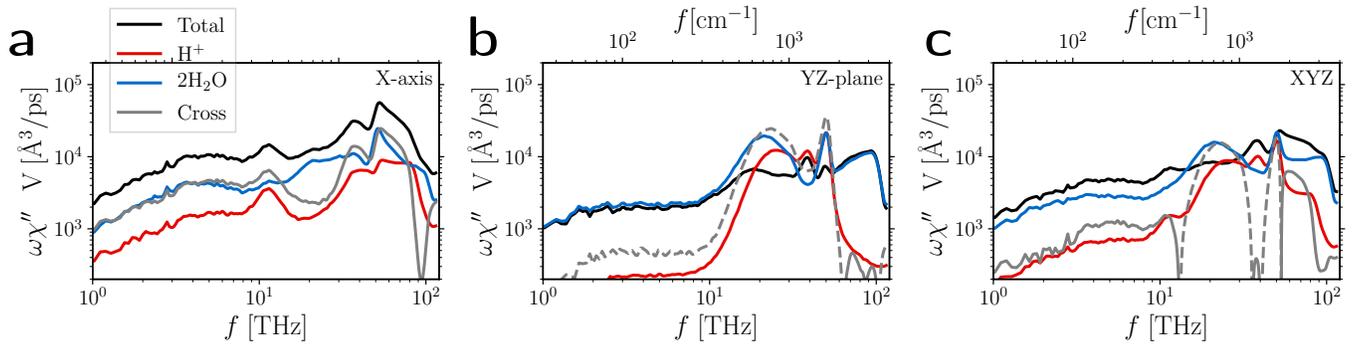

	\centering
	\begin{overpic}[width=\textwidth]{{/../figs/H5O2_specs_X_vs_YZ_vs_orig_rev}.pdf}
	\put(0,22){\huge \bfseries \sffamily  a}
	\put(34,22){\huge \bfseries \sffamily  b}
	\put(68,22){\huge \bfseries \sffamily  c}
	\end{overpic}
	\caption{Absorption spectra including nuclei and electrons of transient H$_5$O$_2{}^+$ complexes consisting of their excess protons (H$^+$, red solid lines), their closest two water molecules (2 H$_2$O, blue solid lines) and cross correlations (Cross, gray solid and broken lines) parallel to the O-O axis (\textbf{a}) and perpendicular to it, i.e. in the yz-plane of the local coordinate system (\textbf{b}). The isotropic spectrum is shown in \textbf{c}. Broken lines indicate negative values.
Note that the spectra are averaged over the spatial dimensions.
	}\label{fig:X_and_YZ}	
\end{figure}

In fig.~\ref{fig:X_and_YZ} the resulting spectra are shown, separated into the contributions along (fig.~\ref{fig:X_and_YZ}A) and perpendicular to the O-O axis (fig.~\ref{fig:X_and_YZ}b) and each averaged over their spatial dimensions. The excess proton spectrum perpendicular to the O-O axis (fig.~\ref{fig:X_and_YZ}b, red solid lines) clearly dominates the isotropic excess proton spectrum (fig.~\ref{fig:X_and_YZ}c, red solid line). Including cross correlations, defined as the difference between the total spectrum and the sum of the proton and water contributions, a different picture emerges. In fig.~\ref{fig:X_and_YZ}a and b the cross correlations between the excess proton and its two closest water molecules are plotted in gray. The cross correlations between proton and water molecules are of the same order of magnitude and of similar shape as the proton spectrum itself and almost entirely positive along the x-axis and negative along the yz-plane, meaning that the water polarization is to some degree amplifying the proton's polarization dynamics along the O-O axis, whereas they cancel perpendicular to the O-O axis. This phenomenon was similarly observed for isolated H$_5$O$_2{}^+$ complexes \cite{Bruenig2021}. Also, \citet{Sauer2005a} found in a molecular dynamics study of H$_5$O$_2{}^+$ that ``[...] the $\mathrm{O}$-$\mathrm{H^+}$-$\mathrm{O_{Y,Z}}$ bends (perpendicular to the O-O axis) have vanishing IR intensities and should not be seen in the spectra [...]''. 

The analysis presented in this section shows, that the \ac{IR} spectra of H$_5$O$_2{}^+$ complexes in \ac{HCl} solutions dominantly represent the motion of the excess proton along the O-O axis. 
This motivates the projection of the protons' movement onto the O-O axis, $d$, that is used for the decomposition in the main text.

\clearpage

\section{Spectral signature resulting from change of coordinate system}
\label{rotationSection}

As presented in section \ref{crossCorrSection}, the dynamics of the transient H$_5$O$_2{}^+$ complexes, including the excess protons, is described in the co-moving internal coordinate systems. 
To estimate the effect of this coordinate transformation on the spectrum of protons and water molecules in the H$_5$O$_2{}^+$ complexes, we compare the decomposed spectra of these complexes in the original coordinate system to those after translation and subsequent rotation, respectively.

\begin{figure}[b]
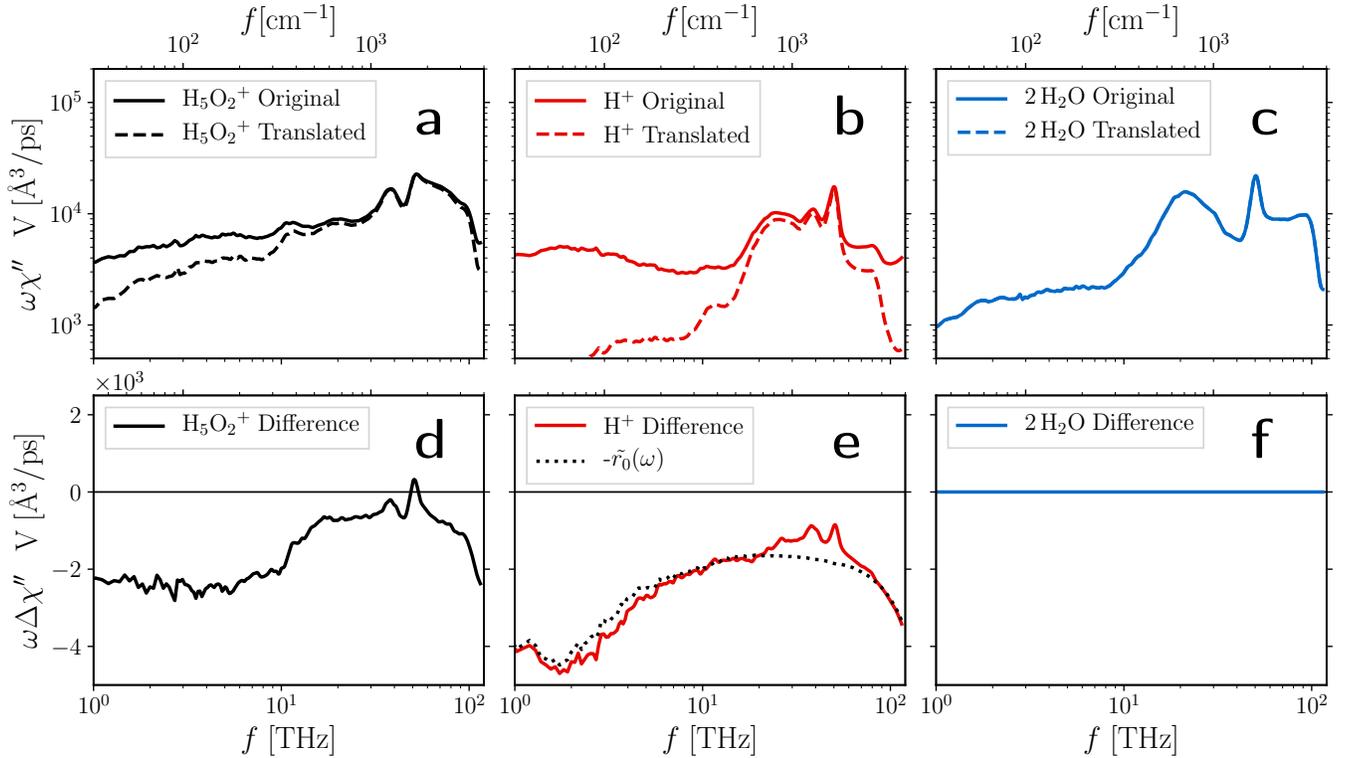

	\centering
	\begin{overpic}[width=\textwidth]{{/../figs/H5O2_specs_effect_translation_rev}.pdf}
	\put(31,47){\huge \bfseries \sffamily  a}
	\put(62,47){\huge \bfseries \sffamily  b}
	\put(93,47){\huge \bfseries \sffamily  c}
	\put(31,23){\huge \bfseries \sffamily  d}
	\put(62,23){\huge \bfseries \sffamily  e}
	\put(93,23){\huge \bfseries \sffamily  f}
	\end{overpic}
	\caption{\textbf{\textup{a--c:}} Averaged non-normalized absorption spectra including nuclei and electrons of transient H$_5$O$_2{}^+$ complexes in their original coordinate system (solid) and translated such that the coordinate origin is the midpoint between the oxygens (broken). \textbf{\textup{d--f:}} The differences between the spectra above. A dotted line in e shows the spectrum of the movement of the oxygen midpoint $\mathbf{r}_0(t)$.}\label{fig:translation}
\end{figure}

\begin{figure}[!ht]
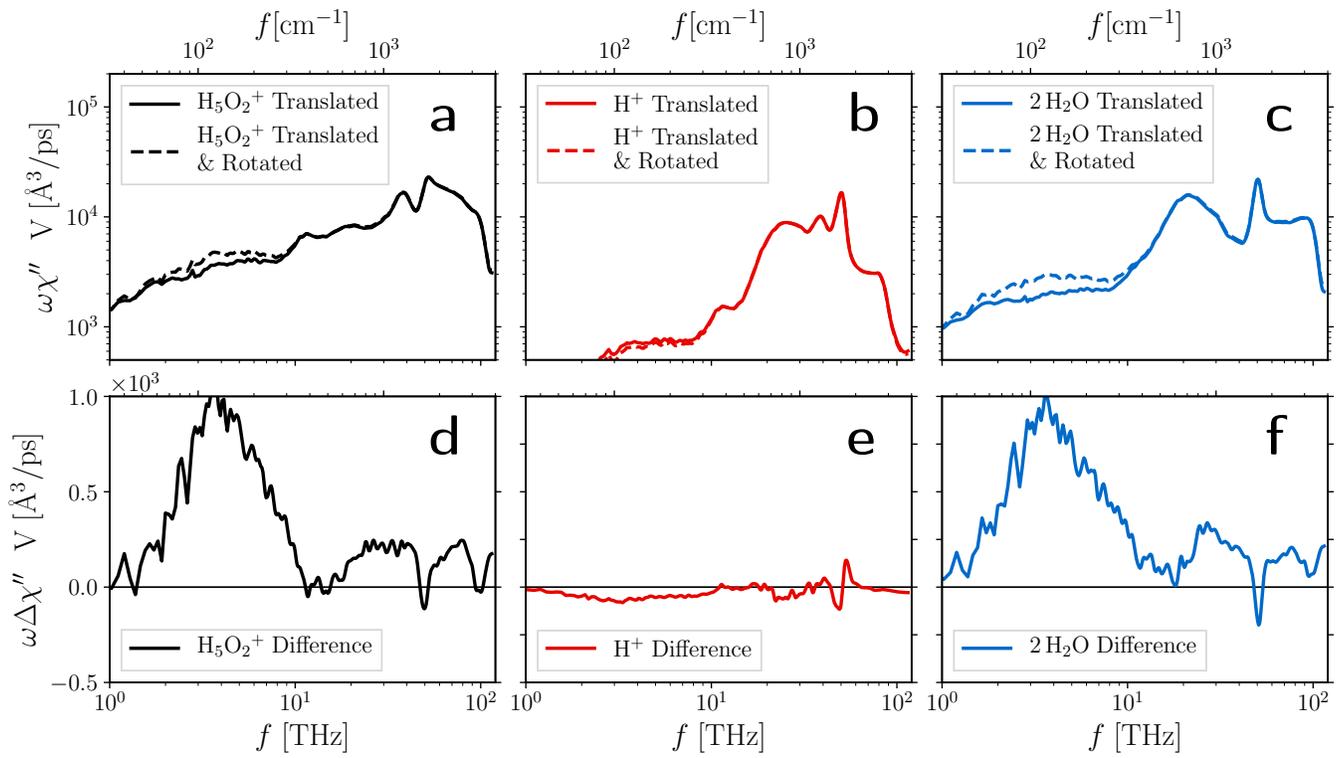

	\centering
	\begin{overpic}[width=\textwidth]{{/../figs/H5O2_specs_effect_rotation_rev}.pdf}
	\put(32,47){\huge \bfseries \sffamily  a}
	\put(63,47){\huge \bfseries \sffamily  b}
	\put(94,47){\huge \bfseries \sffamily  c}
	\put(32,23){\huge \bfseries \sffamily  d}
	\put(63,23){\huge \bfseries \sffamily  e}
	\put(94,23){\huge \bfseries \sffamily  f}
	\end{overpic}
	\caption{\textbf{\textup{a--c:}} Non-normalized absorption spectra including nuclei and electrons of transient H$_5$O$_2{}^+$ complexes that are translated but in their original orientation (solid) and rotated such that the oxygen atoms lie on the O-O axis (broken). \textbf{\textup{d--f:}} The differences between the spectra above.}\label{fig:rotation}
\end{figure}

Figure~\ref{fig:translation} shows a comparison between the spectrum of H$_5$O$_2{}^+$ complexes in their original coordinate system and translated according to $\mathbf{r}_{i,\mathrm{trans}}(t) = \mathbf{r}_i(t) - \mathbf{r}_0(t)$ (compare eq.~\eqref{eq:coordinates}). 
The translation does not have any effect on the spectrum of the two water molecules because they are not charged and the dipole moment of a neutral system does not change upon translation. 
The excess proton, however, has a charge. 
Thus, its dipole moment is affected by this translation. 
The effect can be approximated by the spectrum of all oxygen midpoint trajectories $\mathbf{r}_0(t)$ of the $\mathrm{H_5O_2{}^+}$ complexes using a charge of $\SI{1}{e}$, which is shown in fig.~\ref{fig:translation}e as a dotted line. While the translation leads to  significant
 change in the proton spectrum at low and high frequencies, in the frequency range between 
 \SIrange{20}{80}{THz}
 the effect of the translation is mostly negligible.

In contrast to the translation, the rotation mostly affects the water spectrum, because the translation due to rotation is larger for atoms further away from the rotation center. This can be seen in fig.~\ref{fig:rotation} which shows a comparison between the spectrum of $\mathrm{H_5O_2{}^+}$ complexes that were translated as shown in fig.~\ref{fig:translation} and those that were additionally rotated according to $\mathbf{r'}_i(t) = \mathbf{M}_{\phi(t),\theta(t)}\,\mathbf{r}_{i,\mathrm{trans}}(t)$ (compare eq.~\eqref{eq:coordinates}). The proton spectrum is barely affected and the effect on the water spectrum is limited  to low frequencies.

Since both translation and rotation have at most a minor effect on the transient H$_5$O$_2{}^+$ spectrum in the relevant frequency range, our findings from the decomposition into x- and yz-contribution in section \ref{crossCorrSection} for transient H$_5$O$_2{}^+$ complexes also apply to the lab frame.

\clearpage

\section{Fits of the free energy landscape}
\label{feFitSection}

\begin{figure}[hb]
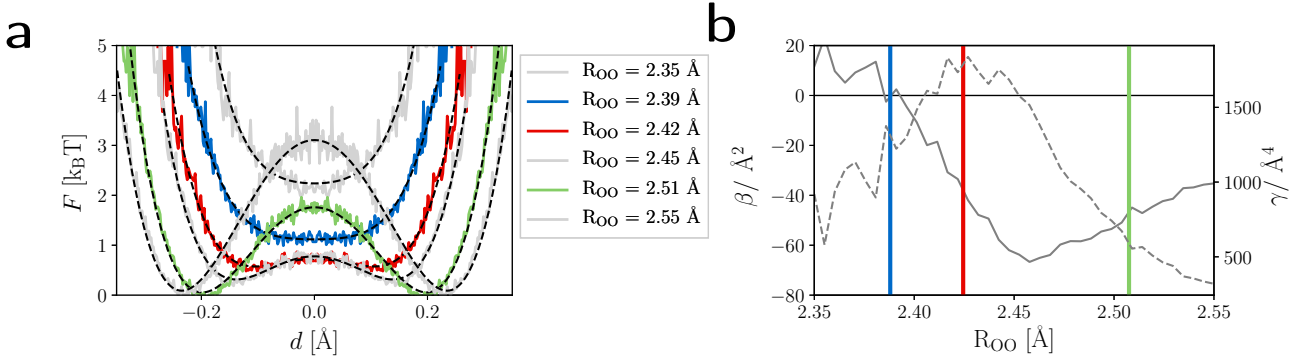

	\centering
	\vspace{3mm}
	\begin{overpic}[height=.25\linewidth]{{/../figs/F_d_fits-eps-converted-to}.pdf}
	\put(-8,48){\huge \bfseries \sffamily  a}
	\end{overpic}
	\begin{overpic}[height=.25\linewidth]{{/../figs/F_d_fits_params-eps-converted-to}.pdf}	
	\put(-4,56){\huge \bfseries \sffamily  b}	
	\end{overpic}
	\caption{Cuts along $d$ through the two-dimensional free energy shown in the main text in fig.~\ref{hcl_f}a and fitted to $F(d)/(k_BT)=F_{d=0}(1+\beta d^2+\gamma d^4)$. (\textbf{a}) Cuts at various values of $R_{\rm OO}$ are shown as solid lines and fits as black broken lines with fit parameters reported in tab.~\ref{tab:fe_fit}. (\textbf{b}) The fit parameters $\beta$, shown as a gray solid line, and $\gamma$, shown as a gray broken line, are given over $R_{\rm OO}$.
}\label{fig:fe_fit}
\end{figure}

The two-dimensional (2D) free energy landscape spanned by $R_{\rm OO}$ and $d$ shown in fig.~\ref{hcl_f}a in the main text, is fitted for cuts through the free energy along $d$.
As in the main text in fig.~\ref{hcl_f}, here in fig.~\ref{fig:fe_fit}a, cuts along $d$ are shown for $R_{\rm{OO}}=\SI{2.39}{\ang}$, where the barrier just vanishes (blue solid line), $R_{\rm{OO}}=\SI{2.42}{\ang}$, where the absolute barrier height is minimal (red solid line), and $R_{\rm{OO}}=\SI{2.51}{\ang}$, for which the global minima of the 2D free energy are obtained (green solid line). 
Additionally, some results for other values of $R_{\rm OO}$  are shown as gray solid lines.
The data is fitted by the quartic expression $F(d)/(k_BT)=F_{d=0}(1+\beta d^2+\gamma d^4)$, shown as black broken lines in fig.~\ref{fig:fe_fit}a with the fit parameters reported in tab.~\ref{tab:fe_fit}.
In fig.~\ref{fig:fe_fit}b the fit parameters $\beta$ (gray solid line) and $\gamma$ (gray broken line) are plotted over $R_{\rm OO}$. 
The vertical colored lines indicated the respective values of $R_{\rm OO}$ shown in fig.~\ref{fig:fe_fit}a. 
The barrier vanishes for $\beta >0$, which defines the blue solid line.
At the position where the absolute barrier height is minimal, indicated by the red solid line, the quartic fit parameter $\gamma$ is maximal.

\begin{table}[htb]
	\centering
	\caption{Fit parameters for the fits shown in fig.~\ref{fig:fe_fit}a using the quartic expression $F(d)/(k_BT)=F_{d=0}(1+\beta d^2+\gamma d^4)$.}
	\label{tab:fe_fit}
	\begin{tabular}{c|c | r | r}
$R_{\mathrm{OO}}$ [\AA] & $F_{d=0}$ [$k_BT$]  & $\beta$ [\AA$^2$] & $\gamma$ [\AA$^4$]  \\ \hline
2.35 & 2.24 & 11.63 & 919.7 \\ 
2.39 & 1.12 & 7.78 & 1099 \\ 
2.42 & 0.73 & -38.29 & 1747 \\ 
2.45 & 0.78 & -62.25 & 1625 \\ 
2.51 & 1.76 & -49.17 & 623.5 \\ 
2.55 & 3.11 & -35.24 & 319.6 
\end{tabular}
\end{table}

\clearpage

\section{Distribution of proton-transfer waiting times}
\label{fptSection}

The mean proton-transfer rate is at the heart of research on excess protons solvated in water, as it is the relevant microscopic time scale that determines the macroscopic large diffusion obtained by the Grotthuss process. 
As outlined in the main text, the mean transfer rate corresponds to the waiting time of a stochastic barrier-crossing process, that is not to be confused with the transfer-path time of the actual transition over the barrier.
The waiting-time distributions that are also shown in fig.~\ref{hcl_decomp}c in the main text are discussed in this section and are compared to  previous results from literature.

\citet{Agmon1995} gives a summary of early experimental results, suggesting a mean proton-transfer time of \SI{1.5}{ps} obtained in NMR studies \cite{Meiboom1961}, which is believed to be correlated with the \ac{HB} rearrangement and water reorientation dynamics on the time scales of \SIrange{1}{2}{ps} determined from a number of other experiments.
These time scales seem to contrast more recent experimental results from 2D \ac{IR} spectroscopy, that report interconversion between different proton hydration structures, i.e. Eigen and Zundel-like structures, on time scales of around \SI{100}{fs} and less \cite{Woutersen2006, Dahms2017, Kundu2019}. 
On the other hand, \citet{Carpenter2018} write, ``the hydrated proton bend displays fast vibrational relaxation and spectral diffusion timescales of $200-300$~fs, however, the transient absorption anisotropy decays on a remarkably long \SI{2.5}{ps} timescale, which matches the timescale for \ac{HB} reorganization in liquid water''. 
Arguing that the latter would be an upper bound, they infer ``the transfer of excess protons in water [...] is an activated process with a timescale of $1-2\,$ps.'' \citet{Kundu2019} confirm that ``during the lifetime of the H$_5$O$_2{}^+$ motif, that is on the order of \SI{1}{ps}, the proton undergoes fluctuating large-amplitude motions exploring essentially all possible positions between the flanking water molecules''. \citet{Yuan2019} measure a concentration-dependent `proton hopping time' in \ac{HCl} solutions using 2D IR chemical exchange spectroscopy with a methyl thiocyanate probe and extrapolate a time of \SI{1.6}{ps} for the dilute limit.

\begin{table}[hb]
\caption{Collection of proton-transfer time scales reported in the literature. }
\label{tab:tws}
\begin{tabular}{c | c |  c | c | c | r  }
 & method & conc. [M] & T [K] & & time \\  \hline
 \citet{Meiboom1961} & NMR exp. &   &     &						 & \SI{1.5}{ps} \\
\citet{Woutersen2006} & 2D IR exp. & 5  & 300  &						 & <\SI{0.1}{ps} \\
\citet{Dahms2017} & 2D IR exp. & <1  & 300  &						 & <\SI{0.1}{ps}  \\ 
\citet{Thaemer2015} & 2D IR exp. & 4  & 300  &						 & >\SI{0.48}{ps} \\
\citet{Carpenter2018} & 2D IR exp. & 2  & 300  &						 & <\SI{2.5}{ps} \\ 
\citet{Kundu2019} & 2D IR exp. &  1 & 300  & 						 & <\SI{0.1}{ps} \\
 &  &  &   &						 & \SI{1}{ps}  \\ 
\citet{Yuan2019} &  2D IR chemical & dillute & 300 & & \SI{1.6}{ps} \\
&  exchange exp. & limit &  & &  \\
\hline
\citet{Fischer2018} & DFT CPMD & 1.7 & 300 & bi-directional & \SI{0.5}{ps} \\
\citet{Fischer2019} & DFT CPMD & 1.7 & 300 & uni-directional & \SI{2.5}{ps} \\
\citet{Calio2020} & MS-EVB & $0.43-3.26$  & 300 & & $0.4-0.5\,$ps \\ &  & &  & & $1.2-2.3\,$ps \\
\citet{Calio2021} & MS-EVB, & $0.22-0.43$  & 300 & & $10-17\,$fs \\ & DFT BOMD & &  & & $0.3-0.5\,$ps \\
&  & &  & & $2.3-3.2\,$ps \\
& + nuc.-quantum   &  &  & & $12-15\,$fs \\ & effects & &  & & $0.83-0.27\,$ps \\
&  & &  & & $1.4\,$ps \\
\citet{Arntsen2021} & DFT BOMD &   & 300 & bi-directional & \SI{0.184}{ps} \\
&  &   &  & uni-directional & \SI{1.69}{ps} \\
\citet{Roy2020} 	   & DFT BOMD & 2 & 300 & 					 & $1-2\,$ps \\
     & DFT BOMD & 8 & 300 & 				     & $2-4\,$ps \\
\end{tabular}
\end{table}

It transpires, that two time scales determine the distribution of proton-transfer waiting times, which was confirmed in various simulation studies and interpreted as stemming from either back-and-forth or uni-directional proton transfer, respectively, in the literature sometimes referred to as `reversible' and `irreversible' proton transfer. 
\citet{Napoli2018} point out that while they find a frequency-correlation time of \SI{1.4 \pm 0.3}{ps} ``corresponding to the pump (\SI{3150}{cm^{-1}}) and probe (\SI{1760}{cm^{-1}}) frequencies used in \cite{Thaemer2015}'', the auto-correlation of the proton asymmetry actually decays on time scales of less than \SI{100}{fs} with a second slower component of \SI{0.8 \pm 0.1}{ps}.
\citet{Fischer2018} find the time scale of proton `hopping' to be around \SI{0.5}{ps}, including back-and-forth events, and deduce a time scale of \SI{2.5}{ps} for uni-directional events in a later study \cite{Fischer2019}.
\citet{Calio2020} extract two timescales of about \SIrange{400}{500}{fs} and \SIrange{1.3}{2.3}{ps} for the concentrations \SIrange{0.43}{3.26}{M} from fits to the long-lived anisotropy decays, which the authors argue ``can correlate experimental time constants to irreversible proton transfer''. 
In a follow-up study \citet{Calio2021} report three timescales of about $10-17\,$fs, $320-490\,$fs and $2.3-3.2\,$ps for the concentrations \SIrange{0.22}{0.43}{M} from fits to the anisotropy decay of the excess-proton dynamics projected on the axis of the two closest oxygen atoms, with a slight acceleration to the values $12-15\,$fs, $83-270\,$fs and \SI{1.4}{ps} when nuclear-quantum effects are considered. 
\citet{Roy2020} find a time scale of $1-2\,$ps for uni-directional proton transfer between two water molecules in \SI{2}{M} \ac{HCl} solutions employing two-dimensional transition state theory and Marcus theory of ion pairing. This number increases significantly to $2-4\,$ps in \SI{8}{M} \ac{HCl} solutions. 
\citet{Arntsen2021} report time constants of the excess-proton identity auto-correlation function, which is elaborated on further below. 
They find a values of \SI{184}{fs}, but after eliminating back-and-forth events from the data the time scale increases to \SI{1.69}{ps}.
Furthermore a couple of studies point out, that the long time scale of proton transport increases significantly with concentration \cite{Carpenter2019,Roy2020}.

A summary of the reported values is given in tab.~\ref{tab:tws}. It transpires that the separation in back-and-forth and uni-directional events is important to distinguish two time scales in the broad distributions of proton-transfer times. For stochastic barrier-crossing processes of highly inertial or non-Markovian coordinates and furthermore for low energy barriers, it is well established that barrier-crossing events exhibit large numbers of subsequent back-and-forth events, due to the slow dissipation of the energy required for the initial barrier-crossing event \cite{Hanggi1990, Ciccotti1990, Roy2016, Kappler2018, Roy2020, Bruenig2021}.
Such events appear especially important with regard to proton-transfer processes and their spectral signatures.

\begin{figure}[tb]
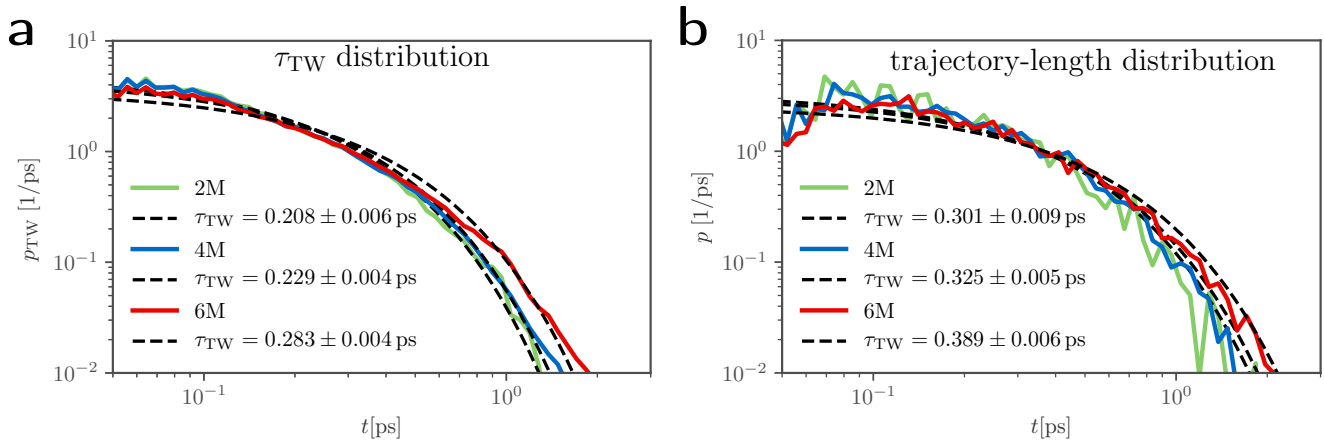

	\centering
	\begin{overpic}[width=.49\linewidth]{{/../figs/FPT_statistics_2ExpFit-eps-converted-to}.pdf}
	\put(0,62){\huge \bfseries \sffamily  a}
	\end{overpic}	
	\begin{overpic}[width=.49\linewidth]{{/../figs/tLen_statistics_2ExpFit}.pdf}
	\put(0,62){\huge \bfseries \sffamily  b}
	\end{overpic}		
	\caption{Distributions of proton-transfer waiting times (a) and of the life times of the transient H$_5$O$_2{}^+$ clusters, i.e. the trajectory lengths, (b) in \ac{HCl} solutions at various concentrations as obtained from ab initio \ac{MD} simulations. The same data as in a is also shown in fig.~\ref{hcl_decomp}c in the main text. For comparison, single exponential decays,  $p(t)=\exp(-t/\tau)/\tau$, using the mean of the distributions $\tau$ reported in the legend, are plotted as black broken lines.
}
\label{fig:mfpts}
\end{figure}

The proton-transfer waiting time distributions of the excess-protons in aqueous \ac{HCl} solutions at various concentrations, reported in the main text in fig.~\ref{hcl_decomp}c, are plotted in fig.~\ref{fig:mfpts}a on a double logarithmic scale in order to investigate the long time behavior. 
The distributions do decay in the picosecond range with a clear concentration dependence, i.e. the distributions have a longer tail for higher concentrations. 
Still, a single exponential fit using the mean of the distribution $\tau_{\mathrm{TW}}$ as the decay time scale, $p_{\mathrm{TW}}(t)=\exp(-t/\tau_{\mathrm{TW}})/\tau_{\mathrm{TW}}$, is sufficient to describe the long time tail of the distributions.
However, since the excess protons are tracked only in the transient H$_5$O$_2{}^+$ cluster, the transfer-waiting times can only capture proton transfer during the life times of these H$_5$O$_2{}^+$ clusters, i.e. the lengths of the trajectories, that are defined according to the procedure that is detailed in section \ref{selectionSection}.
The distributions of trajectory lengths are given in fig.~\ref{fig:mfpts}b. 
They are again compared to single exponential functions,  $p(t)=\exp(-t/\tau)/\tau$, shown as black broken lines, with the mean values of the distributions as the decay constants $\tau$.
For all three concentrations the mean values are about 1.5 times larger than the respective waiting-times reported in fig.~\ref{fig:mfpts}a, indicating that on average one to two transfers occur during the life time of a transient H$_5$O$_2{}^+$ cluster.

\subsection*{Identity auto-correlation functions}

To describe the long time scales of excess-proton diffusion observed in the anisotropy decay of 2D \ac{IR} experiments \cite{Thaemer2015,Carpenter2018}, correlation times of the excess-proton identity have proven useful.
Inspired by previous work \cite{Hassanali2013, Arntsen2021, Calio2021}, we computed auto-correlation functions of the excess-proton and hydronium-oxygen identities from joint trajectories of the excess protons as also prepared for the analysis of the long-time diffusion properties described in section \ref{diffusionSection}.
Following this protocol, the excess protons are given as the remaining protons after the water molecules are assembled for each oxygen atom with the closest two hydrogen atoms at each time step of the simulation.
Hydronium ions are defined by an excess proton together with the water molecule of the closest oxygen atom.
Therefore, at each time step a total number of excess protons $N_{\mathrm{H}^+}$, as well as hydronium ions, equivalent to the number of chloride ions in the simulation is obtained.
The trajectories are then stitched together to give $N_{\mathrm{H}^+}$ trajectories, each of the length of the whole simulation, for the excess protons and hydronium oxygens, respectively. 
Following \citet{Arntsen2021}, rapid back-and-forth fluctuation of hydronium oxygen identities is `filtered' from the trajectories by the following procedure. 
Whenever along a trajectory the identity changes from one oxygen atom to another, we check whether the identity returns to the original nucleus within \SI{0.5}{ps}.
If it returns without passing to a third nucleus in between, the identity remains with the original nucleus as if the identity did not change throughout this time.
For the excess-proton identities this criterion does not suffice since the identity often fluctuates between three candidates within one hydronium ion.
Rather the same procedure as also detailed in section \ref{diffusionSection} and similar to \citet{Calio2021} is used to `filter' the rapid back-and-forth fluctuation of excess-proton identities, i.e. the `special pair dance': 
The candidate proton that either performs the next transfer to another water molecule or is the next to be identified as an excess proton while neighboring a chloride atom, remains the excess proton. 
These procedures define sets of trajectories from which the fast identity fluctuations are `filtered'.
The identity auto-correlation functions are then calculated for excess-proton and hydronium-oxygen identities on both sets of trajectories, `filtered' and `unfiltered'.

We define the identity auto-correlation function as
\begin{align}
\label{eq:corr}
c(t)=\frac{\langle h(t) h(0) \rangle - \langle h \rangle^2}{\langle h \rangle-\langle h \rangle^2},
\end{align}
where $h(t)$ is one if an excess proton or hydronium oxygen atom has the same identity, i.e. is the same nucleus, as at $t=0$, otherwise $h(t)$ is zero. 
In our definition the identity auto-correlation function is designed reach unity for $t=0$ and to decay to zero for long times and is thereby slightly different compared to previous work \cite{Luzar1996, Hassanali2013, Arntsen2021}.
Additionally, the continuous identity auto-correlation function is given as \cite{Arntsen2021, Calio2021}
\begin{align}
\label{eq:corr_break}
C(t)=\frac{\langle H(t) H(0) \rangle}{\langle H \rangle},
\end{align}
where $H(t)$ is one as long as an excess proton or hydronium oxygen atom continuously has the same identity for the entire time interval $[0,t]$ and zero otherwise.

\begin{figure}[tb]
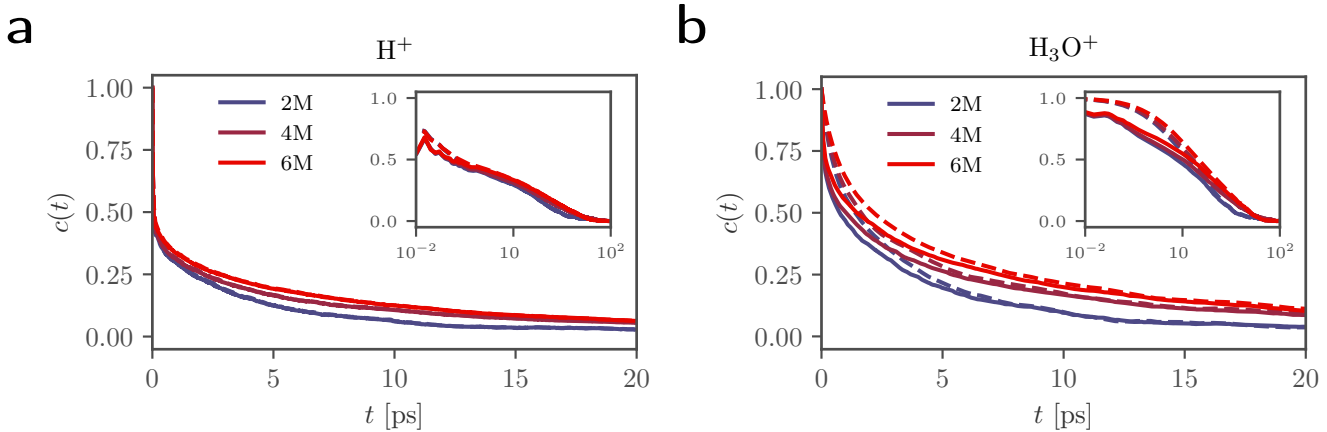

	\centering
	\begin{overpic}[width=.49\linewidth]{{/../figs/protons_corr_all}.pdf}
	\put(-2,63){\huge \bfseries \sffamily  a}
	\end{overpic}	
	\begin{overpic}[width=.49\linewidth]{{/../figs/protons_corr_os_all}.pdf}
	\put(-2,63){\huge \bfseries \sffamily  b}
	\end{overpic}		
	\caption{Identity auto-correlation functions eq.~\eqref{eq:corr} of the excess protons (a) and hydronium oxygens (b) obtained from ab initio \ac{MD} simulations of \ac{HCl} solutions at various concentrations. 
The correlations functions are computed from unfiltered (solid colored lines) and filtered trajectories (broken colored lines), see text for details.
The data is shown on a logarithmic time axis in the insets.
}
\label{fig:corrs}
\end{figure}

The excess-proton and hydronium-oxygen identity auto correlations obtained from our data according to eq.~\eqref{eq:corr} are given in figs.~\ref{fig:corrs}a and b with the same data shown on a logarithmic time axis in the insets.
Auto-correlation functions of such filtered trajectories are shown as broken colored lines for the three \ac{HCl} solutions at various concentrations while the auto-correlation functions of the unfiltered trajectories are shown as solid colored lines.
The curves decay on multiple time scales. 
All show remarkable long-time behavior beyond several picoseconds, indicating that proton as well as hydronium identity is correlated over very long times scales, which occurs from looping of identities over several different nuclei \cite{Hassanali2013}. 
As expected, the two types of data from filtered or unfiltered trajectories converge in these long time regimes.
Both, the excess-proton identity auto correlations in fig.~\ref{fig:corrs}a and the hydronium identities in fig.~\ref{fig:corrs}b show a clear concentration dependence, with longer decay times for higher concentrations.
On short time scales the excess-proton identity auto correlations largely decay within \SI{0.1}{ps} to a value of 0.5, whereas the hydronium identity auto correlations decay to a value of 0.5 only after about \SIrange{1}{2}{ps}.
Furthermore, a slight peak in the hydronium identity auto-correlation of the unfiltered trajectories (solid colored lines in fig.~\ref{fig:corrs}b) at about \SI{0.025}{ps} indicates back-and-forth transfer of the excess proton in the transient H$_5$O$_2{}^+$ cluster occurring at a about twice the transfer-path time, $\tau_{\mathrm{TP}}=$ \SIrange{12.6}{14.3}{fs} reported in the main text.
Similarly, a peak in the excess-proton identity auto correlations of the unfiltered trajectories (solid colored lines in fig.~\ref{fig:corrs}a) at about \SI{0.010}{ps} indicates the time scale of excess-proton rattling within a single hydronium ion, referred to as `special pair dance' in the literature \cite{Markovitch2008, Calio2021}.

\begin{figure}[tb]
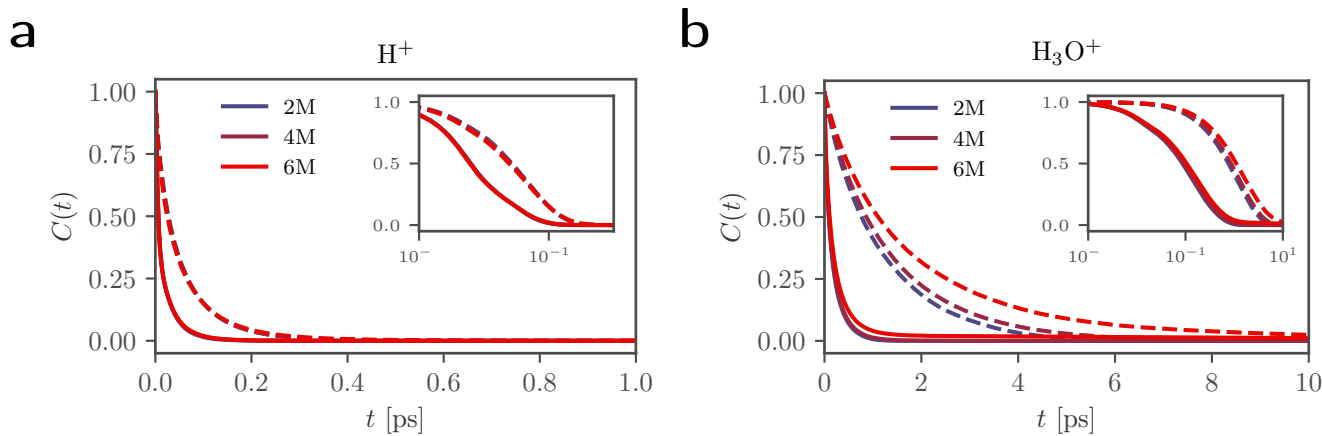

	\centering
	\begin{overpic}[width=.49\linewidth]{{/../figs/protons_corr_Break_all}.pdf}
	\put(-2,63){\huge \bfseries \sffamily  a}
	\end{overpic}	
	\begin{overpic}[width=.49\linewidth]{{/../figs/protons_corr_os_Break_all}.pdf}
	\put(-2,63){\huge \bfseries \sffamily  b}
	\end{overpic}		
	\caption{Continuous identity auto-correlation functions eq.~\eqref{eq:corr_break} of the excess protons (a) and hydronium oxygens (b) obtained from ab initio \ac{MD} simulations of \ac{HCl} solutions at various concentrations. 
The correlations functions are computed from filtered (solid colored lines) and unfiltered trajectories (broken colored lines), see text for details.
The data is shown on a logarithmic time axis in the insets.
}
\label{fig:corrs_Break}
\end{figure} 
 
Next, to focus on the fast time scales of the correlations, the continuous identity auto-correlation functions according to eq.~\eqref{eq:corr_break} of the excess protons and of the hydronium oxygens are given in figs.~\ref{fig:corrs_Break}a and b with the same data shown on logarithmic time axes in the insets. 
Again, the data is shown in each plot as computed from filtered (broken colored lines) and unfiltered trajectories (solid colored lines). 
Note, that for the computation of these continuous identity auto correlations, configurations where the excess-proton is located between the oxygen atom and a chloride ion are excluded.
These configurations obviously produce spurious long-time auto correlations but have been analyzed to make up only 5\% of the total trajectory lengths of all excess protons even at the highest concentration of \SI{6}{M}, which is discussed in detail in section \ref{radialDistSection}.
The excess-proton continuous identity auto correlations decay fully within \SI{0.3}{ps} and the hydronium-oxygen identity continuous auto-correlations within \SI{10}{ps}. 
In contrast to the data presented in fig.~\ref{fig:corrs}a and b, while the hydronium identity auto correlations in fig.~\ref{fig:corrs_Break}b show again a clear concentration dependence, with longer decay times for higher concentrations, such a dependence is not visible for the excess-proton identity auto correlations in fig.~\ref{fig:corrs_Break}a. 

\begin{figure}[tb]
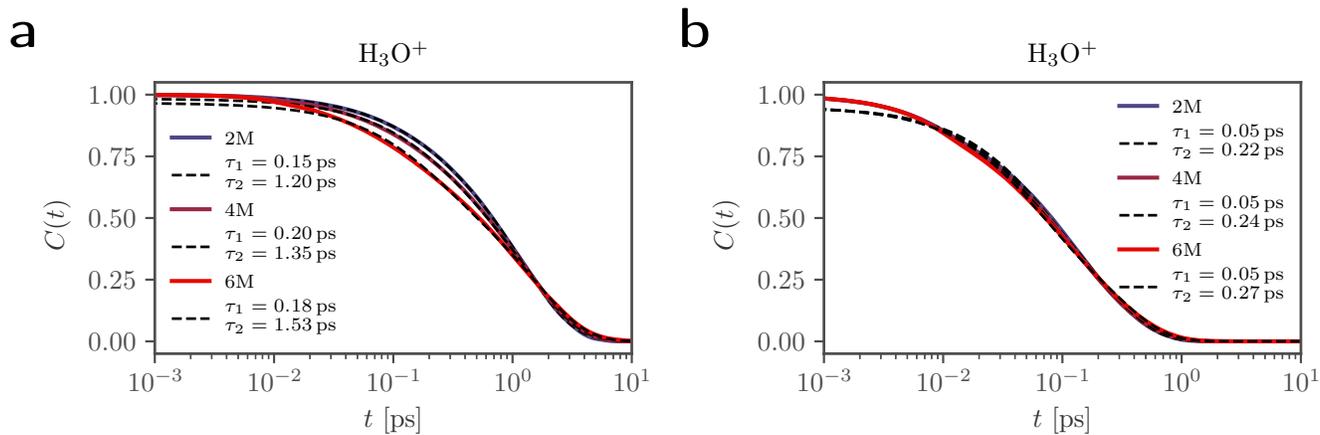

	\centering
 	\begin{overpic}[width=.49\linewidth]{{/../figs/protons_corr_os_clean_clBreak_all_2ExpFit-eps-converted-to}.pdf}
	\put(-2,63){\huge \bfseries \sffamily  a}
	\end{overpic}	
	\begin{overpic}[width=.49\linewidth]{{/../figs/protons_corr_os_clBreak_all_2ExpFit-eps-converted-to}.pdf}
	\put(-2,63){\huge \bfseries \sffamily  b}
	\end{overpic}		
	\caption{Continuous identity auto-correlation functions eq.~\eqref{eq:corr_break} of the hydronium oxygens obtained from ab initio \ac{MD} simulations of \ac{HCl} solutions at various concentrations (already shown in fig.~\ref{fig:corrs_Break}b). 
The correlation functions are computed from filtered (a) and unfiltered trajectories (b), see text for details. 
Each curve is fitted to a sum of two decaying exponentials with the time scales reported in the legends and shown as black broken lines.
}
\label{fig:corrs_Break_fit}
\end{figure}

The time scales of the hydronium continuous identity auto correlations have been interpreted to be consistent with the time scales of the anisotropy decay observed in 2D \ac{IR} experiments \cite{Carpenter2018,Arntsen2021}. 
Fits to these auto correlations with the sum of two decaying exponentials are therefore given as broken black lines, together with the original data from fig.~\ref{fig:corrs_Break}b repeated as solid colored lines in fig.~\ref{fig:corrs_Break_fit}a (for the filtered trajectories) and b (for the unfiltered trajectories).
The long time-scales of the bi-exponential fits to the correlations of the filtered trajectories, shown in \ref{fig:corrs_Break_fit}a, reach from \SIrange{1.2}{1.53}{ps}, increasing with concentration. 
These time scales match the time scales reported for uni-directional proton transfer rather well (see tab.~\ref{tab:tws}).
With regard to the results reported in section \ref{diffusionSection}, where the long-time diffusion constants are found to be too small by a factor of about four when compared to experiment, one would expect the time scales for uni-directional proton transfer to be longer by the same factor. 
However, the present analysis excludes configurations involving chloride ions, which presumably are characterized by longer decay times. 
The diffusion constants on the other hand would have to be split in vehicular diffusion, due to translation of hydronium ions, and the jump diffusion, due to uni-directional proton transfer, to allow for a better comparison to the time scales of the auto correlations.

When regarding the long time-scales of the bi-exponential fits to the correlations of the unfiltered trajectories, shown in \ref{fig:corrs_Break_fit}b, which reach from \SIrange{0.22}{0.27}{ps}, increasing with concentration, we find them to match perfectly the mean proton-transfer waiting times reported in the main text and in fig.~\ref{fig:mfpts}. 
The continuous hydronium identity auto correlation therefore presents an alternative 
and intuitive 
interpretation for the proton-transfer waiting times that we discuss in the main text. 
In case that back-and-forth fluctuations of the excess-protons between two oxygen atoms  within \SI{0.5}{ps} are filtered from the trajectories, the continuous hydronium identity auto correlation in fig.~\ref{fig:corrs_Break_fit}a decays on time scales that match experimental spectroscopic anisotropy decays and have been interpreted as uni-directional proton transfer.

\clearpage

\section{Large-scale excess-proton diffusion}
\label{diffusionSection}

\begin{figure}[b]
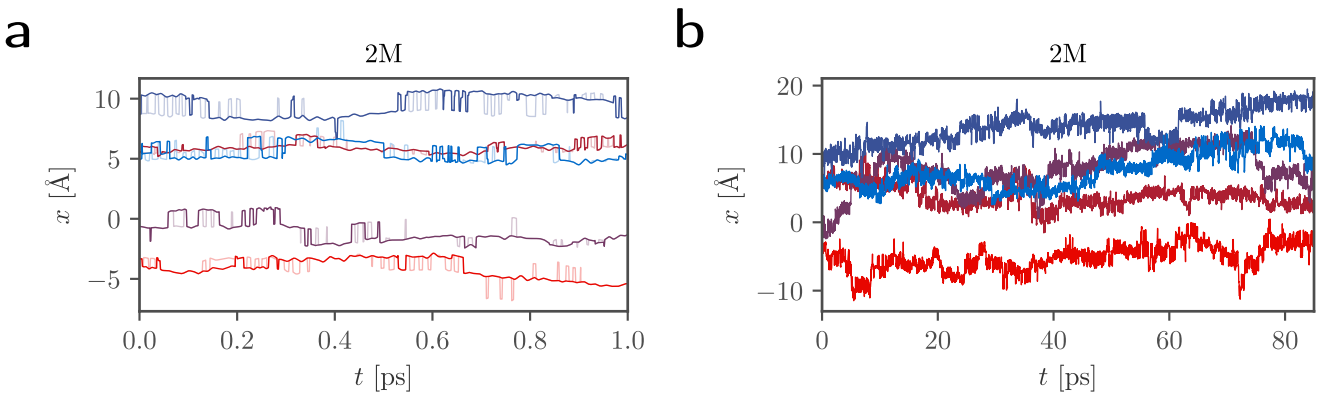

	\centering
	\begin{overpic}[width=.49\linewidth]{{/../figs/protons_traj_2M_n2000-eps-converted-to}.pdf}
	\put(-2,57){\huge \bfseries \sffamily  a}
	\end{overpic}
	\begin{overpic}[width=.49\linewidth]{{/../figs/protons_traj_2M_n170000-eps-converted-to}.pdf}
	\put(-2,57){\huge \bfseries \sffamily  b}
	\end{overpic}		
	\caption{Examples of five joint trajectories of excess protons in \SI{2}{M} \ac{HCl} solution on short (a) and long time scales (b). In a, the pale colored lines indicate the spurious jumps, resulting from the `special pair dance', that are removed from the trajectories (see text for details).}
\label{fig:protons_traj}
\end{figure}

For the analysis of diffusion of the excess protons on long time scales, the principle excess-proton identification and selection scheme detailed in section \ref{selectionSection} needs to be augmented to identify joint excess-proton trajectories throughout the whole simulation trajectory. 
For this, excess protons are first identified based on a geometric criterion: after assembling at each time step the water molecules for each oxygen atom with the closest two hydrogen atoms, the remaining least associated protons form hydronium ions with their respective closest water molecules. 
We thereby obtain at each time step a total number of excess protons $N_{\mathrm{H}^+}$ equivalent to the number of chloride ions in the simulations. 
The trajectories of the excess protons are then stitched together to a total of $N_{\mathrm{H}^+}$ trajectories, each of the length of the whole simulation. 
In contrast to section \ref{selectionSection}, protons that reside between an oxygen and a chloride atom are included in this analysis.
The procedure obviously introduces jumps in the joint trajectories whenever an excess proton changes identity, which is a manifestation of the Grotthuss' process.
However, rapid spurious jumps within the same hydronium ion, the `special pair dance' \cite{Markovitch2008, Calio2021}, are filtered from the trajectories by the following procedure: within each hydronium ion, the candidate proton that either performs the next transfer to another water molecule or is the next to be identified as an excess proton while neighboring a chloride atom, remains the excess proton (a procedure that was also used by \citet{Calio2021}). 
Lastly, the trajectories are unwrapped over the periodic boundary conditions.
Some trajectories of excess protons that produced by this protocol are illustrated in fig.~\ref{fig:protons_traj}a and b along a single Cartesian coordinate and on two different time scales. Additionally in fig.~\ref{fig:protons_traj}a, the pale colored lines show the trajectories before removal of the `special pair dance'.

\begin{figure}[tb]
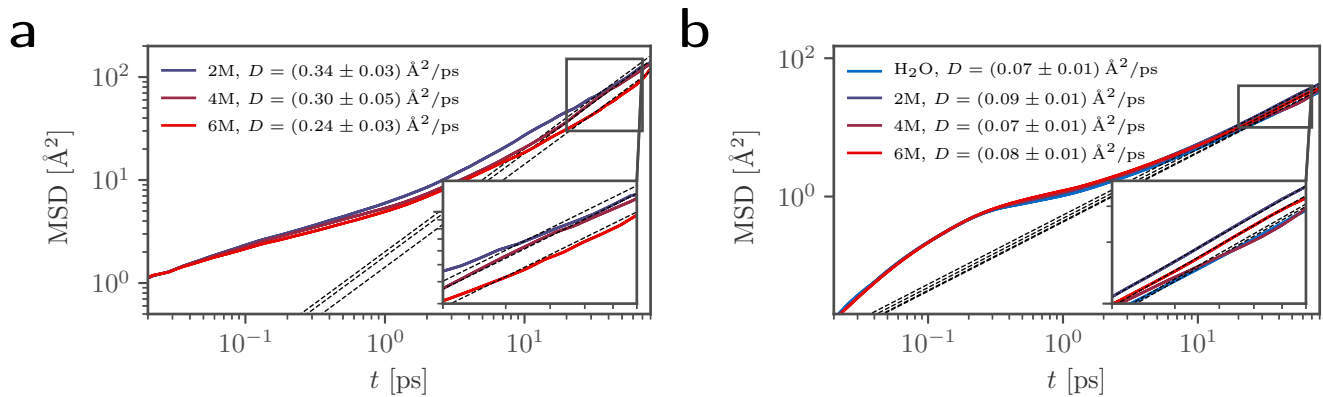

	\centering
	\begin{overpic}[width=.49\linewidth]{{/../figs/protons_msds-eps-converted-to}.pdf}
	\put(-2,57){\huge \bfseries \sffamily  a}
	\end{overpic}
	\begin{overpic}[width=.49\linewidth]{{/../figs/protons_msds_water-eps-converted-to}.pdf}
	\put(-2,57){\huge \bfseries \sffamily  b}
	\end{overpic}		
	\caption{\acsp{MSD} (\aclp{MSD}) in the lab frame, $\langle |\mathbf{r}(t) -\mathbf{r}(0)|^2 \rangle$, of the excess protons (a) and the oxygen atoms (b) computed from the simulation trajectories of \ac{HCl} solutions at three concentrations. 
In b, the MSD of the oxygen atoms in the pure-water simulations is shown as well. 
Errors of the mean are indicated by the line widths and taken from standard deviations computed over the individual excess proton and oxygen atom joint trajectories. 
Linear fits are shown as black broken lines, which are fitted in the long time regimes, \SIrange{20}{80}{ps}, which is the range that is also shown enlarged in the insets.
	}
\label{fig:protons_msds}
\end{figure}

Subsequently, the joint trajectories are used to calculate \acp{MSD}, $\langle |\mathbf{r}(t) -\mathbf{r}(0)|^2 \rangle$, of the excess protons in the lab frame. 
The \acp{MSD} averaged over all excess protons are shown in fig.~\ref{fig:protons_msds}a for simulations of \ac{HCl} solutions at three different concentrations and compared to the \acp{MSD} computed for the oxygen atoms representative of the water molecules in fig.~\ref{fig:protons_msds}b. 
In general, the \ac{MSD} is expected to show inertial scaling, $\rm{MSD} \sim t^2$, for short time scales and diffusive scaling, $\rm{MSD} \sim t$, for long time scales. 
Both regimes are well visible in fig.~\ref{fig:protons_msds}b for the oxygen atoms. 
For the excess protons the inertial regime is perturbed by the jumps in the joint trajectories, caused by changes of the excess proton identities.

The diffusion constant $D$, which is an experimentally observed quantity, is related to the \ac{MSD} by
\begin{align}
D = \frac{1}{6 t} \langle |\mathbf{r}(t) -\mathbf{r}(0)|^2 \rangle,
\end{align}
and computed by a least-squares fit of the slopes in the diffusive regime between \SIrange{20}{80}{ps}. 
The diffusion constants are reported in the legends of figs.~\ref{fig:protons_msds}a and b with statistical errors from the linear fits.

To access the accuracy of these diffusion constants, we first focus on the values for the oxygen atoms. 
Within the error the same diffusion constant of about $D_{\rm O}=\SI{0.08}{\ang^2/ps}$ was obtained in all four simulations. 
While this value is significantly smaller than the experimental value of $D_{\rm O}=\SI{0.23}{\ang^2/ps}$ \cite{Mills1973, Harris1980}, it is well known that specifically the BLYP exchange-correlation functional in our simulation approach produces too small diffusion constants.
In agreement with our results, various studies of water utilizing ab initio \ac{MD} together with the BLYP exchange-correlation functional, reported diffusion constants in the range of \SIrange{0.04}{0.11}{\ang^2/ps} for comparable setups to ours, as reviewed recently \cite{Gillan2016}.

When regarding the excess proton diffusion constant, it has to be noted that it is known to be heavily concentration and temperature dependent in experiments \cite{Dippel1991}. 
In the dilute limit, the experimental diffusion constant of the excess proton, $D_{\rm H^+}=\SI{0.94}{\ang^2/ps}$, is much higher than the experimental diffusion constant of water by a factor of $D_{\rm H^+}/D_{\rm O} = 4.1$, an observation that corroborates Grotthuss' hypothesis \cite{Roberts1972}. However, this factor drops to about $3.0$ at \SI{2}{M} and $1.5$ at \SI{6}{M} \cite{Dippel1991}. 
This trend was qualitatively captured in self-consistent iterative multistate empirical valence bond (SCI-MS-EVB) simulations of HCl \cite{Xu2010, Calio2020}. Both studies obtained values between $D_{\rm H^+} = $ \SIrange{0.2}{0.3}{\ang^2/ps} around \SI{1}{M} and $D_{\rm H^+} =$ \SIrange{0.15}{0.20}{\ang^2/ps} around \SI{3}{M}, compared to a value of $D_{\rm H^+} = \SI{0.37}{\ang^2/ps}$ in the dilute limit (one excess proton in 256 waters) \cite{Biswas2016}.
While a similar trend is also indicated by our simulation data, the errors are too large to draw definite conclusions. 
The obtained diffusion constants for the excess protons, $D_{\rm H^+}=\SIrange{0.24}{0.34}{\ang^2/ps}$, are smaller than the experimental value. 
However, the ratios $D_{\rm H^+}/D_{\rm O} = 0.34/0.09 = 3.8 \pm 0.8 $ for \SI{2}{M}, $D_{\rm H^+}/D_{\rm O} = 0.30/0.07 = 4.3 \pm 1.3$ for \SI{4}{M} and $D_{\rm H^+}/D_{\rm O} = 0.24/0.08 = 3.0 \pm 0.8 $ for \SI{6}{M}, that are observed in our simulations, seem in satisfactory agreement with experiments.
Similar ab initio simulation setups to ours but using a single excess proton in a box of water molecules obtained diffusion constants of $D_{\rm H^+}=$ \SIrange{0.3}{0.6}{\ang^2/ps} \cite{Tse2015} and of $D_{\rm H^+}=$ \SIrange{0.3}{0.8}{\ang^2/ps} \cite{Arntsen2021}. 
A study employing Car-Parrinello molecular-dynamics(CPMD) simulations of \SI{1.7}{M} \ac{HCl} solutions and the PBE exchange-correlation functional found $D_{\rm H^+}=$ \SIrange{0.9}{1.1}{\ang^2/ps} compared to $D_{\rm O}=$ \SIrange{0.04}{0.07}{\ang^2/ps} for the water molecules \cite{Fischer2019}. 
An older study of a single excess proton in a box of 64 water molecules using CPMD with the BLYP functional obtained $D_{\rm H^+}=$ \SIrange{0.05}{0.08}{\ang^2/ps} compared to  $D_{\rm O}=$ \SIrange{0.02}{0.06}{\ang^2/ps} for the water molecules \cite{Izvekov2005}. 
The inclusion of nuclear quantum effects in simulations has been shown to significantly increase the obtained excess-proton diffusion constants \cite{Biswas2016}.

We conclude that the accurate estimation of diffusion constants for the excess proton remains challenging, which is seen from the wide spread of results obtained in previous studies. 
Our results for the diffusion constants appear reasonable when compared to previously reported values and specifically the ratios $D_{\rm H^+}/D_{\rm O} = \SIrange{3.0}{4.3}{}$ are in good agreement with experiment.

\newpage

\section{Hydrogen bond structure}
\label{hbSection}

The \ac{HB} structure of the water molecules in the first hydration shell of the excess proton has been shown to play an important role for when and where the excess proton moves \cite{Tse2015, Biswas2016, Fischer2019, Napoli2018}.
In this section the \ac{HB} structure around the excess protons in \ac{HCl} solutions obtained in our ab initio simulations is discussed using some previously established methods. We follow the standard criterion and define \acp{HB} to be present when the distance between two oxygens, i.e. the donor and the acceptor of the \ac{HB}, is $<\SI{3.5}{\ang}$, and the angle between the vector connecting the two oxygens and the vector connecting the donor oxygen with the hydrogen atom is $<\SI{30}{\degree}$ \cite{Luzar1996}.
\subsection*{Hydrogen-bond asymmetry}
\citet{Napoli2018} introduced the \ac{HB} asymmetry $\phi$ as a measure to identify the excess proton among the candidate protons in a hydronium ion.
First, each water molecule is assigned a coordination number as the difference of the number of acceptor \acp{HB} minus the number of donor \acp{HB}.
Then $\phi$ is defined for each candidate proton at each time step as the coordination number of the closest neighboring water molecule minus the average of the coordination numbers of the closest water molecules of the other candidate protons of the same hydronium ion.
By construction, the sum of $\phi$ within a hydronium ion is zero.
\citet{Napoli2018} found that protons with strongly negative values of $\phi$ show the typical spectral signatures associated with excess protons.
We applied this measure to our simulation data, specifically to the joint trajectories of the excess protons that were prepared for the analysis of the long time diffusion in section \ref{diffusionSection} and from which the spurious `special pair dance' was removed.

\begin{figure}[b]
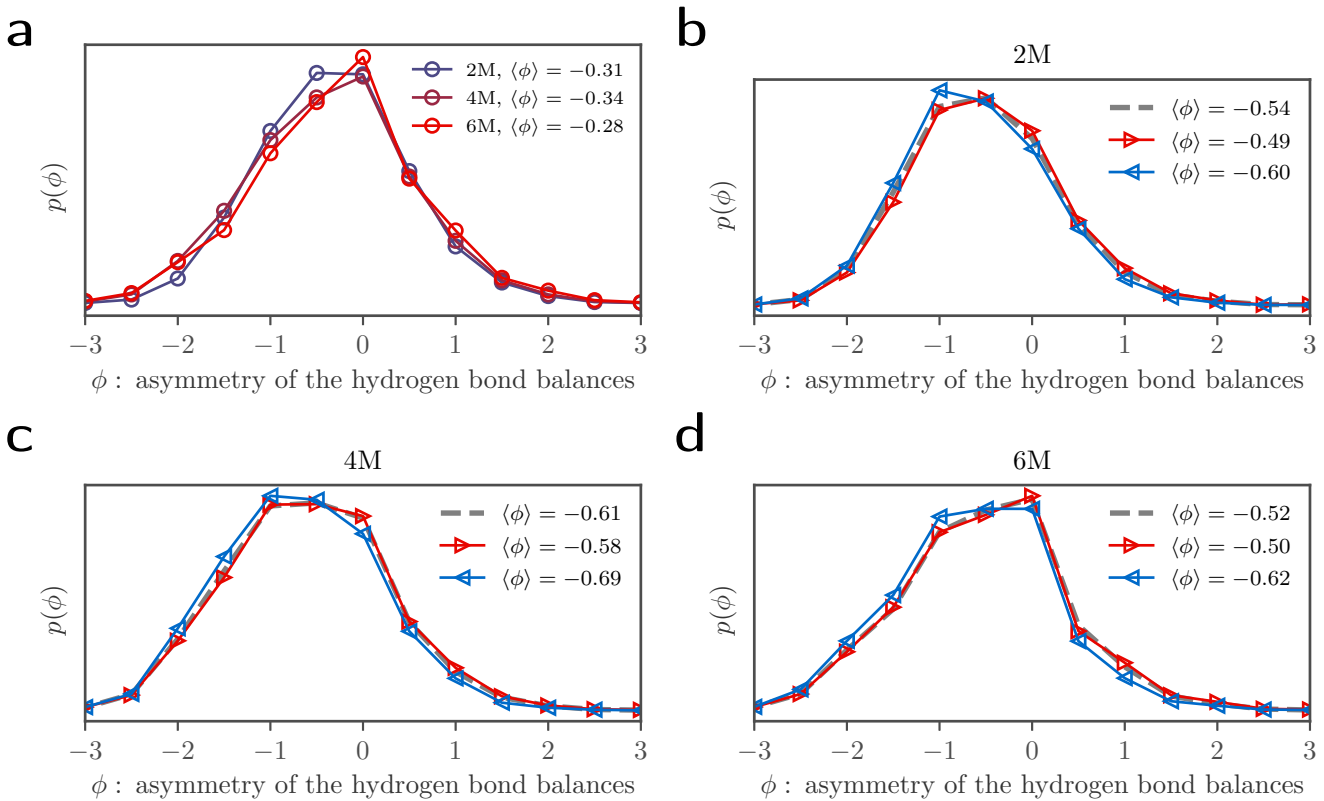

	\centering
	\begin{overpic}[width=.49\linewidth]{{/../figs/protons_hbonds_tp_phi-eps-converted-to}.pdf}
	\put(-2,57){\huge \bfseries \sffamily  a}
	\end{overpic}
	\begin{overpic}[width=.49\linewidth]{{/../figs/protons_hbonds_tp_phi_2M-eps-converted-to}.pdf}
	\put(-2,57){\huge \bfseries \sffamily  b}
	\end{overpic}
	\begin{overpic}[width=.49\linewidth]{{/../figs/protons_hbonds_tp_phi_4M-eps-converted-to}.pdf}
	\put(-2,57){\huge \bfseries \sffamily  c}
	\end{overpic}		
	\begin{overpic}[width=.49\linewidth]{{/../figs/protons_hbonds_tp_phi_6M-eps-converted-to}.pdf}
	\put(-2,57){\huge \bfseries \sffamily  d}
	\end{overpic}	
	\caption{Normalized distributions of the hydrogen bond asymmetries, denoted as $\phi$, of the excess proton trajectories from ab initio \ac{MD} simulations of \ac{HCl} solutions at various concentrations. 
See text for definition and details. Mean values of the distributions are reported in the legends.
{\bfseries \sffamily  a:} Distributions over the whole trajectories.
{\bfseries \sffamily  b--d:} Distributions of $\phi$ during \SI{20}{fs} before and after any transfer event (grey broken lines), i.e. when the excess proton changes the closest oxygen and distributions of $\phi$ around each uni-directional transfer event, which are split into the \SI{20}{fs} before, corresponding to the donor oxygen of the proton transfer event (blue and left pointing triangles) and the \SI{20}{fs} after, corresponding to the acceptor oxygen (red and right pointing triangles).}
\label{fig:protons_phi}
\end{figure}

Normalized distributions of $\phi$ over the whole simulations are shown in fig.~\ref{fig:protons_phi}a with mean values given in the legend. 
The mean values are clearly negative and we therefore conclude that this criterion agrees with our excess proton identification scheme.
To study the relation of hydrogen bond structure with proton transfer, we filter the trajectories in time and regard the normalized distributions of $\phi$ values around the transfer events, when an excess proton changes its closest oxygen. 
Distributions of $\phi$ during \SI{20}{fs} (40 time steps) before and after a transfer event are given in fig.~\ref{fig:protons_phi}b--d as grey broken lines with mean values reported in the legends. 
To further elucidate the data, we applied the same filter before and after each uni-directional transfer event. 
Uni-directional here refers to transfer events, that are not followed by a fast return back to the original oxygen atom within the following \SI{50}{fs} (100 time steps).
By only regarding these uni-directional events, we can discriminate different \ac{HB} configurations around the donor oxygen of the proton-transfer event, i.e. before the uni-directional transfer (shown in blue with left pointing triangles), and the acceptor oxygen, i.e. after the uni-directional transfer (shown in red with right pointing triangles), which show distinct distributions in fig.~\ref{fig:protons_phi}b--d.
The mean values of the distributions before a transfer are more negative than the corresponding averages in fig.~\ref{fig:protons_phi}a, indicating that on average a more negative value of $\phi$ precedes an imminent transfer event. 
Therefore, $\phi$ actually seems to be a useful predictor for proton transfer.

\begin{figure}[b]
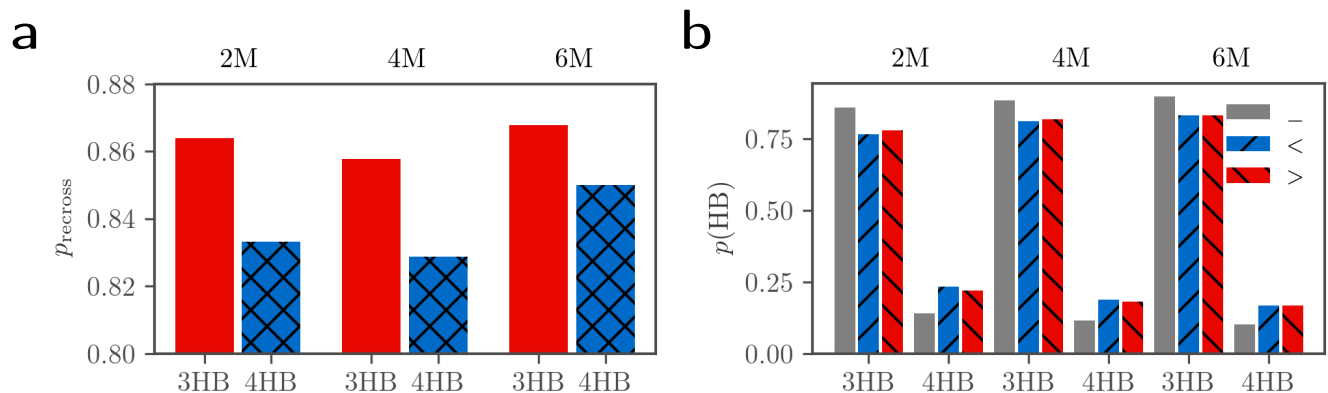

	\centering
	\begin{overpic}[width=.49\linewidth]{{/../figs/protons_hbonds_tp_hb4th_prob_all}.png}
	\put(-2,57){\huge \bfseries \sffamily  a}
	\end{overpic}	
	\begin{overpic}[width=.49\linewidth]{{/../figs/protons_hbonds_tp_hb4th_all}.png}
	\put(-2,57){\huge \bfseries \sffamily  b}
	\end{overpic}	
	\caption{Correlations between the existence of a fourth hydrogen bond of a hydronium ion (4HB) and the return probability of the excess proton, as observed in ab initio \ac{MD} simulations of \ac{HCl} solutions at various concentrations. 
{\bfseries \sffamily  a:} Probability of the excess proton to return across the mid plane between the oxygens within the following \SI{50}{fs}.
{\bfseries \sffamily  b:} Time-averaged probabilities of all hydronium ions for the fourth hydrogen bond to exist (4HB) or not (3HB), as obtained from the whole simulation (grey, no hatching) and during the \SI{20}{fs} before (blue) and after (red) each uni-directional transfer event.}
\label{fig:protons_hb4th}
\end{figure}

\subsection*{The fourth hydrogen bond}

Inspired by \citet{Tse2015}, \citet{Biswas2016} and \citet{Fischer2019} investigated the role of a fourth water molecule hydrogen-bonded to the hydronium ion. 
Presumably, its presence determines whether an excess proton would transfer to a different oxygen atom \cite{Biswas2016} and in particular whether it is likely to return to the original oxygen atom, i.e. to perform back-and-forth transfers, or not  \cite{Fischer2019}.
In their simulations of \SI{1.7}{M} \ac{HCl} solutions, \citet{Fischer2019} observed a higher return probability (67 \%) if the hydronium ion was coordinated with only three donor \acp{HB} as compared to four \acp{HB} (53 \%), the additional one being an accepted \ac{HB} from a fourth water molecule. 
In accordance with previous works, the fourth \ac{HB} is defined to be present if the vector connecting the hydronium oxygen and the hydrogen atom of the fourth water molecule has a length of less than \SI{2.6}{\ang} and points at an angle of less than \ang{35} with respect to the normal of the plane spanned by the three hydrogen nuclei of the hydronium ion \cite{Tse2015,Fischer2019}.
The probability of a return within the following \SI{50}{fs} (100 time steps) upon any transfer across the mid plane between the oxygens obtained in our simulations is illustrated in fig.~\ref{fig:protons_hb4th}a depending on whether the donor hydrogen ion is accepting a \ac{HB} from a fourth water molecule (4HB, shown in blue with hatching) or not (3HB, shown in red without hatching). 
The data shows the clear trend that the return probability is reduced if a fourth \ac{HB} is present, in agreement with previous studies \cite{Fischer2019}.
However, even though this correlation is discernible in our data, the mechanism does not appear to be a dominant driver for proton transfer in \ac{HCl} solutions. 
This follows from fig.~\ref{fig:protons_hb4th}b, where the probability of observing the fourth \ac{HB} is plotted for three different time averages; the averages over the whole simulations of all hydronium ions are given in grey, the averages over the \SI{20}{fs} (40 time steps) before and after each uni-directional transfer event, i.e. without return, are given in blue and red respectively.
Throughout the data, the time-averaged probability of a fourth \ac{HB} is only about 10--20\%. 
Before and after uni-directional transfer events this probability is significantly increased but it is still smaller than the probability of a uni-directional transfer event happening without the presence of a fourth \ac{HB}.

We confirm that \ac{HB} structure is highly correlated with the excess-proton transfer dynamics and the presented comparison with previous studies strengthens the existing hypotheses, that the \ac{HB} asymmetry $\phi$ or the fourth water molecule are useful descriptors.

\clearpage

\section{Wiener-Kintchine theorem}
\label{WienerKintchineSection}

The correlation function $C_{xy}(t)$ of two stochastic processes $x(t)$ and $y(t)$ limited to the time interval $[0,L_t]$ is efficiently computed from the Fourier-transformed expressions $\tilde x(\omega)$ and $\tilde y(\omega)$ according to
\begin{align}
\label{WienerKintchine}
C_{xy}(t) = \frac{1}{2 \pi (L_t-t)} \int_{-\infty}^{\infty} d\omega\ e^{-i\omega t} \tilde x(\omega) \tilde y^*(\omega),
\end{align}
where the asterix denotes the complex conjugate. This is known as the Wiener-Kintchine theorem \cite{Wiener1930}. Both sides of eq.~\eqref{WienerKintchine} are Fourier-transformed to give
\begin{align}
\int_{-\infty}^{\infty} dt\ e^{i\omega t}\ 2 \pi L_t \left( 1-\frac{t}{L_t}\right) C_{xy}(t) &= \tilde x(\omega) \tilde y^*(\omega),
\end{align}
which in the limit of large $L_t$ reduces to
\begin{align}
\label{WienerKintchineFT}
\widetilde C_{xy}(\omega) &= L_t^{-1} \tilde x(\omega) \tilde y^*(\omega).
\end{align}
Eq.~\eqref{WienerKintchine} can be derived starting off with the definition of the correlation function
\begin{align}
C_{xy}(t) = \frac{1}{L_t-t}\int_{0}^{L_t-t} dt'\ x(t'+t) y(t').
\end{align}
Defining $x(t), y(t)=0$ for $t \in [0,L_t]$, the integral bounds can formally be extended
\begin{align}
C_{xy}(t) = \frac{1}{L_t-t}\int_{-\infty}^{\infty} dt'\ x(t'+t) y(t'),
\end{align}
and making use of the convolution theorem
\begin{align}
C_{xy}(t) &= \frac{1}{4 \pi^2 (L_t-t)} \int_{-\infty}^{\infty}dt' \int_{-\infty}^{\infty} d\omega\ e^{-i\omega (t+t')} \tilde x(\omega) \int_{-\infty}^{\infty} d\omega'\ e^{-i\omega' t'} \tilde y(\omega') \nonumber \\
 &= \frac{1}{4 \pi^2 (L_t-t)} \int_{-\infty}^{\infty} d\omega\ e^{-i\omega t} \tilde x(\omega) \int_{-\infty}^{\infty} d\omega'\ \tilde y(\omega') \int_{-\infty}^{\infty} dt'  e^{-it' (\omega+\omega')} \nonumber \\
 &= \frac{1}{4 \pi^2 (L_t-t)} \int_{-\infty}^{\infty} d\omega\ e^{-i\omega t} \tilde x(\omega) \int_{-\infty}^{\infty} d\omega'\ \tilde y(\omega')\ 2 \pi \delta(\omega+\omega') \nonumber \\
 &= \frac{1}{2 \pi (L_t-t)} \int_{-\infty}^{\infty} d\omega\ e^{-i\omega t} \tilde x(\omega) \tilde y(-\omega),
\end{align}
noting that $\tilde y(-\omega)=\tilde y^*(\omega)$ for a real function $y(t)$ in order to obtain eq.~\eqref{WienerKintchine}.

\clearpage

\begin{acronym}[Bash]
\acro{FTIR}{Fourier-transform infrared}
\acro{HB}{hydrogen bond}
\acro{HCl}{hydrochloric acid}
\acro{IR}{infrared}
\acro{MD}{molecular dynamics}
\acro{MSD}{mean squared displacement}
\acro{RDF}{radial distribution function}
\end{acronym}

\bibliography{bibliography.bib}